\documentclass[10pt, final]{york-thesis}

\usepackage{amsmath}
\usepackage{amssymb}
\usepackage{cite}
\usepackage{indentfirst}
\usepackage{verbatim} 
\usepackage{graphicx}
\usepackage{slashed}
\usepackage{epigraph}
\usepackage{verbatim}
\usepackage{fancyhdr}
\usepackage{rotating}
\usepackage[vcentremath]{youngtab}
\renewcommand\headheight{24pt}

\def\bra{{\langle}}
\def\ket{{\rangle}}
\def\e{\textrm{e}}
\def\F!{\;\;\;\;\;\;\,\,}
\def\H!{\;\;\;\,}
\def\cL{{\cal L}}      
\def\cH{{\cal H}}      
\def\hH{\hat{H}}       
\def\hP{\hat{P}}
\def\ds{\displaystyle}
\def\di{{\partial}}
\def\xp{{{ x}^{\prime}}}
\def\xpp{{{ x}^{\prime\prime}}}
\def\xppp{{{ x}^{\prime\prime\prime}}}
\def\zp{{{ z}^{\prime}}}
\def\zpp{{{ z}^{\prime\prime}}}
\def\zppp{{{ z}^{\prime\prime\prime}}}
\def\x{{\bf x}}        
\def\z{{\bf z}} 
\def\r{{\bf r}}
\def\w{{\bf w}}
\def\v{{\bf v}}
\def\p{{\bf p}}
\def\q{{\bf q}}
\def\pp{\p^\prime}
\def\nn{\nonumber}
\def\cY{{\cal Y}}           
\def\cC{{\cal C}}           
\def\cQ{{\cal Q}}
\def\cF{{\cal F}}
\def\cG{{\cal G}}
\def\de{\delta}
\def\tx{\textrm}
\def\E1{\textrm{E}_{1}}     
\def\vr{\varrho}            
\def\b{\beta}
\def\erf{\text{erf}}
\def\bs{\boldsymbol}
\def\ub{{\bar u}}
\def\vb{{\bar v}}
\def\psib{{\bar \psi}}
\def\ss{s^{\prime}}
\def\rr{r^{\prime}}
\def\tt{t^{\prime}}

\setboolean{masters}{false}     
\setboolean{hasfigures}{false}   
\setboolean{hastables}{false}   
\title{Inter-Particle Potentials in Non-Linear Quantum Field Theories}   
\author{Alexander Chigodaev}  
\department{Physics and Astronomy}   
\masterof{Arts}           
\degreename{Doctorate}    
\date{May 2012}         

\abstractfile{abstract.tex}
\dedicationfile{dedication.tex}
\acknowledgementsfile{acknowledgements.tex}
\committeememberslist{
  \begin{enumerate}
  \item Wendy Taylor
  \item Roman Koniuk
  \item Marko Horbatsch
  \item Radu Campeanu
  \item Manu Paranjape 
\end{enumerate}}

\begin{document}

\makefrontmatter

%
%

\chapter{Introduction}
\spacing{1}
\epigraph{Physics - where the action is.}{Unknown}
\spacing{2}
\section{The Present State of Affairs in Particle Physics}
The science of Particle Physics is concerned with the description of nature at the most fundamental level. The foundation, on which it rests, was laid down with the invention of Quantum Mechanics (QM) and Quantum Field Theory (QFT). These two frameworks provide, as it is believed, the correct description of the physical laws which govern the microscopic world. 

QM mechanics is based on principles which are non-intuitive in the macroscopic world. A particle or a system of particles is described by state vectors; in coordinate space they are known as wavefunctions. If a state vector is known then the measurable observables, such as position, momentum and energy, can be determined as statistical averages. The concept of a trajectory of a particle in Classical Mechanics is replaced by probability density, derived from state vectors, of a particle to be at a specific point in space and time. QM can be extended to the relativistic realm for highly energetic systems in a natural way. A primary application of QM is in the treatment of atoms and molecules which has given chemistry its modern appearance.

QFT is an extension of QM to situations where the number of particles, not necessarily elementary, may not remain conserved, i.e. particle creation and annihilation is permitted. Its novelty is to treat the mediating fields on par with particle fields. It is motivated by the belief that the vacuum is filled with virtual particles which can become physical (i.e. go on-the-mass-shell) provided that all conservation laws remain stringent. The experimental study of elementary particles requires large amounts of energy, therefore particle physicists are appropriately interested in relativistic QFTs. The structure of QFT comprises many branches of mathematics and to have a grasp of it requires many years of dedicated learning. 

The Standard Model (SM) of Particle Physics is a multi-component QFT which describes the elementary particles of nature and their interactions. The interactions among elementary particles are mediated by the electromagnetic, weak, and strong mediating fields which are included in the SM. 
The SM has been established by conducting experiments which inter-played with theory to exalt the SM to its current form. The objective of the present day experiments, for instance the ones at the Large Hadron Collider, is to go beyond the domain of validity of the SM and possibly discover contradictions to the SM predictions. Discoveries violating the SM would signal the existence of the highly anticipated physics beyond the SM (BSM). More importantly such discoveries could lead to possible resolutions of the shortcomings of the SM.

The first successful QFT was Quantum Electrodynamics (QED) where electrically charged particles interact via the mediating photon field. QED paved the way for the perturbative method of calculating quantum amplitudes for scattering cross sections and decay rates. The famous Feynman diagrams, historically stemming from perturbative QED, have become an indispensable tool in particle physics. In addition, the method of renormalization, the process of removing infinities in a systematic and meaningful manner, was also initially proposed in the context of QED. Basically, QED has become a template for all subsequent QFTs in particle physics. 

The early scattering experiments in the 1940s and 1950s discovered numerous, then thought to be elementary, particles known as hadrons. At that time, it was dubbed the ``particle zoo`` for the lack of explanation of their large number. As more and more hadrons were being discovered, physicists began to suspect that they are not truly elementary (i.e. structureless)~\footnote{In fact, it is not certain whether the particles believed to be elementary today are really such.}. They organized these particles into various categories and introduced new quantum numbers for this purpose. Shortly, it was realized that symmetries played an essential role in categorizing hadrons. The Parton Model, in which constituent partons are treated as being entirely free inside of hadrons, and the advent of gauge symmetries led to the invention of Quantum Chromodynamics (QCD) where the fundamental massive quarks interact with each other via the massless gluons. 

Simultaneously, particle theory was readily extended beyond QED to account for the existence of neutrinos. More importantly, it unified the electromagnetic and weak interactions into a new Electroweak Theory (EWT). In this theory, neutrinos are electrically neutral elementary particles that interact via the weak interaction only. Presently, experiments indicate that there are three flavours of nearly massless neutrinos. The full incorporation of neutrinos into the SM remains problematic because of their non-zero mass. This predicament, along with many others, motivates to seek BSMs. In addition, the $W$ and $Z$ bosons, the quanta of the weak interaction fields, were discovered, as predicted by the EWT. There is much debate about the SM in the particle physics community but, despite some significant shortcomings and inconsistencies, it remains the most successful description of nature. For recent comprehensive summaries of the SM, see references~\cite{Djouadi:2005gi,PDBook}.

The shortcomings of the SM may be resolved by the BSMs. A principal question is the lack of explanation as to why elementary particles have masses. It appears that to preserve the local gauge invariance of the theory all particles must be massless. The as yet undiscovered Higgs boson, and the spontaneous symmetry breaking mechanism associated with it, circumvents this requirement~\cite{HHG}. However, as the experimentalists sift through more data at the Large Hadron Collider, the window on the Higgs boson mass is being gradually reduced. As an additional challenge, it appeals to theorists to explain the mass hierarchy of particles - the mass range from neutrinos to the top quark spans $11$ orders of magnitude. Moreover, the experimental observation of the rotation of galaxies suggests that there are new heavy particles, the so-called dark matter, which do not interact via the electromagnetic or strong interactions and have not been produced in colliders to date. There is no dark matter candidate in the SM as it does not address the dark matter problem. Lastly, neutrinos are assumed to be massless in the SM, which is contrary to the experimental observation. 

A long-standing problem of theoretical physics is to reconcile General Relativity, the classical theory of gravitation, with QFT. The desire for a Quantum Theory of Gravity (QTG) originates from the ambitious endeavour to describe all physics in one unified theory. However, given the energy of the unification scale\footnote{It is believed that the arrangement of the fundamental constants $\ds\sqrt\frac{\hbar \, c^5}{G}$, which has the dimension of energy, is actually related to the scale where the gravity effects on the microscopic world become large.} $\Lambda_\tx{Planck} \approx 10^{19} \, \textrm{GeV}$ and the collision energy of today's colliders $\Lambda_\tx{Collider} \approx 10^3 \, \tx{GeV}$, it seems that in the near and intermediate future no experimental testing of QTGs is foreseen. Presently, the only possible arena of testing QTGs is in observational astronomy and cosmology. Now, it is plausible there exist new interactions and particles below $\Lambda_\tx{Plank}$ thereby conceivably thwarting the present idea of unification in particle physics alone~\cite{Georgi:1974sy,Georgi:1974yf}. It can not be excluded altogether that the framework of current QFT itself becomes obsolete at some high energy scale, and consequently the idea of unification could become erroneous. In the author's view, it is rather premature to address the problem of unification given the present state of technology and knowledge.

Returning back to the current state of affairs, the majority of particle physicists are nonetheless concerned with phenomenology and model-building, which are still very important fields. However, in the author's opinion, it is even more fundamental to be able to solve QFT fully. This might be the key to understanding QFT better. Perhaps, the next interaction among new particles to be discovered is non-perturbative as well as non-linear. One needs to know how to calculate cross-sections, decay rates and energy spectra reliably in such cases. These issues are already noticeable in QCD where the strong coupling and the non-linear mediating field have triggered new ideas and techniques of calculation. In the view of the author, the fundamental task is to develop universal methods of calculation, regardless of the coupling strength and the interaction types. If such tools are invented, it would be a major step forward.

\section{The Bound State Problem in QFTs}
The analytic solution of the Schr\"odinger equation for the hydrogen-like atom problem in QM served as one of the primary confirmations of quantum theory. The Dirac equation (and to a lesser extent the Klein-Gordon equation) being relativistic generalization of the Schr\"odinger equation, incorporates the ideas of anti-matter and spin (the intrinsic angular momentum of a particle) which have become an integral part of physics. This equation serves as a starting point in the development of realistic QFT.  The Dirac equation has provided the correct ${\cal O}(\alpha^4)$ contributions\footnote{The perturbative expansion parameter $\alpha  \approx \ds\frac{1}{137}$ is the coupling constant of QED.} to the energy spectrum of hydrogen~\cite{Bjorken1964} which can not be accounted for in QM alone. The Dirac and Klein-Gordon equations have become the standard pedagogy.

The inter-particle interactions, entering into these bound state equations, arise from classical electromagnetism. The Coulombic potential describes such bound states adequately in the non-relativistic limit. The number of particles in these equations are fixed; yet there can be an arbitrary number of them. For instance, the Breit equation~\cite{PhysRev.39.616}, being a many-particle Dirac-like equation, describes systems with more than one electron in the vicinity of a heavy nucleus. It includes a Coulombic potential along with spin-orbit, spin-spin and the other ${\cal O}(\alpha^4)$ corrections. The main weakness of the Breit equation is that it is not invariant with respect to Lorentz transformations (i.e is not Lorentz covariant), and therefore it describes the relativistic effects perturbatively to order ${\cal O}(\alpha^4)$ (excluding field theoretical effects such as virtual annihilation). 

In QED, the mediators of the electromagnetic interaction are photons. Photons are massless, chargeless, spin one vector bosons which are classically described by the Maxwell equation. The Proca equation~\cite{Pauli:1941zz} is an extension of the Maxwell equation to include the case of massive vector bosons. However, the principle of local gauge invariance (i.e. that is the QED Lagrangian must have no photon mass term in order to be gauge invariant) provides a sufficient theoretical explanation as to why photons should be massless. Thus, even though one can write the Proca equation for massive photons, it currently has little practical use. In addition, no bound states of photons are observed in nature. This is consistent with the absence of self-interaction terms for photons in the QED Lagrangian. In contrast, electrically charged massive particles can exist as free or in bound states.

The search for a quantum field theoretic description of bound states of massive particles in QED has led to the Bethe-Salpeter (BS) equation~\cite{PhysRev.84.1232, Nakanishi:1969ph}. It is a fully relativistic covariant equation which can be cast in integral or differential form. Being such, it is not surprising that the equation can not be solved analytically for any realistic problem. The usual perturbative expansion of the kernel of the BS equation (i.e. the interaction) and its truncation to the given order makes it only an approximate description. Furthermore, it contains negative energy solutions as well as relative time coordinates pertaining to the constituent particles. The BS equation produces accurate perturbative results for bound systems of two massive particles in QED. However, its applicability to three and more particles becomes increasingly difficult. This is all the more so for QCD~\cite{Lucha:1991vn}.

In the early days of particle physics, before the discovery of the strong interactions, there was no particular reason to study bound states in strongly interacting theories with the notable exception of heavy atoms. The establishment of QCD has profoundly changed this attitude because bound states are an important aspect in this theory. The interactions among the quarks (i.e. massive spin one-half quanta of the strong matter fields) and gluons (i.e. massless spin one quanta of the strong mediating field) originate from the non-Abelian gauge symmetry of the theory. It means that, in addition to the QED-like linear interaction between quarks and gluons, there are two non-linear gluon self-interaction terms of order three and four respectively. In light of these new terms, the effect on the running of the coupling constant in QCD is opposite to that in QED~\cite{Gross:1973id, Politzer:1973fx}. In QED, the running of the coupling constant is attributed to screening by virtual pairs hence the opposite behaviour in QCD is suggestive of anti-screening. At large energies the coupling strength decreases - a property known as asymptotic freedom. On the other hand, the coupling strength increases at lower energies and becomes non-perturbative at the scale $\Lambda_\tx{QCD} \approx 200 \, \textrm{MeV}$ - this is known as infrared slavery. 

The discovery of asymptotic freedom  was deemed worthy of the Nobel Prize in 2004 (Gross, Politzer, Wilczek) thus it deserves to be reviewed in some detail. The beyond-leading-perturbative-order corrections to the correlation functions\footnote{Correlation functions are the expectation values of time-ordered products of field operators.} are ultravioletly divergent in most QFTs. Nonetheless, for certain types of QFT, the ultraviolet infinities can be absorbed into the parameters of the theory (i.e. coupling constants and mass parameters) in a consistent and meaningful manner by introducing field renormalization. The behaviour of the correlation functions and the parameters of the theory as the energy scale varies is described by the Callan-Symanzik differential equation~\cite{Callan1970,Symanzik:1970rt}. The Callan-Symanzik equation for massless QCD is
\begin{align}
  \left[\frac{\di}{\di \, \log \,\mu} + \beta(g) \, \frac{\di}{\di g} + n \, \gamma_A (g) + m \, \gamma_\psi(g)  \right] \, G^{(n, m)} \, = 0,
\end{align}
where $n$ and $m$ are the number of the quark and gluons fields in the correlation function, the parameter $\mu$ is the renormalization scale, the function $\beta(g)$ is simply entitled the ``beta'' function, whereas $\gamma_A(g)$ and $\gamma_\psi(g)$ are called the anomalous dimensions of the quark and gluon fields respectively. These functions can be calculated to a given order in perturbation theory.

The massless QCD Callan-Symanzik equation has a solution provided the following subsidiary condition holds:
\begin{equation}
  \frac{\partial \, g(\lambda)}{\partial \log\lambda} = \beta(g),
  \label{EQ:RENORM_GROUP}
\end{equation}
where $\lambda = \ds\frac{\Lambda_R}{\Lambda_0}$ is the ratio of the renormalized $\Lambda_R$ and bare $\Lambda_0$ scales. This equation is known as the group renormalization equation and asserts that the beta function is just the rate of change of the coupling constant with respect to the change of scale.  The beta function can be obtained by evaluating Feynman diagrams using a regularization scheme (such as dimensional regularization) to isolate the ultraviolet divergences. There are seven diagrams which must be calculated in order to extract the leading term of the beta function in QCD. The result of a detailed calculation yields~\cite{Gross:1973id, Politzer:1973fx}:
\begin{align}
  \beta(g) = - \frac{7 \, g^3}{(4\, \pi)^2} + {\cal O}(g^4),
  \label{EQ:BETA}
\end{align}
where the negative sign reflects the behaviour of the coupling constant as stated above. One can solve the differential equation (\ref{EQ:RENORM_GROUP}) with the approximation (\ref{EQ:BETA}) to obtain 
\begin{align}
  g^2(\lambda) = \frac{g_0^2}{1 + 7 \, g^2_0 \, \ds{\log(\lambda) / (8 \, \pi^2)}},
  \label{EQ:ALPHA_S}
\end{align}
where $g_0$ is the value of $g$ measured at $\lambda = 1$. According to this solution, the coupling constant is stronger for small $\lambda$. Furthermore, it conforms with the experimental observation that the interaction is stronger at the lower energy scale. After all, the masses of the known hadrons formed by composite quarks and gluons seem to be all below about $10 \, \tx{GeV}$~\cite{PDBook} . The running of the coupling is a theoretical hint that quarks and gluons should be confined. 

The phenomenon of confinement of quarks and gluons in bound states is an experimental fact. It should be emphasized that no free quarks and gluons have been found in nature. It is no exaggeration, then, to claim that the QCD Lagrangian is written in terms of the ``wrong'' degrees of freedom (i.e. quark and gluon fields instead of meson and baryon fields). Indeed, it is a big mystery why it should be like that. Moreover, the masses of the bound states of quarks are greater than the sum of its constituent quark masses. This is a characteristic of confinement and a defining property of QCD (but not QED). The theory indicates that it is important to develop non-perturbative techniques of calculation. A very ingenious method called Lattice QCD (LQCD) has been developed to study the hadron spectrum of the strong interaction precisely~\cite{Wilson:1974sk, Rothe:1992}.

In LQCD, continuous space-time is replaced by a finite grid of space-time points. Consequently, the action becomes discretized with the quark fields defined at each lattice point connected by the link variables representing the gauge bosons. The discretization requires a momentum cutoff on the order of the inverse lattice spacing $a^{-1}$ to regularize the theory. The continuum limit can be approached as the spacing $a$ is reduced to zero which retrieves the exact QCD. The LQCD formulation is in principle gauge invariant but the practical implementation requires an extensive computational effort. To calculate the mass spectrum of hadrons one numerically calculates the correlation functions for given quantum numbers and then fits them to exponential curves~\cite{Brambilla:1999ja}. The results of LQCD are highly successful and are presently considered the best tool for describing the QCD bound spectrum.

The confined bound states in QCD are a feature that still awaits its ultimate explanation. There is no rigorous and unambiguous explanation why quarks are confined. Identically, there is no satisfactory explanation of the phenomenon of ``string breaking'' (i.e. the observation that hadrons do not separate into free quarks but rather create more hadrons if provided with sufficient energy). Non-relativistically, the existence of confinement must correspond to an appropriate shape of the quark-antiquark potential. The potential must rise at large separations to accord with the strong attraction and consequently explain the confining bound states. 

LQCD provides a way to calculate the potential for a static quark-antiquark pair by calculating the expectation value of the Wilson loop\footnote{In the early days, due to the complications induced by dealing with the quark fields, the quark fields were absent in the Wilson loop.}~\cite{Bali:2000gf}:
\begin{align}
  \bra W_C \ket  = Z^{-1} \int {\cal D} U \, {\cal D} {q} \, {\cal D} \bar{q} \; W_C \, \e^{-S}, \H! \textrm{where} \H! W_C = \textrm{Tr} \prod_{x, \mu \in P} U_{x, \mu}
\end{align}
where $Z$ is the QCD generating functional, $U$ is the link variable connecting neighbouring sites on a lattice, $q$ and $\bar{q}$ are the quark and anti-quark fields respectively, $S$ is the discretized Euclidean action, $P$ denotes a particular enclosed path on the lattice and the trace is calculated over the gauge indices of $U$s. The static quark-antiquark potential can be then calculated via
\begin{align}
  V(r) = - \lim_{T \rightarrow \infty} \, \frac{1}{T} \, \log\bra W_C \ket. 
  \label{EQ:VOFW}
\end{align}
The calculation of the static quark-antiquark potential using equation (\ref{EQ:VOFW}) is a laborious numerical task. Nonetheless, it has been done and the resulting curves can be found in references~\cite{Rothe:1992, Brambilla:1999ja, Bali:2000gf}. 

The form of the quark-antiquark potential has also been postulated in phenomenological models. The original model~\cite{Eichten:1978tg}, which has become known as the Cornell potential, serves as a good description for the low energy bound states of heavy quarks. A more general treatment of the mesonic bound states is presented in the work by Godfrey and Isgur~\cite{PhysRevD.32.189}. They confirmed the existence of known and predicted hundreds, then-undiscovered, bound states of mesons by incorporating relativistic effects in the lowest order. Another extensive discussion on the nature of the quark-antiquark potential along with its spin, angular momentum and delta function corrections is provided by Lucha, Schoberl and Gromes in reference~\cite{Lucha:1991vn}.  A more recent and realistic model~\cite{Li:2009nr} accounts for the string breaking effects by requiring the potential to soften at large distances (i.e. asymptotically reach a constant as the separation increases). Although such phenomenological models are effective, they all suffer from the fact that the underlying interactions are inserted by hand. As already mentioned, it is an outstanding problem in particle physics to derive the quark-antiquark potential in QCD from first principles. It is a challenge that keeps theorists preoccupied. The numerous attempts and techniques of various complexity to derive the potential are discussed in the book by Greensite~\cite{greensite2011an}. However, even Greensite's book is not a complete review of the subject; as the author says ``to include every proposal would require an encyclopedia'' (\cite{greensite2011an} p. 207). There is a biannual conference devoted to the subject of ``Quark Confinement and the Hadron Spectrum''. Proceedings of the conferences are published regularly, the most recent one being ~\cite{Ribeiro:2008zz}; these present current updates of the status of the field.


%
The research presented in this dissertation is concerned with the development of a technique for calculating the inter-particle potentials in QFTs containing non-linear mediating fields. The inter-particle potential provides a means of calculating the bound state spectrum and may indicate whether a QFT is confining or not. The existing method of determining the inter-particle potential based on LQCD, in the view of the author, is rather abstract and requires prolonged numerical calculations. It is then prudent to seek alternative methods to acquire the inter-particle potentials in QFTs. The basis of a such method, used in this dissertation, are discussed in the next section. 

\section{The Method and Dissertation Outline}
The derivation of the inter-particle potential is implemented in the Hamiltonian formalism. The classical equations of motion and their formal solutions for the mediating fields are used to reformulate the original Hamiltonian. The resultant Hamiltonian has a dependence on the mediating field via the Green functions only. There are two advantages to performing this reformulation. First, there is no requirement to quantize the mediating field since only bound states of particles are of concern. Second, the derived relativistic few-particle equations contain functions describing particle quanta only; without the reformulation the equations would be coupled in the fields that describe both particles and mediators. Thus, the complexity of the principal equations is reduced by the reformulated Hamiltonian.

The primary subject of interest are theories in which the mediating fields are non-linear; in particular a scalar model with a Higgs-like mediating field as well as QCD. To study the effect of the non-linear interaction terms on the inter-particle potential the following scheme is implemented. The matrix elements of the Hamiltonian operator are calculated by using suitable trial states which describe few-particle bound systems or their elastic scattering (the latter is not addressed in this dissertation). It is possible to probe some or all interaction terms in the Hamiltonian by selecting different trial states. Once the matrix elements for the desired bound systems are found, the variational method is applied and relativistic equations, in momentum space, for the few-particle systems under consideration are obtained. These equations provide an approximate description of the bound state systems in question. In the Hamiltonian formalism, the bound state energy is bounded from below~\cite{Ballentine1989} and this provides the possibility for systematic improvement of the variational approximation. The interaction terms of these equations are examined as is described below.

A scalar model containing a Higgs-like non-linear mediating field is studied in the second chapter. The results of reformulation and quantization are presented in some detail. The bound systems of a particle-antiparticle pair, three-particle, and four-particle systems are investigated at first. The trial states for these systems are mono-component in Fock-space and consequently relativistic equations contain only one unknown function (i.e. the approximate wavefunction of the system being studied). The interaction terms of these equations are reduced to the non-relativistic limit and Fourier-transformed to the coordinate representation. Hence, these terms are just the usual non-relativistic inter-particle potentials and those corresponding to the non-linear terms are in the form of multi-dimensional integrals. It is shown that the inter-particle potentials are dependent on the inter-particle distances only. Unfortunately, there are no simple closed-form expressions for arbitrary positions of the particles. However, considering particular cases with restrictions on the particle positions, it is possible to obtain a general picture of the full inter-particle potential. 

A trial state containing a one-pair (particle-antiparticle) and a two-pair components is investigated thereupon. This trial state is capable, at least in principle, of describing the process of string breaking which has been mentioned in the context of QCD. The inclusion of the second component in the trial state leads to coupled relativistic equations for the functions separately describing the one- and two-pair components of the bound state. The non-relativistic limit and Fourier-transform, along with an ansatz for the four-component wavefunction, yield an intricate expression for the inter-particle potential between the particle and antiparticle of the two-component of the bound state where the dependence on the four-component coordinates gets integrated out. This expression is in the form of a multi-dimensional integral. It is solved numerically to obtain an inter-particle potential which, for the case where the mediating Higgs field is massive, shows no evidence of confinement. 

The third chapter is devoted to QCD. It commences with a review of the Dirac equation and the quantization of QED followed by a succinct review of the QCD Lagrangian and its non-Abelian gauge symmetry. A similar scheme for quantization is followed for QCD as in QED, however, it is complicated by the addition of the various extra indices. The reformulation of QCD is performed and the reformulated Hamiltonian is presented in a fixed gauge. Armed with the knowledge gained from the scalar Higgs-like model, a mono-component trial state describing a three quark system (i.e. a baryon) is considered first. This trial state has the right number of ladder operators to probe the cubic term of the QCD Hamiltonian. Surprisingly, it turns out that the summation over the colour indices (i.e. the colour factors) makes the cubic contribution of the matrix element vanish in the lowest order of iterative approximation. Next, a multi-component trial state describing a quark-antiquark system (i.e. a meson), and possibly string breaking, is considered. The summation over the colour indices indicates that only the non-linear cross terms of the interactions do not vanish. Again, the resulting potential energy terms in the equation that describes a non-relativistic meson are in the form of multi-dimensional integrals. Here, they are further complicated by the presence of momentum scalar vector products in the numerator. Unfortunately, it turns out that the only available technique to solve such multi-dimensional integral expressions, the Monte Carlo method, is incapable of producing stable numerical results. Thus, the best that can be done is to approximate these multi-dimensional integrals by upper bounds, for which the results are stable.

It is worth mentioning that the variational method has been used effectively in the study of few-boson states in the scalar Yukawa theory~\cite{Ding:1999ed, EmamiRazavi:2006yx}, in the Higgs theory~\cite{DiLeo:1994xc}, in few-fermion bound states in relativistic quantum mechanics~\cite{Grandy1991} and QED~\cite{Terekidi:2003gp, Barham:2007vd, Terekidi:2006fq}, and even in QCD~\cite{DiLeo:2000yh, 9971505002}. Many results that appear in chapter two have been published in the preprint~\cite{Chigodaev:2011yr}.

The long and cumbersome derivations and results have been relegated to the appendices to keep the contents more presentable. The summation convention on the repeated indices, where no summation sign is explicitly included, is implied throughout.

\chapter{A Scalar Model with a Higgs-like Mediating Field}
\spacing{1}
  \epigraph{We can't solve problems by using the same kind of thinking we used when we created them.}{Albert Einstein}
\spacing{2}
\section{Reformulation}
Inter-particle potentials in few-particle systems described by a theory with the following non-linear scalar Lagrangian density\footnote{Often in the literature a Lagrangian density is referred to as simply a Lagrangian. The correct terminology will be maintained throughout this dissertation.} shall be addressed in the current chapter (units: $\hbar=c=1$):
\begin{equation} 
  \cL=\di^{\nu}\phi^{\ast} \, \di_{\nu}\phi - m^2 \, \phi^{\ast} \, \phi - g \, \chi \, \phi^{\ast} \phi - \lambda \left(\phi^{\ast} \phi\right)^2 + \frac{1}{2}\left(\di^{\nu}\chi \, \di_{\nu}\chi - \mu^2 \, \chi^2\right) - \frac{1}{3} \, \eta \, \chi^3 - \frac{1}{4} \, \sigma \, \chi^4.
  \label{EQ:SCLLAGR}
\end{equation}
The quantities $g, \eta$ are coupling constants with dimensions of mass, while $\sigma, \lambda > 0$ are dimensionless coupling constants. The parameters $m$ and $\mu$ represent the bare masses of the scalar and mediating field quantum respectively. 

This Lagrangian density is Lorentz covariant and the mediating field $\chi$ mimics the gluon field of QCD if $\mu = 0$. In QCD, the cubic term contains a contraction of the gluon field and its derivative thus the entire term is covariant. There is no such contraction in the cubic term of the scalar Lagrangian density (\ref{EQ:SCLLAGR}). Thus the coupling constant $\eta$ is required to have dimensions of mass to maintain the correct dimensionality of the Lagrangian density. Undeniably, the $\chi$ field is of the form of the Higgs field of the Standard Model. The $\lambda$ term in the Lagrangian density is required to keep the classical ground state energy bounded from below. It leads to a repulsive contact (delta function) inter-particle interaction for few-particle systems~\cite{Beg:1984yh, Darewych:1997uc}. Thus, one can set $\lambda = 0$ since it has negligible effect on the results.

The reformulation corresponds to a partial solution of the equations of motion. The equations of motion corresponding to the Lagrangian density (\ref{EQ:SCLLAGR}), derived from the action principle $\delta\, \ds\int dx\, {\cal L} = 0$, are 
\begin{gather}
  \left(\di^{2} + m^2 \right) \, \phi(x) = \, -g\, \phi(x) \, \chi(x)  
  \label{EQ:PHI}\\
  \left(\di^{2} + \mu^2 \right) \, \chi(x) = \, -g \, \phi^{*}(x) \, \phi(x) - \eta \, \chi^2(x) - \sigma \, \chi^3(x). 
  \label{EQ:CHI}
\end{gather}
Equation (\ref{EQ:CHI}) has the integral representation (i.e. ``formal solution'')
\begin{equation} 
  \chi(x) = \chi_0(x) + \int d\xp \,D(x-\xp)\, \rho(\xp),
  \label{EQ:CHI_FORMAL_SOL}
\end{equation}
where $x = (t, \x)$, $dx = dt \, d\x $, $\rho(x) =  - g \phi^*(x) \phi (x) - \eta \, \chi^2(x) - \sigma \chi^3(x)$ is the ``source term'' of the inhomogeneous equation, $\chi_0 (x)$ satisfies the homogeneous (or free field) equation with the right hand side of (\ref{EQ:CHI}) equal to zero, while $D(x-\xp)$ is a covariant Green function (or the Feynman propagator) for the mediating field $\chi$, such that
\begin{equation}
  \left ( \di^2  + \mu^2 \right ) D(x-\xp) = \delta^{(4)}(x-\xp).
\end{equation}

Recall that the Green function can be expressed as
\begin{equation}
  D(x-\xp)=D(x,\xp)=\lim_{\epsilon \rightarrow 0} \int \frac{dk}{(2\pi)^{4}} \frac{1}{\mu^{2}-k^{2}+i\epsilon} \textrm{e}^{-i k(x-\xp)}.
  \label{EQ:GREEN} 
\end{equation}
where $\epsilon$ specifies the prescription for handling the singularities when integrating over $k^0$.

No processes involving free quanta of the mediating field will be studied, henceforth $\chi_0$ will be left out. Then, substituting (\ref{EQ:CHI_FORMAL_SOL}) into (\ref{EQ:PHI}), omitting terms containing $\chi_0$, one obtains the equation  
\begin{multline}
    \left(\di^2 + m^2\right) \phi(x) = g^2 \phi(x) \int d\xp \phi^*(\xp) \phi (\xp) D(x,\xp) \\ + g \eta \, \phi(x)  \int d\xp \, \chi^2(\xp) \, D(x,\xp) + g \sigma \, \phi(x) \int d\xp \chi^3(\xp) \, D(x,\xp). 
  \label{EQ:MOD_CHI}
\end{multline}
Equation (\ref{EQ:MOD_CHI}) is derivable from the following Lagrangian density 
\begin{multline}
  \cL = \di_{\mu} \phi^{\ast}(x) \, \di^{\mu} \phi (x) - m^2 \, \phi^{\ast}(x) \phi (x) + \frac{1}{2}\,  g^2 \int d\xp \phi^{\ast}(x) \phi(x) D(x,\xp) \phi^{\ast}(\xp) \phi(\xp) \\ +  g\eta \, \phi^{\ast}(x) \phi(x) \int d\xp D(x,\xp)\chi^2(\xp) +   g \sigma \, \phi^{\ast}(x) \phi(x) \int d\xp D(x,\xp)\chi^3(\xp)  
  \label{EQ:MODLAGR}
\end{multline}
provided that the Green function is symmetric, i.e. $D(x-\xp) = D(\xp-x)$.

For the case of a linear mediating field, i.e. when $\eta = \sigma =0$, the reformulated Lagrangian density (\ref{EQ:MODLAGR}) gives field equations that involve the particle fields only; the mediating field $\chi$ appears only through the propagator $D(x,\xp)$. Such a reformulated Lagrangian density, with $\eta = \sigma =0$, is convenient for the study of few-boson relativistic bound states in the scalar Yukawa theory as will be pointed out below.

However, for the case of a non-linear mediating field, i.e. $\eta, \sigma \neq 0$, it is not possible to obtain a closed-form solution of (\ref{EQ:CHI_FORMAL_SOL}) for the field $\chi$ in terms of $D(x,x')$ and $\phi$, thus one requires to resort to approximations. An obvious one is an iterative sequence based on (\ref{EQ:CHI_FORMAL_SOL}). The first-order iterative approximation corresponds to substituting the formal solution (\ref{EQ:CHI_FORMAL_SOL}), with $\rho = - g\, \phi^{\ast} \phi$ (with $\chi_0$ left out), into the Lagrangian density (\ref{EQ:MODLAGR}). This yields the first-order approximate expression for the Lagrangian density
\begin{gather}
  \cL = \di_{\mu} \phi^{\ast}(x) \, \di^{\mu} \phi (x) - m^2 \, \phi^{\ast}(x) \phi (x) + \frac{1}{2}\,  g^2 \int d\xp \phi^{\ast}(x) \phi(x) D(x,\xp) \phi^{\ast}(\xp) \phi(\xp) \nn
  \\ + \;  g\eta \, \phi^{\ast}(x) \phi(x) \int d\xp \, d\zp \, d\zpp  \; D(x,\xp) \, D(\xp, \zp) \, D(\xp, \zpp) \; \rho(\zp) \, \rho(\zpp) \nn 
  \\ + \; g\sigma \, \phi^{\ast}(x) \phi(x) \int d\xp \, d\zp \, d\zpp \, d\zppp \; D(x,\xp) \, D(\xp, \zp) \, D(\xp, \zpp) \, D(\xp, \zppp) \, \rho(\zp) \, \rho(\zpp) \, \rho(\zppp).  
  \label{MOD_II_LAGRANGIAN}
\end{gather}
The interaction terms in the Lagrangian density (\ref{MOD_II_LAGRANGIAN}) do not contain the mediating field $\chi$ explicitly but rather implicitly through the mediating field propagators. The advantage of reformulating the Lagrangian density is that it enables one to use simple few-particle trial states free of the $\chi$ field quanta while still probing the effects of all interaction terms, including the non-linear terms (i.e. those with $\eta, \sigma \neq 0$).  

The Hamiltonian and momentum densities corresponding to (\ref{MOD_II_LAGRANGIAN}) are obtained from the energy-momentum tensor in the usual way:
\begin{equation}
  T^{\mu\nu} = \frac{\di \cL}{\di (\di_{\mu} \phi_i)}\, \di^{\nu}\phi_i - g^{\mu\nu} \,\cL
\end{equation}
where the index $i = 1,2$ stands for the fields $\phi_1 = \phi$ and $\phi_2 = \phi^{\ast}$. The $T^{00}$ component of the energy-momentum tensor is the reformulated Hamiltonian density
\begin{equation}
  \cH = {\dot \phi} \, \Pi_\phi + {\dot \phi}^{\ast} \, \Pi_{\phi^{\ast}}  - \cL =  \cH_{\phi} + \cH_{I_{1}} + \cH_{I_{2}} + \cH_{I_{3}}  
    \label{EQ:MODHAMI}
\end{equation}
where
\begin{gather}
  \cH_{\phi} = \; \Pi_{\phi^{\ast}} \, \Pi_\phi + (\nabla {\phi^{\ast}}) \cdot (\nabla \phi) + m^2 \, \phi^{\ast} \phi,  
  \label{EQ:MODHAMF}\\
  \cH_{I_{1}} = - \, \frac{1}{2}\,  g^2 \phi^{\ast}(x) \phi(x)\int d\xp   \phi^*(\xp) \phi(\xp)D(x-\xp),   
  \label{EQ:MODHAMY} \\
  \cH_{I_{2}} = - g\eta \, \phi^{\ast}(x) \phi(x) \int d\xp \, d\zp \, d\zpp  \; D(x,\xp) \, D(\xp, \zp) \, D(\xp, \zpp) \; \rho(\zp) \, \rho(\zpp), 
  \label{EQ:MODHAMC} \\
  \cH_{I_{3}} = - g\sigma \, \phi^{\ast}(x) \phi(x) \int d\xp \, d\zp \, d\zpp \, d\zppp \; D(x,\xp) \, D(\xp, \zp) \, D(\xp, \zpp) \, D(\xp, \zppp) \, \rho(\zp) \, \rho(\zpp) \, \rho(\zppp)
  \label{EQ:MODHAMQ}.
\end{gather}
The conjugate momenta of the fields are defined in the usual way
\begin{align}
  \Pi_\phi = \, \ds\frac{\di \cal L}{\di {\dot \phi}} =  {\dot \phi}^{\ast} \F!
  \Pi_{\phi^{\ast}} = \, \ds\frac{\di \cal L}{\di {\dot \phi}^{\ast}} = {\dot \phi}.
\end{align}
Observe that every term of the reformulated Hamiltonian density has dimensions of $M^4$ (as it should). Notice that the Hamiltonian density is non-local as it contains integrals over space-time variables. Equations (\ref{EQ:MODHAMI})-(\ref{EQ:MODHAMQ}) specify the Hamiltonian density that will be used to study few-particle bound states in this chapter.

The $T^{0i}$ components of the energy-momentum tensor are the momentum density components:
\begin{equation}
  {\cal P}^i = - \Pi_\phi \, \di_i \, \phi - \Pi_{\phi^{\ast}} \, \di_i \, \phi^{\ast}.
  \label{EQ:MOMENTUM}
\end{equation}
Note that the interaction terms do not contribute to the momentum density above. This expression will be used when the total momentum of the few-particle systems will be discussed.

\section{Formalism and Quantization}
In the Hamiltonian formalism of QFT, the basic equation to be solved is the 4-momentum eigenvalue equation
\begin{equation}
  \hP^{\mu} \,|\Psi \ket = Q^{\mu}\, |\Psi\ket,
  \label{EQ:EV}
\end{equation}
where $\hP^{\mu}$ is the energy-momentum operator of the quantized theory which follows from equations (\ref{EQ:MODHAMI}) and (\ref{EQ:MOMENTUM}), $Q^{\mu} = (E, {\bf Q})$ is the energy-momentum eigenvalue and $|\Psi\ket$ are the corresponding eigenfunctions. It is not possible to obtain exact solutions for the $\mu = 0$ component of equation (\ref{EQ:EV}). The exact solution would comprise all the possible basis states and would be expressed as an infinite sum over the components; such an approach is evidently intractable. Instead, the variational method will be used to obtain approximate results. For $\mu = 1,2,3$ the solutions of (\ref{EQ:EV}) are free-field-like as is pointed out below.

Variational approximations to the $\mu = 0$ component of the eigenvalue equation (\ref{EQ:EV}), i.e. the Hamiltonian component, require the evaluation of
\begin{equation}
  \delta \bra \Psi_{t}|{\hat H} - E|\Psi_{t} \ket =  0,
  \label{EQ:VARPRIN}
\end{equation}
where $|\Psi_{t} \ket$ are suitably chosen trial states that contain adjustable features (functions, parameters). The accuracy of a variational approximation depends on the choice of the trial states. In general, the more flexible features a trial state possesses the better are the approximate results that it yields.  

The canonical quantization of the theory with the Hamiltonian density (\ref{EQ:MODHAMI}) is performed next. First, one replaces the matter fields with their Fourier mode components:
\begin{align}
  \phi(x) = & \int d\p\ds\left[(2\pi)^{3} \, 2 \omega_{\p}\right]^{-\frac{1}{2}} \Big( \, a\left(\p\right)\e^{-i p \cdot x}+b^{\dagger}\left(\p\right)\e^{i p \cdot x}\Big), 
  \label{EQ:PHIQUAN1} \\
  \phi^{\ast}(x) = & \int d\p\ds\left[(2\pi)^{3} \, 2 \omega_{\p}\right]^{-\frac{1}{2}}\Big( \, a^{\dagger}\left(\p\right)\e^{i p \cdot x}+b\left(\p\right)\e^{-i p \cdot x}\Big),
  \label{EQ:PHIQUAN2}
\end{align}
where $\omega^2_{\p} = m^2 + \p^2$. Note that since the mediating field $\chi$ does not appear in the Hamiltonian (\ref{EQ:MODHAMI}) explicitly, one doesn't have to write out its Fourier mode representation. The next step is to promote the fields to the status of operators and impose the following non-vanishing, equal time commutation relations:
\begin{equation}
  \left[\phi(x), \Pi_{\phi}(y)\right] = \left[\phi^{\ast}(x), \Pi_{\phi^{\ast}}(y)\right] = i \, \delta(\x-{\bf y}).
\end{equation}
All other commutators of the field and conjugate momentum operators vanish. The operators $a(\p)$ and $b(\p)$, in equations (\ref{EQ:PHIQUAN1}) and (\ref{EQ:PHIQUAN2}), are the free particle and antiparticle annihilation operators respectively whereas the operators $a^{\dagger}(\p)$ and $b^{\dagger}(\p)$ are the free particle and antiparticle creation operators. The vacuum state $|0 \ket$ is defined by $a(\p) | 0 \ket = b(\p) | 0 \ket = 0$. Note that the canonical quantization is performed in the interaction picture. 

Expressing the creation and annihilation operators in terms of the fields and its derivatives, the commutation rules become
\begin{equation}
   \left[a({\bf p}), a^{\dagger}({\bf q})\right] = \left[b({\bf p}), b^{\dagger}({\bf q})\right] = \delta({\bf p}-{\bf q}).
\end{equation}

The vacuum energy question is not essential to the bound state problem addressed in this dissertation, thus the creation and annihilation operators in the Hamiltonian are normal-ordered. To obtain the Hamiltonian operator the spatial coordinates are integrated out from the Hamiltonian density (\ref{EQ:MODHAMI}):
\begin{equation}
    \hat{H}(t) = \int d\x \, :{\cal H}(t, \x):,
\end{equation}
Next, the Hamiltonian is expressed in terms of the particle and antiparticle operators $a$, $b$, $a^\dagger$ and $b^\dagger$. This is straightforward for the free-field part of the Hamiltonian as can be seen in standard books on QFT, see for instance ~\cite{Peskin1995, book_Srednicki}. However, obtaining the analogous expressions for the interaction terms is tedious and this has been accomplished with the use of a computer code. 

The momentum operator is similarly obtained by normal-ordering and integrating out the spatial coordinates from the momentum density (\ref{EQ:MOMENTUM}):
\begin{equation}
  {\bf \hP} = \int d\x \, : \vec{\cal P}(\x):.
  \label{EQ:MO}
\end{equation}
Notice that since the momentum density operator does not involve the interaction terms, it is time independent. 

For the description of stationary (i.e time-independent) few-particle bound states, it is convenient to switch to the Schr\"odinger picture in which the time-dependence of the Hamiltonian is removed. The two pictures are related by 
\begin{equation}
  |\Psi_{I}(t) \ket =\e^{i \, H_{0}\, t} \, |\Psi_{S}(t) \ket
  \label{EQ:PICTURES}
\end{equation}
where $H_0$ is the free-field part of the Hamiltonian which in this theory is given by equation (\ref{EQ:MODHAMF}).

\section{Particle-Antiparticle State}
The simplest particle-antiparticle trial state can be expressed in terms of Fock-states as follows:
\begin{equation}
  |\Psi_2\ket = \int d\p_{1,2} \; F(\p_{1,2})\; a^{\dag} (\p_1) \, b^{\dag} (\p_2) \; | 0 \ket,
  \label{EQ:TRIALTWO}
\end{equation}
where $F(\p_{1,2})$, is an adjustable coefficient function which is determined variationally \footnote{The subscript notation means that $d\p_{1,2} = d\p_{1} \, d\p_{2}$ and $F(\p_{1,2}) = F(\p_1, \p_2)$. This notation is used throughout the dissertation.}. 

\begin{figure}[t]
  \center{
    \includegraphics[scale=0.7]{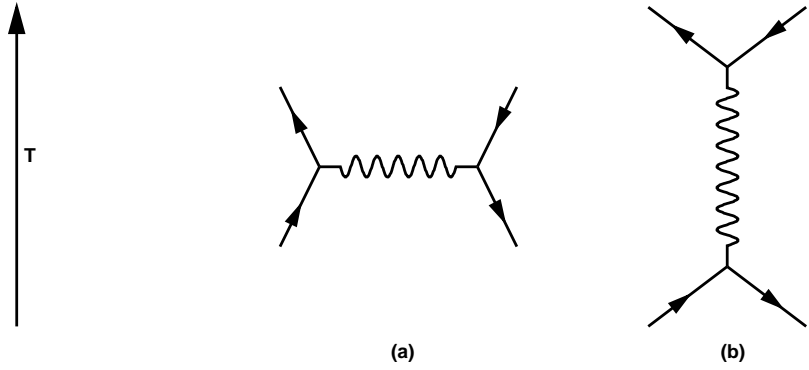}
  }
  \spacing{1}
  \caption{The one-chion exchange (a) and virtual annihilation (b) diagrams of the Yukawa kernel equation (\ref{EQ:Y_22}).}
  \label{FIG:FD1}
\end{figure}

To implement the variational principle (\ref{EQ:VARPRIN}) the matrix element $\bra \Psi_2 | \, \hat{H} \, - E \, | \Psi_2 \ket$ is worked out and is varied with respect to the adjustable coefficient function $F^{\ast}$.  This leads to the following particle-antiparticle relativistic wave equation in momentum space:
\begin{equation}
  F(\p_{1,2}) \, \big(\omega_{\p_1} + \omega_{\p_2} - E\big) = \int d\p^{\prime}_{1,2} \; \cY_{2,2}(\p^{\prime}_{1,2}, \p_{1,2}) \, F(\p^{\prime}_{1,2}),
  \label{EQ2}
\end{equation}
where $\cY_{2,2}$ is the relativistic Yukawa particle-antiparticle interaction kernel (inter-particle ``momentum-space'' potential). In the interaction picture, it is given by
\begin{align}
  \cY_{2,2} & \;(\p^{\prime}_{1,2}, \p_{1,2}) = \frac{g^{2}}{8\,(2\pi)^{3}} \, \frac{\e^{i \, (\omega_{\p^{\prime}_1} + \omega_{\p^\prime_2}-\omega_{\p_1}-\omega_{\p_2}) \, t}}{\ds\sqrt{\ds\omega_{\p^{\prime}_{1}}\ds\omega_{\p^{\prime}_{2}}\ds\omega_{\p_{1}}\omega_{\p_{2}}}} \, \ds\delta(\p^{\prime}_{1}+\p^{\prime}_{2}-\p_{1}-\p_{2}) \nonumber \\
  & \times \left[\frac{1}{\mu^{2}-(p^{\prime}_1-p_1)^{2}}+\frac{1}{\mu^{2}-(p^{\prime}_2-p_2)^{2}} + \frac{1}{\mu^{2}-(p_{1}+p_{2})^{2}} + \frac{1}{\mu^{2}-\ds(p^{\prime}_{1}+p^{\prime}_{2})^{2}}\right].
  \label{EQ:Y_22}
\end{align}
The first two terms in the square brackets in equation (\ref{EQ:Y_22}) correspond to one-mediating-field-quantum exchange (or one-``chion'' exchange) and the last two to virtual annihilation Feynman diagrams. These diagrams are shown in Figure \ref{FIG:FD1}.

%
The time dependence of the particle-antiparticle interaction kernel $\cY_{2,2}$ can be ``rotated away'' by the use of equation (\ref{EQ:PICTURES}). If the trial state $| \Psi_2^\prime \ket = \e^{i \, H_0 \, t} | \Psi_2 \ket$ is used in the above, then the corresponding interaction kernel in the Schr\"odinger picture will be time independent, i.e. one obtains equation (\ref{EQ:Y_22}) with $t=0$. Therefore, the Schr\"odinger picture (time independent) interaction kernels can be found, in effect, by setting $t = 0$ in all interaction picture kernels. 

Unfortunately, the quantized version of the non-linear terms (\ref{EQ:MODHAMC}) and (\ref{EQ:MODHAMQ}) of the Hamiltonian density are not probed by the particle-antiparticle trial state (\ref{EQ:TRIALTWO}), i.e. the matrix elements $\bra \Psi_2 | \, \hat{H}_{I_i} \, | \Psi_2 \ket$ vanish for $i = 2, 3$. The vanishing occurs because of a mismatch in the number of the ladder operators between the trial state and the interaction terms. More elaborate trial states than (\ref{EQ:TRIALTWO}) must be used in order to examine the effects of non-linear terms. As it is, with the trial state (\ref{EQ:TRIALTWO}), the problem reduces to the scalar Yukawa model~\cite{Ding:1999ed}.

For the particle-antiparticle trial state (\ref{EQ:TRIALTWO}) to be an eigenstate of the momentum operator (\ref{EQ:MO}) we require that ${\bf \hP} \, |\Psi_2 \ket =  {\bf Q} \, |\Psi_2 \ket$. In the rest frame, where ${\bf Q} = 0$, this requires that $F(\p_{1,2}) = \delta(\p_1 + \p_2) \, f(\p_1)$. Consequently, an energy calculation yields the rest energy (i.e. the mass) of the particle-antiparticle system. The centre of mass motion separates and the relativistic momentum-space particle-antiparticle wave equation simplifies to
\begin{multline}
  f(\p) \, \big(2 \, \omega_\p - E\big) = \frac{g^{2}}{4\,(2\pi)^{3}} \int \frac{d\p^{\prime}}{\ds\omega_{\p^{\prime}}\ds\omega_\p} \, f(\p^{\prime}) \, \left[\frac{1}{\mu^2+(\p^{\prime} - \p)^2 - (\omega_{\p^{\prime}} - \omega_{\p})^2} \right. \\ \left.+ \frac{1}{2}\left(\frac{1}{\mu^2 - 4 \, \omega^2_{\p}}\right) + \frac{1}{2}\left(\frac{1}{\mu^2 - 4 \, \omega^2_{\p^\prime}}\right)\right].
  \label{EQ2A}
\end{multline}
%

%
In the non-relativistic limit (i.e. $\p^2 << m^2$) the Fourier transform of equation (\ref{EQ2A}) yields the expected Schr\"odinger equation in coordinate space for the relative motion of the particle-antiparticle system:
\begin{equation}
  -\frac{1}{m}\nabla^2 \, \psi(\x) + V(x) \, \psi(\x) = \varepsilon \, \psi(\x),
  \label{EQ:SCHROD}
\end{equation}
where $\varepsilon = E - 2 \, m$ is the non-relativistic energy, and the potential energy depends on the particle-antiparticle distance: 
\begin{equation}
  V (x) = -\alpha_g \; \left(\frac{\e^{-\mu x}}{x} - \frac{4 \, \pi }{\mu^2 - 4m^2} \, \delta(\x) \right), 
  \label{PE2}
\end{equation}
where $\alpha_g = \ds\frac{g^2}{16\pi m^2}$ is a dimensionless coupling constant. The first term, due to the one-chion exchange, is the usual Yukawa potential and is always attractive (i.e. gravity-like) in this scalar theory. The second term, due to the virtual annihilation, can be either attractive or repulsive depending on the values of the parameters $\mu$ and $m$. It is a correction to the Yukawa potential that is a feature of the quantum field theory and has no analog in relativistic QM.

The relativistic equation (\ref{EQ2A}) is not analytically solvable, so approximation methods must be used. Perturbative and variational approximations are presented in the paper by Ding and Darewych~\cite{Ding:1999ed}, along with a comparison to results obtained using the ladder Bethe-Salpeter equation and various quasi-potential approximations. 

The particle-antiparticle trial state has served as a preamble. In the following sections, trial states that probe the non-linear terms are examined.

\section{Three-Particle State}
To observe the effects of the non-linear terms of the Hamiltonian density (\ref{EQ:MODHAMI}) on the inter-particle potential, one must consider trial states with either more particle content or more Fock-space components. The easier task is to consider a three identical particle trial state given by 
\begin{equation}
  |\Psi_3\ket = \int d\p_{1,2,3} \; F(\p_{1,2,3}) \; a^{\dag} (\p_1) \, a^{\dag} (\p_2) \, a^{\dag} (\p_3) \; | 0 \ket,
  \label{EQ:TRIALTHREE}
\end{equation}
where $F(\p_{1,2,3})$ is a three-particle function to be determined variationally. Note that this trial state can be taken to be an eigenstate of the momentum operator (\ref{EQ:MO}), i.e. ${\bf \hP} \, |\Psi_3 \ket = {\bf Q} \,  |\Psi_3 \ket$ with the choice $F(\p_{1,2,3}) = \de(\p_1 + \p_2 + \p_3 - {\bf Q}) \, f(\p_{1,2})$, where ${\bf Q}$ is the constant total momentum of the three-particle system. Therefore, the wavefunction will be of the form where the centre of mass motion is completely separable, just as was the case for the particle-antiparticle system. If ${\bf Q} = 0$, the the corresponding eigenenergy will represent the rest mass of the three-particle system.
The addition of an extra third particle operator to the trial state significantly complicates the evaluation of the matrix element. However, the implicit symmetry of equation (\ref{EQ:TRIALTHREE}) with respect to interchanges of momentum variables due to the identity of the particles makes the task easier. It permits one to carry out the evaluation of the matrix element for the three-particle trial state in terms of the symmetrized function $F_S$:
\begin{equation}
  F_S(\p_{1,2,3}) = \sum_{i_1, i_2, i_3}^6 F(\p_{i_1,i_2,i_3}),
\end{equation}
where the summation is on the six permutations of the indices $1, 2$ and $3$. Making use of this symmetrization enables one to keep the arguments of the kernels non-permuted while storing all information about the symmetry under interchanges of the momentum variables in $F_S$.

\begin{figure}[t]
  \center{
    \includegraphics[scale=0.6]{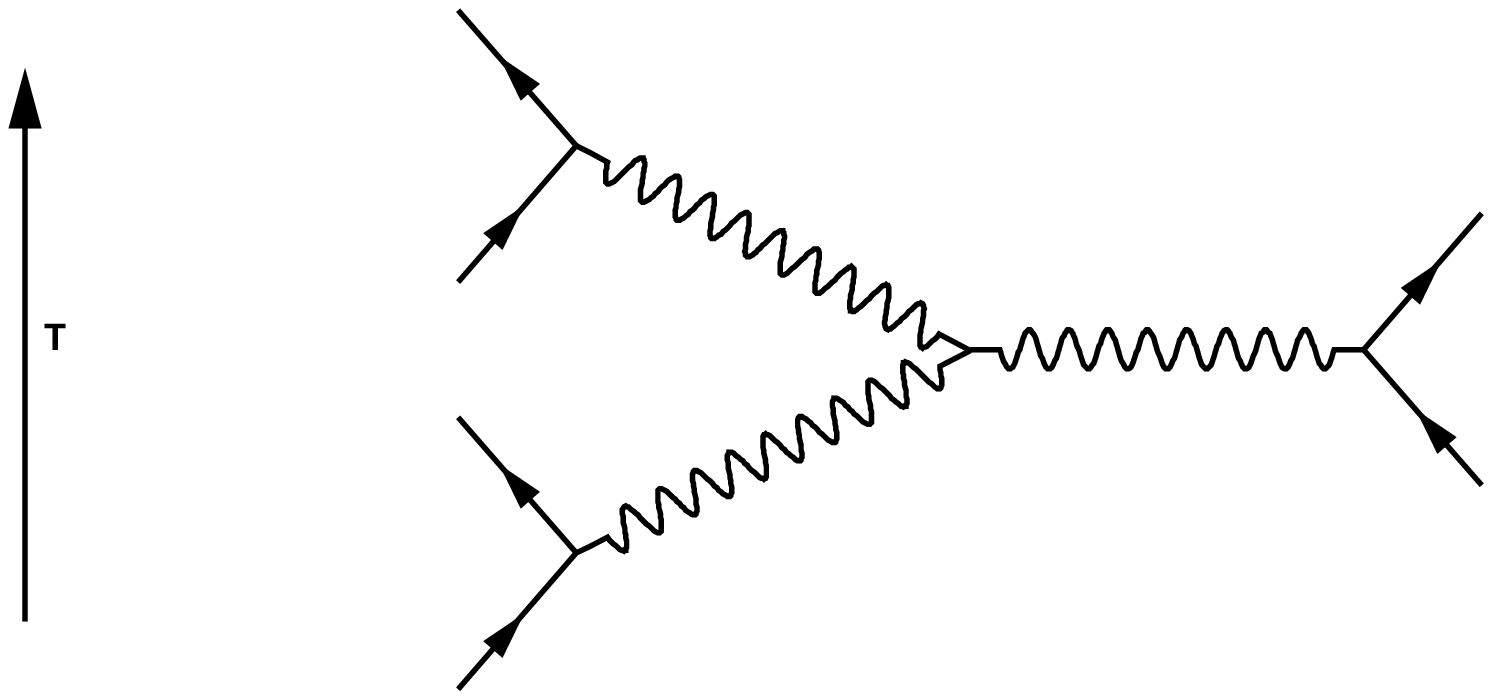}
  }
  \spacing{1}
  \caption{The three-chion propagator vertex corresponding to the cubic interaction kernel $\cC_{3,3}$ equation (\ref{EQ:C_33}). The two propagators on the left should actually overlap (impossible to draw) such that they are perpendicular to the direction of time.}
  \label{FIG:FD2}
\end{figure}

The matrix element $\bra \Psi_3 | \, \hat{H} \, - E \,| \Psi_3 \ket$ is calculated by commuting the ladder operators and turning them into momentum delta functions. Once the commutation is completed, the momentum integrals can be found in a straightforward, although involved, manner. Next, the variational derivative with respect to $F^{\ast}$ is carried out and set to zero (The complete expressions for the matrix element as well as other intermediate steps in the calculations are given in Appendix A section \ref{SEC:scalar_three}). This yields the following relativistic momentum space integral wave equation for stationary states of three identical particles:
\begin{multline}
  F_S(\p_{1,2,3}) \, \big(\omega_{\p_1} + \omega_{\p_2} + \omega_{\p_3} - E\big) = \int d\p^{\prime}_{1,2} \; \cY_{3,3}(\p^{\prime}_{1,2,3}, \p_{1,2,3}) \, F_S(\p^{\prime}_{1,2,3}) \\ + \int d\p^{\prime}_{1,2,3} \; \cC_{3,3}(\p^{\prime}_{1,2,3}, \p_{1,2,3}) \, F_S(\p^{\prime}_{1,2,3}).
  \label{EQ3}
\end{multline}
The relativistic interaction kernels (i.e. the relativistic momentum-space inter-particle potentials) are given by 
\begin{align}
  \cY_{3,3} & \, (\p^{\prime}_{1,2,3}, \p_{1,2,3}) = -\frac{g^2}{8\,(2\pi)^{3}} \nn \\ 
  & \F! \times\sum_{i_1, i_2, i_3}^6 \frac{\delta(\p^{\prime}_1 + \p^{\prime}_2 - \p_{i_1} - \p_{i_2}) \, \delta(\p^{\prime}_3 - \p_{i_3})}{\ds\sqrt{\omega_{\p^{\prime}_{1}}\omega_{\p^{\prime}_{2}}\omega_{\p_{i_1}}\omega_{\p_{i_2}}}} \, \left[\frac{1}{\mu^2-(p^{\prime}_1 - p_{i_1})^2}\right] 
  \label{EQ:Y_33} \\
  \cC_{3,3} & \, (\p^{\prime}_{1,2,3}, \p_{1,2,3}) = - \frac{g^{3}\eta}{8(2\pi)^{6}} \sum^{6}_{i_1, i_2, i_3} \frac{\delta(\p^{\prime}_1 + \p^{\prime}_2 +\p^{\prime}_3 - \p_{i_1} - \p_{i_2} - \p_{i_3})}{\ds\sqrt{\omega_{\p^{\prime}_{1}}\omega_{\p^{\prime}_{2}}\omega_{\p^{\prime}_{3}}\omega_{\p_{i_1}}\omega_{\p_{i_2}}\omega_{\p_{i_3}}}} \nonumber \\
  & \F!\times\left[\frac{1}{\mu^2-(p^{\prime}_1 + p^{\prime}_2 - p_{i_1} - p_{i_2})^2}\frac{1}{\mu^2-(p^{\prime}_1 - p_{i_1})^2}\frac{1}{\mu^2-(p^{\prime}_2 - p_{i_2})^2}\right], 
  \label{EQ:C_33}
\end{align}
where the summation is on the six permutations of the three indices $1, 2$ and $3$. It is evident from the covariant factors of equation (\ref{EQ:Y_33}), that the kernel $\cY_{3,3}$ corresponds to the three inter-particle one-chion exchange interactions. There are no virtual annihilation terms since only particle operators (and no antiparticle operators) are present in the trial state (\ref{EQ:TRIALTHREE}). The kernel $\cC_{3,3}$, equation (\ref{EQ:C_33}), corresponds to the non-linear interaction term $\cH_{I_2}$ (see section \ref{SEC:scalar_three} of Appendix A for details). It is similarly evident, from the covariant factors of equation (\ref{EQ:C_33}), that the Feynman diagram corresponding to the cubic kernel $\cC_{3,3}$ contains a three-chion propagator vertex. This diagram is shown in Figure \ref{FIG:FD2}. Note that the three-particle trial state (\ref{EQ:TRIALTHREE}) does not probe the quartic interaction term $\cH_{I_3}$ equation (\ref{EQ:MODHAMQ}), i.e. $\bra \Psi_3 | \, \hat{H}_{I_3} \, | \Psi_3 \ket = 0$.
Equation (\ref{EQ3}), together with the kernels (\ref{EQ:Y_33}) and (\ref{EQ:C_33}), is a relativistic wave equation for stationary states of a system of three identical scalar particles. It can describe purely bound states of the three particles (spinless bosons) or elastic scattering among them, but not processes involving the emission or absorption of chions. The relativistic kinematics (i.e. kinetic energy) of the system is described without approximation, but the relativistic dynamics (i.e. potential energy) is described only at the level of one-chion exchange between the particle pairs by the relativistic kernel $\cY_{3,3}$, along with a three-particle iterative first-order correction $\cC_{3,3}$ due to the non-linear interaction terms. 

In principle, one would wish to solve the relativistic three-particle equation (\ref{EQ3}). However, this is a formidable task which cannot be done exactly. Even the determination of approximate solutions is a very challenging task which will not be undertaken in this dissertation. Instead, it is of interest to consider the non-relativistic limit of equation (\ref{EQ3}), which in the coordinate representation is just a non-relativistic three-particle Schr\"odinger equation and the Fourier-transforms of $\cY_{3,3}$ and $\cC_{3,3}$ are the inter-particle potentials. Among other things, these inter-particle potentials can be used to calculate the bound state energy for systems of three heavy spinless particles. 

The non-relativistic limit of equations (\ref{EQ3})-(\ref{EQ:C_33}), just as for the particle-antiparticle system, is obtained by assuming $\p^2 << m^2$ and then Fourier-transformed to coordinate space. From equation (\ref{EQ:Y_33}), the non-relativistic inter-particle potential for the three-particle system due to the Yukawa interaction is, as expected,  
\begin{equation}
  V_Y(\x_{1,2,3}, \mu) = V_Y(x_{ij}, \mu) = - \alpha_g \Bigg\{\frac{\e^{- \mu x_{12}}}{x_{12}} + \frac{\e^{- \mu x_{13}}}{x_{13}} + \frac{\e^{- \mu x_{23}}}{x_{23}}\Bigg\}.
  \label{EQ:V_Y123} 
\end{equation}
Unlike in the particle-antiparticle case (\ref{PE2}), no virtual annihilation delta function contributions appear in this expression since there are no antiparticles present in the three identical particle system.

The non-relativistic ``cubic'' potential term that follows from equation (\ref{EQ:C_33}) is
\begin{equation}
  V_C(\x_{1,2,3}, \mu) = - \alpha_{\eta} \, \pi^3 \int d\q_{1,2,3} \, \frac{\prod_{i}^{3}\e^{-i \, \q_{i}\cdot\x_{i}} \, \delta(\q_1+\q_2+\q_3)}{\left(\mu^2+\q_1^2\right)\left(\mu^2+\q_2^2\right)\left(\mu^2+\q_3^2\right)},
  \label{EQ:V_C123}
\end{equation}
where $\alpha_{\eta} = \ds\frac{3 \, g^3 \, \eta}{256\pi^6m^3}$ is a coupling constant with dimensions of mass (see section \ref{SEC:scalar_three} of Appendix A for details). $V_C$ can be simplified to a three-dimensional quadrature (see section \ref{SEC:scalar_three} of Appendix A for details):
\begin{equation}
  V_C(\x_{1,2,3}, \mu) = - \alpha_\eta \, \pi^3 \int d\x \, \frac{\e^{-\mu|\x_1+\x|}}{|\x_1+\x|} \, \frac{\e^{-\mu|\x_2+\x|}}{|\x_2+\x|} \, \frac{\e^{-\mu|\x_3+\x|}}{|\x_3+\x|}.
  \label{EQ:V_C123A}
\end{equation}
This is an overall well behaved convergent integral for any $\mu > 0$. However, it cannot be evaluated analytically in general. The expression for $V_C$, equation (\ref{EQ:V_C123A}), with the substitution $\v = \x + \x_1$, can be written as 
\begin{equation}
  V_C(\x_{1,2,3}, \mu) = - \alpha_\eta \, \pi^3\int d\v \, \frac{\e^{-\mu |\v|}}{|\v|} \, \frac{\e^{-\mu |\v + \x_{21}|}}{|\v + \x_{21}|} \, \frac{\e^{-\mu |\v + \x_{31}|}}{|\v + \x_{31}|},
  \label{EQ:V_C123B}
\end{equation}
where $\x_{21} = \x_2 - \x_1$ and $\x_{31} = \x_3 - \x_1$. The expression for $V_C$, as shown in Appendix A section \ref{SEC:scalar_three}, can also be written $V_C$ in the form
\begin{multline}
  V_C(\x_{ij}, \mu) = - \alpha_{\eta} \, \pi^3 \int_{0}^{\infty}d\beta_{1,2,3} \; \frac{\e^{-\mu^{2}(\beta_{1}+\beta_{2}+\beta_{3})}}{(\beta_{1}\beta_{2}+\beta_{1}\beta_{3}+\beta_{2}\beta_{3})^{3/2}}  \\
  \times\exp\left(-\frac{\beta_{1}\x_{21}^{2}+\beta_{2}\x_{31}^{2}+\beta_{3}\x_{32}^{2}}{4\left(\beta_{1}\beta_{2}+\beta_{1}\beta_{3}+\beta_{2}\beta_{3}\right)}\right),  
  \label{EQ:V_C123C}
\end{multline}
which shows explicitly that $V_C$ depends only on inter-particle distances $x_{ij} = x_{ji} =|\x_i - \x_j|$. In light of equation (\ref{EQ:V_C123C}), one can conclude that the total potential energy function $V = V_Y + V_C$ depends only on the inter-particle distances, and it is invariant under 3D rotations and translations of coordinates as is expected of a closed system.  Therefore, to emphasize this fact the notation $V_C(x_{ij}, \mu)$ is used in what follows. In addition, the representation (\ref{EQ:V_C123C}) is suitable for numerical evaluations of $V_C$ for arbitrary values of $x_{ij}$. 

Unfortunately, it is impossible to render multi-dimensional plots of inter-particle potentials of three independent variables $x_{12}$, $x_{23}$ and $x_{13}$, even if they can be worked out numerically for arbitrary $x_{ij}$. The best one can do is to plot one and two dimensional sections, corresponding to certain restrictions on the coordinates, to gain some idea of the shape of the cubic term $V_C$. 


%

%
\subsubsection*{Case 1: $V(\x_{1,2,2}, \mu) = V(x_{21}, \mu)$}
\begin{figure}[h!] 
  \begin{center}
    \includegraphics[scale = 0.6]{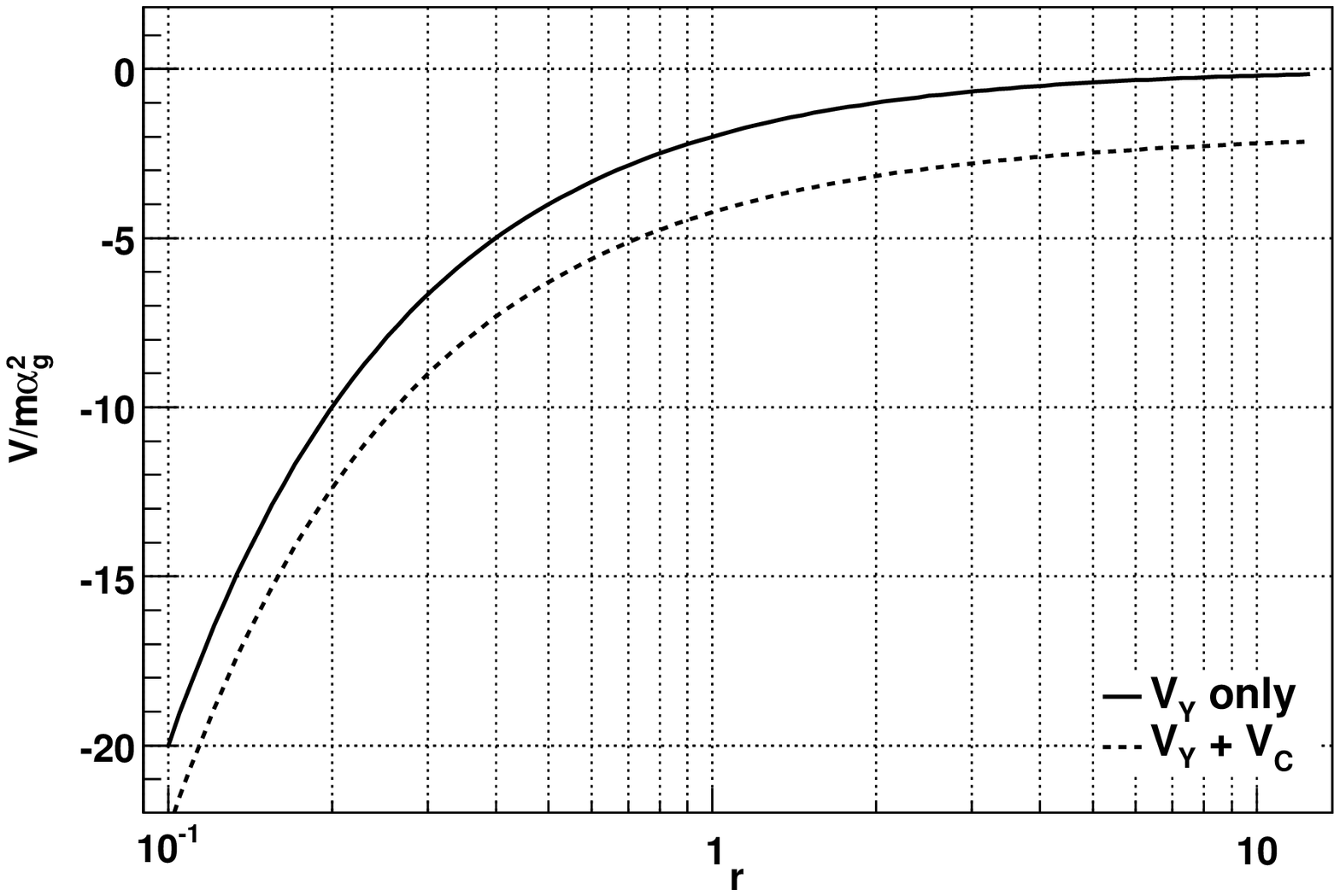}
    \spacing{1}
    \caption{The inter-particle potential $V(r, M) \, / \, m\alpha^2_g$ of equation (\ref{EQ:V_122MU}) as a function of $r = x_{21}\, m\alpha_g$ for $M = 10^{-10}$ with $\kappa_1 = 0$ (i.e. $V_Y$ only) and $\kappa_1 = 0.1$ (i.e. $V_Y + V_C$).}
    \label{FIG:V_122SEPARATE}
%
    \includegraphics[scale = 0.6]{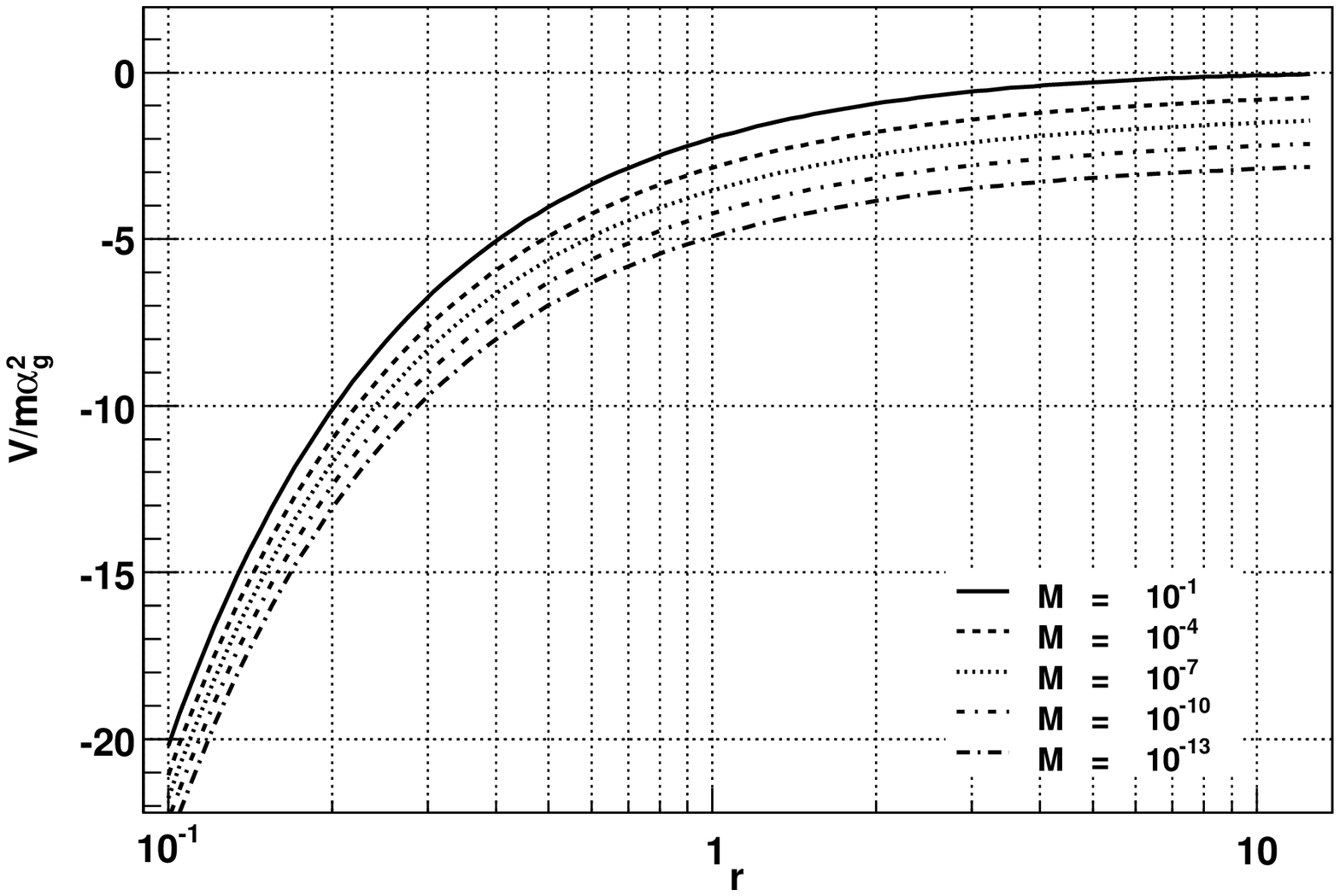}
    \spacing{1}
    \caption{The inter-particle potential $V(r, M) \, / \, m\alpha^2_g$ of equation (\ref{EQ:V_122MU}) as a function of $r = x_{21}\, m\alpha_g$ for $\kappa_1 = 0.1$ and various values of $M$.}
    \label{FIG:V_122MU}
  \end{center}
\end{figure}

The first case is for $\mu > 0$ and $\x_2 = \x_3$ (i.e. two particles overlap) so that there is only one inter-particle distance, $x_{21} = x_{12} = |\x_2 - \x_1| = |\x_3 - \x_1|$, to be concerned with. The integral for $V_C$, equation (\ref{EQ:V_C123B}), in this case is expressible in terms of the special function, the exponential integral, defined by 
\begin{equation}
  \E1(z) = \int^{\infty}_z \frac{\e^{-t}}{t} \, dt,
  \label{EQ:EXPINT}
\end{equation}
namely
\begin{multline}
  V_C(\x_{1,2,2} ,\mu) = \\ V_C(x_{21}, \mu) = -\frac{2 \, \pi^4\alpha_{\eta}}{x_{21} \mu} \Bigg\{\e^{-x_{21} \mu}\ds\left[\ln\left(3\right) - \E1\left(x_{21}\mu\right)\right] + \e^{x_{21}\mu} \, \E1\left(3x_{21}\mu\right) \Bigg\}.
  \label{EQ:V_C122MU}
\end{multline}
The length is chosen to be expressed in units of the Bohr radius $\ds\frac{1}{m \, \alpha_g}$ and, correspondingly, the energy in units of $m\alpha^2_g$. Equation (\ref{EQ:V_C122MU}) can be written in terms of the dimensionless variables $r = x_{21} \, m \, \alpha_g$ and $M=\ds\frac{\mu}{m \, \alpha_g}$. Suppressing the singularity at $\x_2 = \x_3$ in the Yukawa term (\ref{EQ:V_Y123}), the total inter-particle potential $V =V_Y + V_C$ for the three-particle trial state in the case when $\mu > 0$ and $\x_2 = \x_3$ is 
\begin{multline}
  V(\x_{1,2,2}, \mu > 0) = V(r,M) \\
  = m \, \alpha^{2}_g \; \Bigg\{ - \, 2 \, \frac{\e^{- r \, M}}{r} - \frac{\kappa_1}{2}\ds\left(\frac{\e^{-r \, M}}{r \, M} \, \ln(3) 
  - \frac{\e^{-r \, M}}{r \, M} \, \E1(rM)
  + \frac{\e^{r \, M}}{r \, M} \, \E1(3 \, r \, M) \right) \Bigg\},
  \label{EQ:V_122MU}
\end{multline}
where $\kappa_1 = \ds\frac{4 \, \pi^4 \, \alpha_\eta}{m \, \alpha^2_g}=\ds\frac{12 \, \eta}{g}$ is a dimensionless constant. Figures (\ref{FIG:V_122SEPARATE}) and (\ref{FIG:V_122MU}) are plots of $V(r,M)/ m \, \alpha_g^2$, equation (\ref{EQ:V_122MU}), as a function of $r$ for the different values of $M$, with $\kappa_1 = 0$ and $\kappa_1 = 0.1$ respectively. It is apparent that the contribution of $V_C$ lowers the overall potential with decreasing values of $M$.

%

For the case where $\mu = 0$, $V_C$, equation (\ref{EQ:V_C123}), becomes logarithmically divergent and requires regularization. There are various ways of regularizing $V_C$ with $\mu = 0$. One way is by subtracting off an infinite constant and absorbing it in the bound state energy, namely:
\begin{gather}
  \widetilde{V}_C(\x_{ij}, \mu) = \, V_C(\x_{ij}, \mu) - V_C({\bf a}_{ij}, \mu) 
  \label{EQ:REC1} \\
  {\cal \varepsilon} \, \rightarrow \; \widetilde{\varepsilon} = \, \varepsilon - V_C({\bf a}_{ij}, \mu),
  \label{EQ:REC2}
\end{gather}
where ${\bf a}_{ij} = {\bf a}_i - {\bf a}_j$ are reference (i.e. constant) inter-particle vectors and $\widetilde{\varepsilon}$ is the eigenenergy corresponding to $\varepsilon$ in equation (\ref{EQ:SCHROD}). In the limit $\mu = 0$, such a procedure renders $\widetilde{V}_C$ to be finite.

The inter-particle potential for the case when $\x_2 = \x_3$ and $\mu = 0$ was originally derived using a different method by Darewych and Duviryak~\cite{Darewych:2009wk}. One can retrieve this expression from equation (\ref{EQ:V_122MU}) by expanding in the limit $M \rightarrow 0$, namely:
\begin{gather}
  V_C(\x_{1,2,2}, \mu > 0) = V_C(r,M) = 4 \, \pi^4 \, \alpha_\eta \left( \, \ln(3) - 1 + \gamma + \ln(r \, M) \, \right), 
\end{gather}
where $\gamma = 0.5772$ is the Euler's constant. Consequently, implementing a regularization, as prescribed by equations (\ref{EQ:REC1}) and (\ref{EQ:REC2}), with the reference separation $a_{21} \, m \, \alpha_g = 1$ (where $a_{21} = | {\bf a}_2 - {\bf a}_1 |$) yields:
\begin{equation}
  V_{\tx{regularized}}(\x_{1,2,2} , \mu = 0) = V_{\tx{regularized}}(r, \mu = 0) = m\alpha^2_g \, \Bigg\{-\frac{2}{r} + \kappa_1 \, \ln r\Bigg\}.
  \label{EQ:V_122}
\end{equation}
Notice that $V(r,\mu = 0)$ is dominated by the Coulombic $\ds-\frac{2}{r}$ term for small $r$ but by the logarithmic term for larger $r$. 

However, the results for the $\mu = 0$ case are regulator-dependent. Within a perturbative calculation, it is known that there are other infrared singular effects which lead to a cancellation of all regulator-dependent terms. It is expected that a more inclusive ansatz would do the same. 
Therefore, the results of equation (\ref{EQ:V_122}) are most likely unphysical. Nevertheless, in the theory with $\eta = 0$ (i.e. without the cubic term in the Lagrangian density (\ref{EQ:SCLLAGR})), infrared-divergences do not arise, as is pointed out in Section \ref{SEC:FOUR}.

\subsubsection*{Case 2: $V(x_{21} = x_{31} = x_{23} = \Delta, \mu) = V(\Delta, \mu)$}
\begin{figure}[t] 
  \begin{center}
    \includegraphics[scale = 0.7]{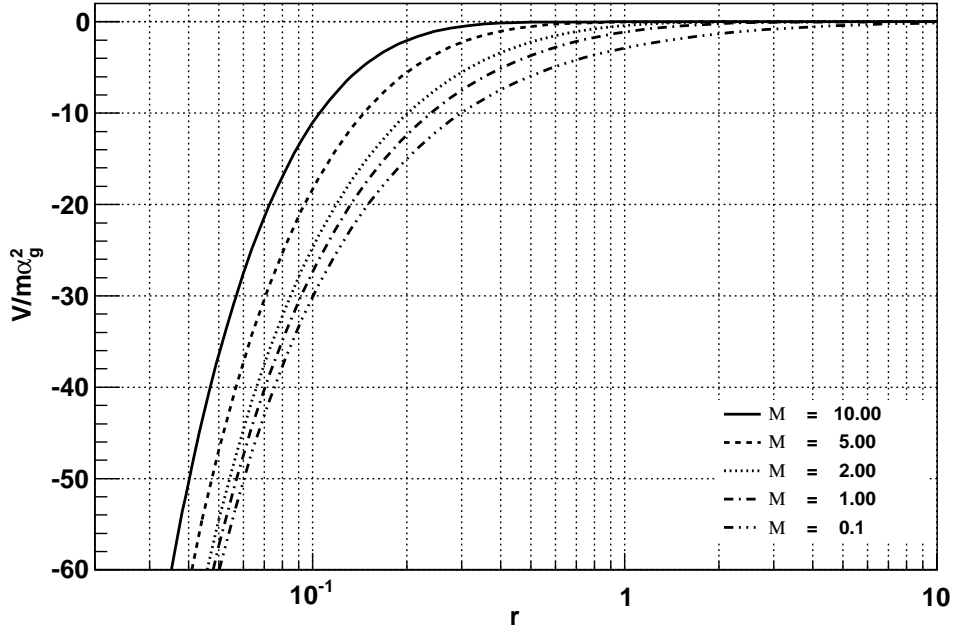}
    \spacing{1}
    \caption{The inter-particle potential $V(r, M)\, / \, m\alpha^2_g$ of equation (\ref{EQ:V_DMU}) for $\kappa_1 = 0.1$ and the indicated values of  $M = \ds\frac{\mu}{m\alpha_g}$, where $r = \Delta \, m \alpha_g$ and $\Delta = x_{21} = x_{31} = x_{23}$.}
      \label{FIG:V_DMU}
  \end{center}
\end{figure}

Another case where there is only one distance argument in the inter-particle potential is when the coordinates are at the vertices of an equilateral triangle. Unfortunately, no analytical solution is possible in this case and one has to carry out a numerical integration. The numerical integration was performed with the GNU Scientific Library\cite{gslgnu}. With the previous choice for the units of length and energy, the expression for $V_C$ is
\begin{multline}
  V(\Delta, \mu ) = V(r, M) = - \, m \, \alpha_g^2 \,  \Bigg\{ \, 3 \, \frac{\e^{-r M}}{r} \\
  + \frac{\kappa_1}{4\pi}\int_0^\infty d\beta_{1,2,3} \; \frac{\e^{-M^{2}(\beta_{1}+\beta_{2}+\beta_{3})}}{(\beta_{1}\beta_{2}+\beta_{1}\beta_{3}+\beta_{2}\beta_{3})^{3/2}} \, \exp\left(-\frac{r^2 \, (\beta_1 + \beta_2 + \beta_3)} {4(\beta_1\beta_2 + \beta_1\beta_3 + \beta_2\beta_3)}\right) \Bigg\}
  \label{EQ:V_DMU}
\end{multline}
where $r = \Delta \, m \alpha_g$ is the dimensionless inter-particle distance, $\Delta = |\x_2 - \x_1| = |\x_3 - \x_1| = |\x_3 - \x_2|$ and $M = \ds\frac{\mu}{m \, \alpha_g}$ is the dimensionless mass parameter of the mediating field. A plot of equation (\ref{EQ:V_DMU}) is given in Figure (\ref{FIG:V_DMU}). The overall character of the potential retains the same features as in the particular case $\x_{23} = 0$, equation (\ref{EQ:V_122MU}). The curves show no hint of confinement as they all asymptotically approach zero for this $M > 0$ case.

%
\subsubsection*{Case 3: Arbitrary inter-particle distances}
\begin{figure}
  \begin{center}$ 
    \begin{array}{cc}
      \includegraphics[scale = 0.35]{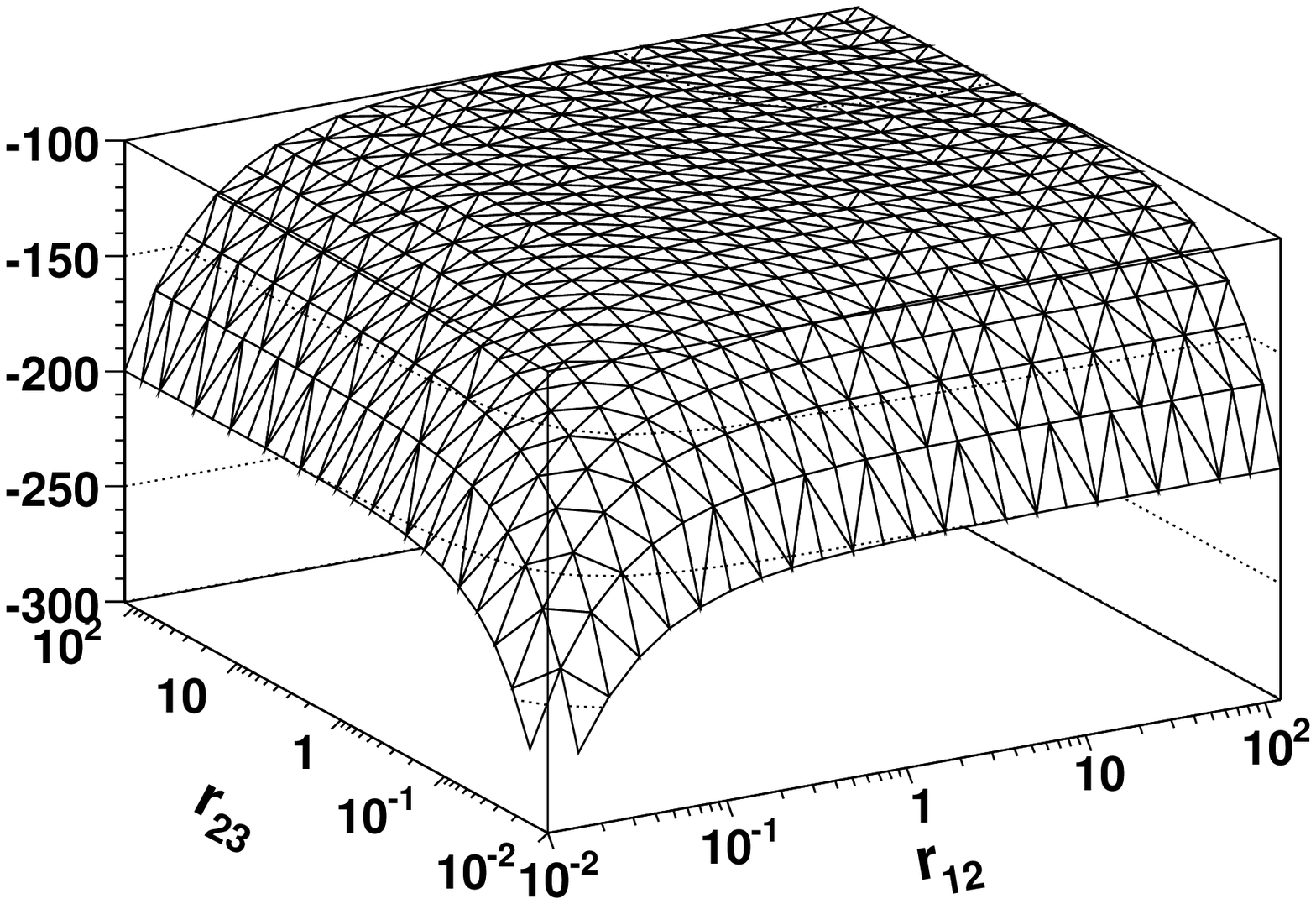} & \includegraphics[scale = 0.35]{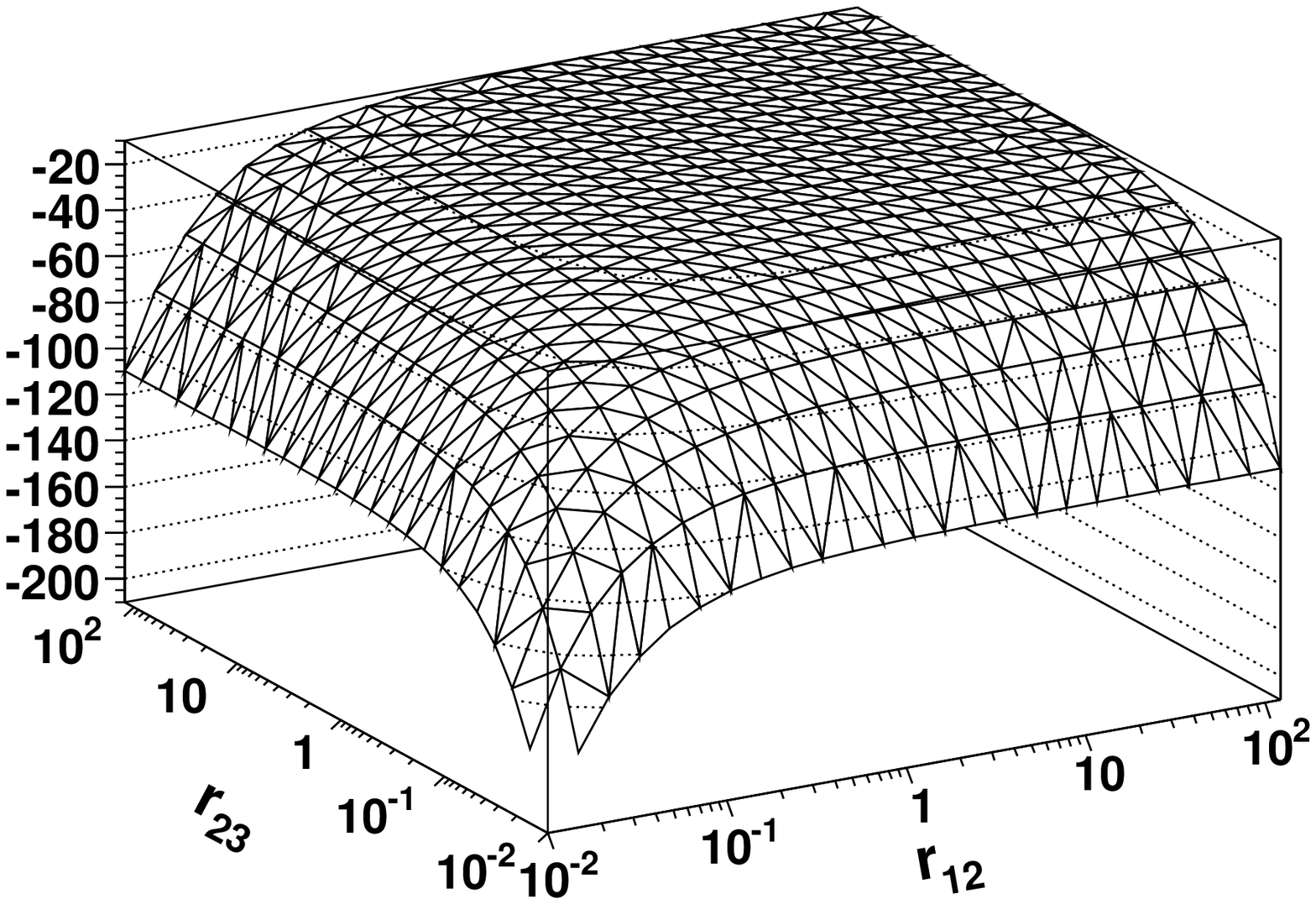} \\
      \includegraphics[scale = 0.35]{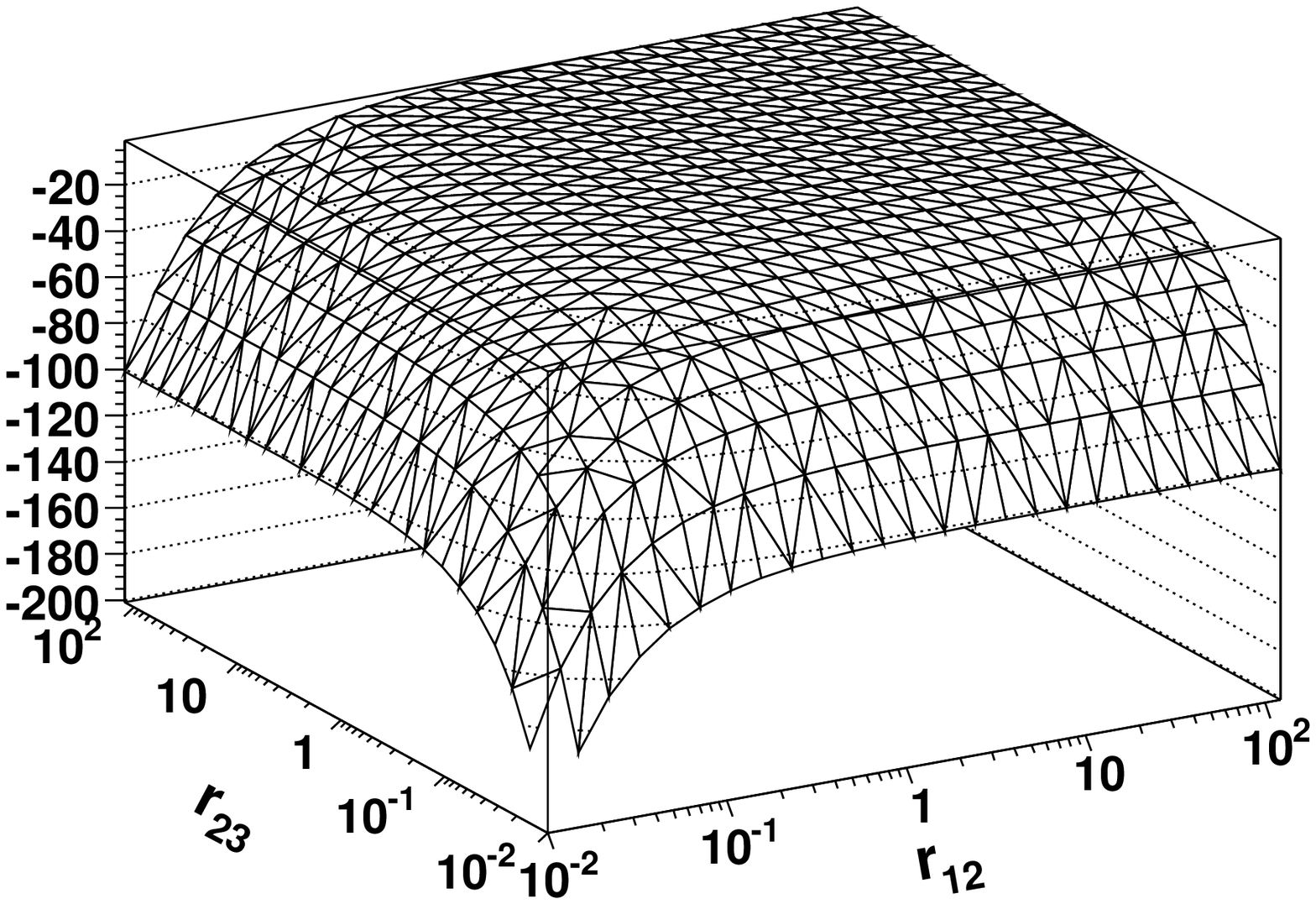} & \includegraphics[scale = 0.35]{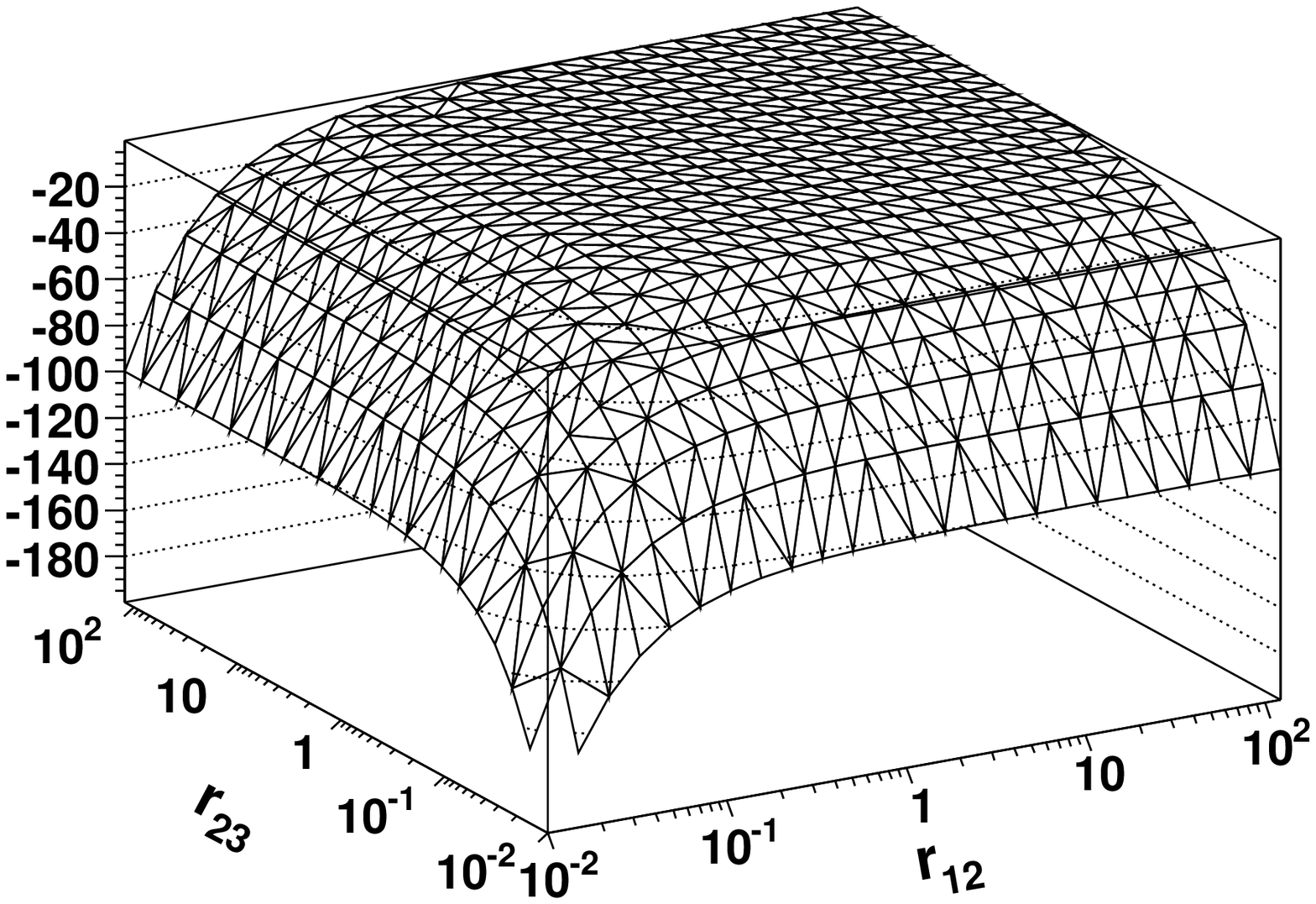} 
    \end{array}$
  \end{center}
  \spacing{1}
  \caption{The inter-particle potential $V(r_{ij}, M)\, / \, m\alpha^2_g$ of equation (\ref{EQ:V_arbitrary}) for $\kappa_1 = 0.1$, $M = \ds\frac{\mu}{m\alpha_g} = 0.1$ and $r_{13} = 0.01 (\tx{top left}), 0.1(\tx{top right}), 1.0 (\tx{bottom left}), 10 (\tx{bottom right})$ where $r_{ij} = \Delta_{ij} \, m \alpha_g$.}
  \label{FIG:V_surface}
\end{figure}

Finally, the inter-particle potential for arbitrary inter-particle distances is discussed in this section. There are no analytical solutions available for $V_C$, equation (\ref{EQ:V_C123A}), and the potential is calculated numerically with the help of the GNU Scientific Library. Using the Gaussian parametrization given by equation (\ref{EQ:V_C123C}) and with the previous choice of units, one obtains
\begin{multline}
  V(x_{ij}, \mu )  = V(r_{ij}, M) = - \, m \, \alpha_g^2 \, \Bigg\{ \, \frac{\e^{-r_{12} M}}{r_{12}} + \frac{\e^{-r_{23} M}}{r_{23}} + \frac{\e^{-r_{13} M}}{r_{13}}  \\
    + \frac{\kappa_1}{4 \, \pi}\int_0^\infty  \, d\beta_{1,2,3} \; \frac{\e^{-M^{2}(\beta_{1}+\beta_{2}+\beta_{3})}}{(\beta_{1}\beta_{2}+\beta_{1}\beta_{3}+\beta_{2}\beta_{3})^{3/2}} \, \exp\left(-\frac{r_{12}^2 \beta_1 + r_{23}^2 \beta_2 + r_{13}^2\beta_3} {4(\beta_1\beta_2 + \beta_1\beta_3 + \beta_2\beta_3)}\right) \Bigg\} 
  \label{EQ:V_arbitrary}
\end{multline}
where $r_{ij} = x_{ij} \, m \alpha_g$ are the dimensionless inter-particle distances and $x_{ij} = |\x_i - \x_j|$. Four surface plots of equation (\ref{EQ:V_arbitrary}) for different values of $r_{13}$ are given in Figure \ref{FIG:V_surface}. Note how the inter-particle potential rises with increasing separations among particles. However, these plots show that the inter-particle potential is of non-confining nature as the curves do not cross the zero plane. 

\section{Four-Particle State}
\label{SEC:FOUR}
It has been shown above that the cubic $\cH_{I_2}$ interaction term of the Hamiltonian affects the inter-particle interaction and consequently the energy spectrum for a three-particle system. However, the three-particle trial state (\ref{EQ:TRIALTHREE}) does not probe the quartic $\cH_{I_3}$ interaction term of the Hamiltonian. Thus, one must consider a system of four identical particles and examine how both non-linear terms of the Hamiltonian, $\cH_{I_2}$ and $\cH_{I_3}$, affect the inter-particle potential. 

\begin{figure}[t]
  \center{
    \includegraphics[scale=0.6]{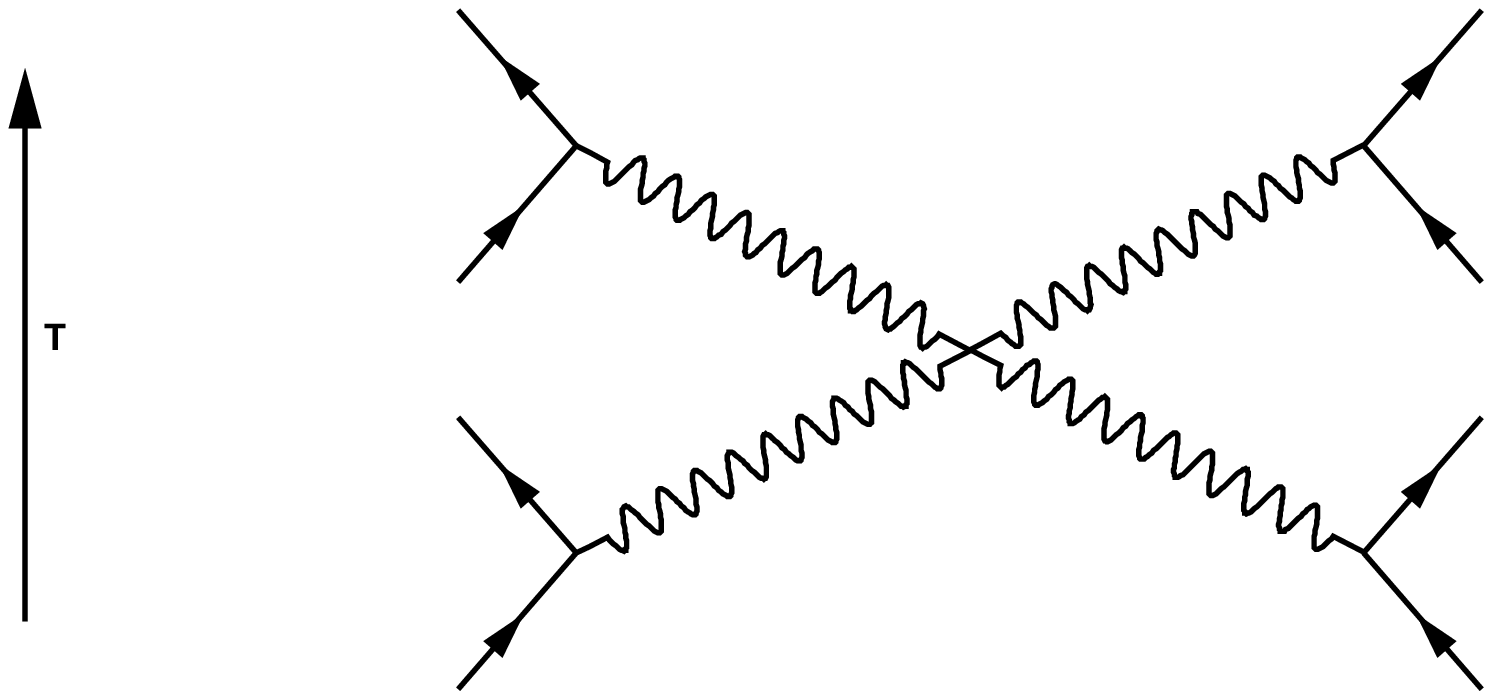}
  }
  \spacing{1}
  \caption{The four-chion propagator vertex corresponding to the quartic interaction kernel $\cQ_{4,4}$ equation (\ref{EQ:Q44}). The two propagators on both sides should actually overlap (impossible to draw) such that they are perpendicular to the direction of time.}
  \label{FIG:FD3}
\end{figure}

The four identical particle trail state analogous to equation (\ref{EQ:TRIALTHREE}) is given by 
\begin{equation}
    |\Psi_{4}\ket = \ds \int d\p_{1..4} \; F(\p_{1..4}) \; a^{\dagger}(\p_1) \, a^{\dagger}(\p_2) \, a^{\dagger}(\p_3) \, a^{\dagger}(\p_4) \, |0\ket.
    \label{EQ:TRIALFOUR}
\end{equation}
For this four-particle trial state to be an eigenvector of the momentum operator (\ref{EQ:MO}), one requires that $\hat{\bf P} \, | \Psi_4 \ket = {\bf Q} \,| \Psi_4 \ket$ which can be achieved with the choice $F(\p_{1..4}) = \de(\p_1 + \p_2 + \p_3 + \p_4 - {\bf Q}) \, f(\p_{1,2,3})$. Thereupon, as for the particle-antiparticle and the three-particle cases, the wavefunctions will be of the form where the centre of mass motion is completely separable for this identical four-particle system. It is convenient to write everything in terms of the completely symmetrized function $F_S$ because the trial state (\ref{EQ:TRIALFOUR}) is completely symmetric under interchanges of the momentum variables. The symmetrized function is 
\begin{equation}
  F_S(\p_{1..4})  = \sum_{i_1, i_2, i_3, i_4}^{24} \,   F(\p_{i_1, i_2, i_3, i_4}) 
\end{equation}
where the summation is on the 24 permutations of the indices $1, 2, 3$ and $4$. The matrix element $\bra \Psi_4 | \, \hat{H} \, - E \,| \Psi_4 \ket$ is calculated and the variational derivative with respect to $F^{\ast}$ is found and set to zero (refer to section \ref{SEC:scalar_four} of Appendix A for details). This leads to the following four identical particle relativistic equation for the function $F_S$ (in momentum space): 
\begin{multline}
  F_S(\p_{1..4}) \, \big(\omega_{\p_1} + \omega_{\p_2} + \omega_{\p_3} + \omega_{\p_4} - E\big) = \int d\p^{\prime}_{1,2} \; \cY_{4,4}(\p^{\prime}_{1..4}, \p_{1..4}) \, F_S(\p^{\prime}_{1..4}) \\ + \int d\p^{\prime}_{1..4} \; \cC_{4,4}(\p^{\prime}_{1..4}, \p_{1..4}) \, F_S(\p^{\prime}_{1..4}) + \int d\p^{\prime}_{1..4} \; \cQ_{4,4}(\p^{\prime}_{1..4}, \p_{1..4}) \, F_S(\p^{\prime}_{1..4}).
  \label{EQ4}
\end{multline}
The Yukawa and cubic interaction kernels $\cY_{4,4}$ and $\cC_{4,4}$ are similar in structure to those of the three-particle trial state except there is dependence on an extra momentum coordinate (details in section \ref{SEC:scalar_four} of Appendix A). The relativistic quartic interaction kernel is 
\begin{multline}
  \cQ_{4,4}(\p^{\prime}_{1..4}, \p_{1..4}) = \frac{g^4\sigma}{16}\frac{1}{(2\pi)^9} \sum^{24}_{i_1, i_2, i_3, i_4} \, \frac{ \de(\p^{\prime}_1 + \p^{\prime}_2 + \p^{\prime}_3 + \p^{\prime}_4 - \p_1 -\p_2 - \p_3 -\p_4)}{\ds\sqrt{\omega_{\p^{\prime}_1}\omega_{\p^{\prime}_2}\omega_{\p^{\prime}_3}\omega_{\p^{\prime}_4}\omega_{\p_1}\omega_{\p_2}\omega_{\p_3}\omega_{\p_4}}} \\
  \times \left[\frac{1}{\mu^2-(p^{\prime}_1+p^{\prime}_2+p^{\prime}_3-p_{i_1}-p_{i_2}-p_{i_3})^2} \right.\\ 
    \times\left.\frac{1}{\mu^2-(p^{\prime}_{1}-p_{i_1})}\frac{1}{\mu^2-(p^{\prime}_2-p_{i_2})^2}\frac{1}{\mu^2-(p^{\prime}_3-p_{i_3})^2}\right].
  \label{EQ:Q44}
\end{multline}
where the summation is on the 24 permutations of the indices $1, 2, 3$ and $4$. This contribution to the potential corresponds to a four-chion propagator vertex shown in Figure \ref{FIG:FD3}. The non-relativistic limit of the interactions in the coordinate representation is examined below.
In the non-relativistic limit, the Yukawa and cubic kernel reduce just as their three-particle trial state counter-parts. There are six Yukawa interaction terms which can be worked out in analytical form:
\begin{equation}
  V_Y(\x_{1..4}, \mu)= - \alpha_g \Bigg\{\frac{\e^{-\mu x_{12}}}{x_{12}}+\frac{\e^{-\mu x_{13}}}{x_{13}}+\frac{\e^{-\mu x_{14}}}{x_{14}}+\frac{\e^{-\mu x_{23}}}{x_{23}}+\frac{\e^{-\mu x_{24}}}{x_{24}}+\frac{\e^{-\mu x_{34}}}{x_{34}}\Bigg\},
\end{equation}
where $\alpha_g = \ds\frac{g^2}{16 \, \pi \, m^2}$ is the dimensionless coupling constant and $x_{ij} = x_{ji} = |\x_j - \x_i|$ are the inter-particle distances as before. 
The cubic interaction kernel for the non-relativistic four-particle case reduces to a sum of four terms, i.e. one for every three-way interaction:
\begin{multline}
  V_C(\x_{1..4}, \mu) = -\alpha_{\eta} \, \int d\q_{1..4}\, \\
  \Bigg\{\frac{\prod_{i}^{4}\e^{-i \, \q_{i}\cdot\x_{i}} \, \delta(\q_1+\q_2+\q_3) \, \delta(\q_{4})}{\left(\mu^2+\q_{1}^2\right)\left(\mu^2+\q_{2}^2\right)\left(\mu^2+\q_{3}^2\right)} +  \frac{\prod_i^4 \e^{- i \, \q_i\cdot\x_i} \, \delta(\q_1+\q_2+\q_4) \, \delta(\q_3)}{\left(\mu^2+\q_1^2\right)\left(\mu^2+\q_2^2\right)\left(\mu^2+\q_4^2\right)} \\
  + \frac{\prod_{i}^{4}\e^{- i \, \q_{i}\cdot\x_{i}} \, \delta(\q_1+\q_3+\q_4) \, \delta(\q_{2})}{\left(\mu^2+\q_1^2\right)\left(\mu^2+\q_3^2\right)\left(\mu^2+\q_4^2\right)} +  \frac{\prod_i^4 \e^{- i \, \q_i\cdot\x_i} \, \delta(\q_2+\q_3+\q_4) \, \delta(\q_1)}{\left(\mu^2+\q_2^2\right)\left(\mu^2+\q_3^2\right)\left(\mu^2+\q_4^2\right)} \Bigg\} 
  \label{EQ:V_C1234},
\end{multline}
where $\alpha_{\eta}=\ds\frac{3 \, g^3 \, \eta}{256\pi^6 m^3}$ is a coupling constant with dimensions of mass (refer to section \ref{SEC:scalar_four} of Appendix A for details). The delta function with a single momentum variable in every term indicates that the particle carrying that momentum is a ``spectator'' of the three-way interaction. This expression is obtained by following similar steps as those leading to equation (\ref{EQ:V_C123}). 
The quartic interaction kernel for the four-particle system in the non-relativistic limit reduces to
\begin{equation}
  V_Q(\x_{1..4}) = \alpha_{\sigma} \int d\q_{1..4} \; \frac{\prod_i^4 \e^{-\i \q_i\cdot\x_i} \, \de(\q_{1}+\q_2+\q_3+\q_4)}{\left(\mu^2+\q_1^2\right)\left(\mu^2+\q_2^2\right)\left(\mu^2+\q_3^2\right)\left(\mu^2+\q_4^2\right)},
  \label{EQ:V_Q1234}
\end{equation}
where $\alpha_\sigma=\ds\frac{3 \, g^4 \, \sigma}{1024\pi^9 m^4}$ is a dimensionless coupling constant. This expression is obtained by using similar steps as those leading to equation (\ref{EQ:V_C1234}). The cubic $V_C$ and quartic $V_Q$ contributions to the inter-particle potential are both symmetric under 3D rotations and translations even though it is not readily evident from equations (\ref{EQ:V_C1234}) and (\ref{EQ:V_Q1234}). In Appendix A, it is explicitly shown, using Gaussian parametrization, that these equations indeed depend only on the inter-particle distances, i.e. $V_C = V_C(x_{ij})$ and $V_Q = V_Q(x_{ij})$.
The expressions for $V_C$ and $V_Q$ can be simplified to three-dimensional quadratures:
\begin{align}
  V_C(\x_{1..4}) = & - \alpha_{\eta} \, \pi^3 \int d\x \, \Bigg\{\frac{\e^{-\mu|\x_1 + \x|}}{|\x_1 + \x|}\frac{\e^{-\mu|\x_2 + \x|}}{|\x_2 + \x|}\frac{\e^{-\mu|\x_3 + \x|}}{|\x_3 + \x|} + \frac{\e^{-\mu|\x_1 + \x|}}{|\x_1 + \x|}\frac{\e^{-\mu|\x_2 + \x|}}{|\x_2 + \x|}\frac{\e^{-\mu|\x_4 + \x|}}{|\x_4 + \x|} \nonumber \\
  & \F! + \frac{\e^{-\mu|\x_1 + \x|}}{|\x_1 + \x|}\frac{\e^{-\mu|\x_3 + \x|}}{|\x_3 + \x|}\frac{\e^{-\mu|\x_4 + \x|}}{|\x_4 + \x|} + \frac{\e^{-\mu|\x_2 + \x|}}{|\x_2 + \x|}\frac{\e^{-\mu|\x_3 + \x|}}{|\x_3 + \x|}\frac{\e^{-\mu|\x_4 + \x|}}{|\x_4 + \x|}\Bigg\} ,
  \label{EQ:V_C1234A} \\
  V_Q(\x_{1..4}) = & \, \alpha_{\sigma} \, \pi^4 \int d\x \; \frac{\e^{-\mu|\x_1 + \x|}}{|\x_1 + \x|}\frac{\e^{-\mu|\x_2 + \x|}}{|\x_2 + \x|}\frac{\e^{-\mu|\x_3+\x|}}{|\x_3 + \x|}\frac{\e^{-\mu|\x_4 + \x|}}{|\x_4 + \x|}.
  \label{EQ:V_Q1234A}
\end{align}
Here, analogously to the three-particle case (\ref{EQ:V_C123A}), the integral expression for $V_C$, equation (\ref{EQ:V_C1234A}), is convergent for $\mu > 0$. However, if $\mu=0$ the integrand of $V_C$ goes as $|\x|^{-1}$ for large $|\x|$ and the integral diverges. The integrand in $V_Q$, equation (\ref{EQ:V_Q1234A}), behaves as $|\x|^{-2}$ for large $|\x|$ and the integral remains finite for $\mu \ge 0$. It is shown below that the contribution of $V_Q$ to the total inter-particle potential of the four identical particle system is a diminishing correction as the inter-particle separations increases. On the other hand, for small separations, depending on the values of the coupling constants, it can have a significant effect.

The expressions for $V_C$ and $V_Q$, equations (\ref{EQ:V_C1234A}) and (\ref{EQ:V_Q1234A}), analogously to equation (\ref{EQ:V_C123B}), can be written as
\begin{align}
  V_C(\x_{1..4}, \mu) = & -\pi^3 \alpha_{\eta} \, \int d\v \, \frac{\e^{-\mu|\v|}}{|\v|} \, \Bigg\{\frac{\e^{-\mu|\v + \x_{21}|}}{|\v + \x_{21}|}\frac{\e^{-\mu|\v + \x_{31}|}}{|\v + \x_{31}|} + \frac{\e^{-\mu|\v + \x_{21}|}}{|\v + \x_{21}|}\frac{\e^{-\mu|\v + \x_{41}|}}{|\v + \x_{41}|} \nonumber \\
  & \F!\F!\F!\F!\H! + \frac{\e^{-\mu |\v + \x_{31}|}}{|\v + \x_{31}|}\frac{\e^{-\mu |\v + \x_{41}|}}{|\v + \x_{41}|} + \frac{\e^{-\mu|\v + \x_{32}|}}{|\v + \x_{32}|}\frac{\e^{-\mu|\v + \x_{42}|}}{|\v + \x_{42}|}\Bigg\},
  \label{EQ:V_C1234B} \\
  V_Q(\x_{1..4}, \mu) = & \, \pi^4 \alpha_{\sigma} \, \int d\v \; \frac{\e^{-\mu|\v|}}{|\v|}\frac{\e^{-\mu|\v + \x_{21}|}}{|\v + \x_{21}|}\frac{\e^{-\mu|\v +\x_{31}|}}{|\v + \x_{31}|}\frac{\e^{-\mu|\v + \x_{41}|}}{|\v + \x_{41}|},
  \label{EQ:V_Q1234B}
\end{align}
where $\x_{ij} = \x_i - \x_j$. It is impossible to obtain analytical expressions for $V_C$ and $V_Q$ in general and one must resort to numerical evaluation of these integrals. However, as for the three-particle system, there is a particular solvable case, namely when $\x_1 = \x_3$ and $\x_2 = \x_4$, which shows the general features of the total potential energy. With this restriction on the coordinates there is only one inter-particle distance $x_{21} = |\x_2 - \x_1| = |\x_4 - \x_3|$ to be concerned with.
\subsubsection*{Special Case : $V(\x_{1,2,1,2} ,\mu > 0) = V(x_{21}, \mu > 0)$.}
For the case when $\x_1 = \x_3$ and $\x_2 = \x_4$ with $\mu > 0$ the cubic potential term $V_C$ of equation (\ref{EQ:V_C1234B}) leads to four copies of what has been previously obtained for the three-particle case, namely, 
\begin{multline}
  V_C(\x_{1,2,1,2} ,\mu > 0) = V_C(x_{21}, \mu > 0) \\ 
  = -\frac{8 \, \pi^4 \, \alpha_{\eta}}{x_{21} \mu} \Bigg\{\e^{-x_{21} \mu}\ds\left[\ln\left(3\right) - \E1\left(x_{21}\mu\right)\right] + \e^{x_{21}\mu} \, \E1\left(3x_{21}\mu\right) \Bigg\}.
  \label{EQ:V_C1212MU}
\end{multline}
The single difference is in the numerical factor of four in front. 
%

\begin{figure}
  \begin{center}$
    \begin{array}{c}
      \includegraphics[scale = 0.6]{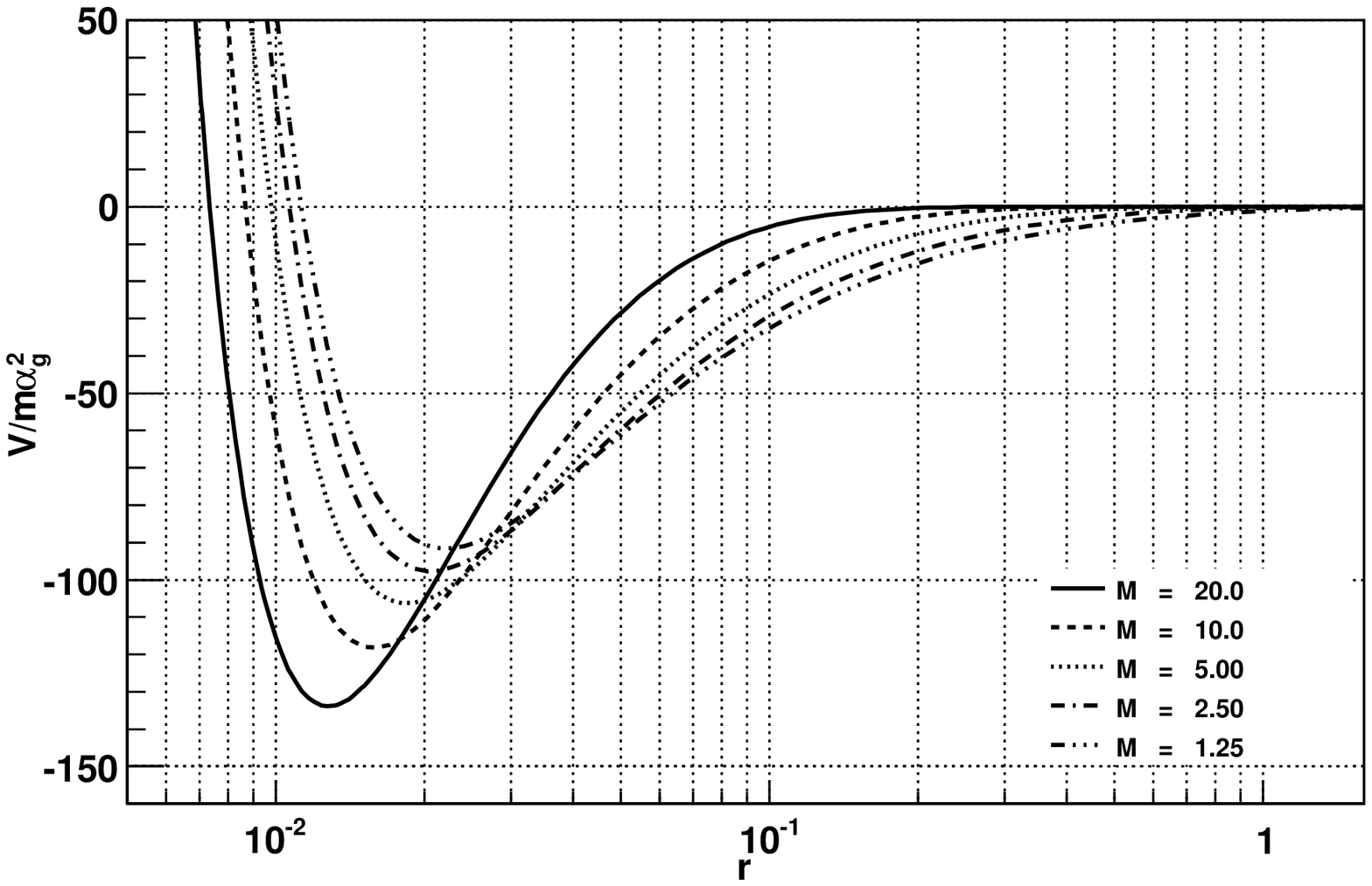}\\
      \includegraphics[scale = 0.6]{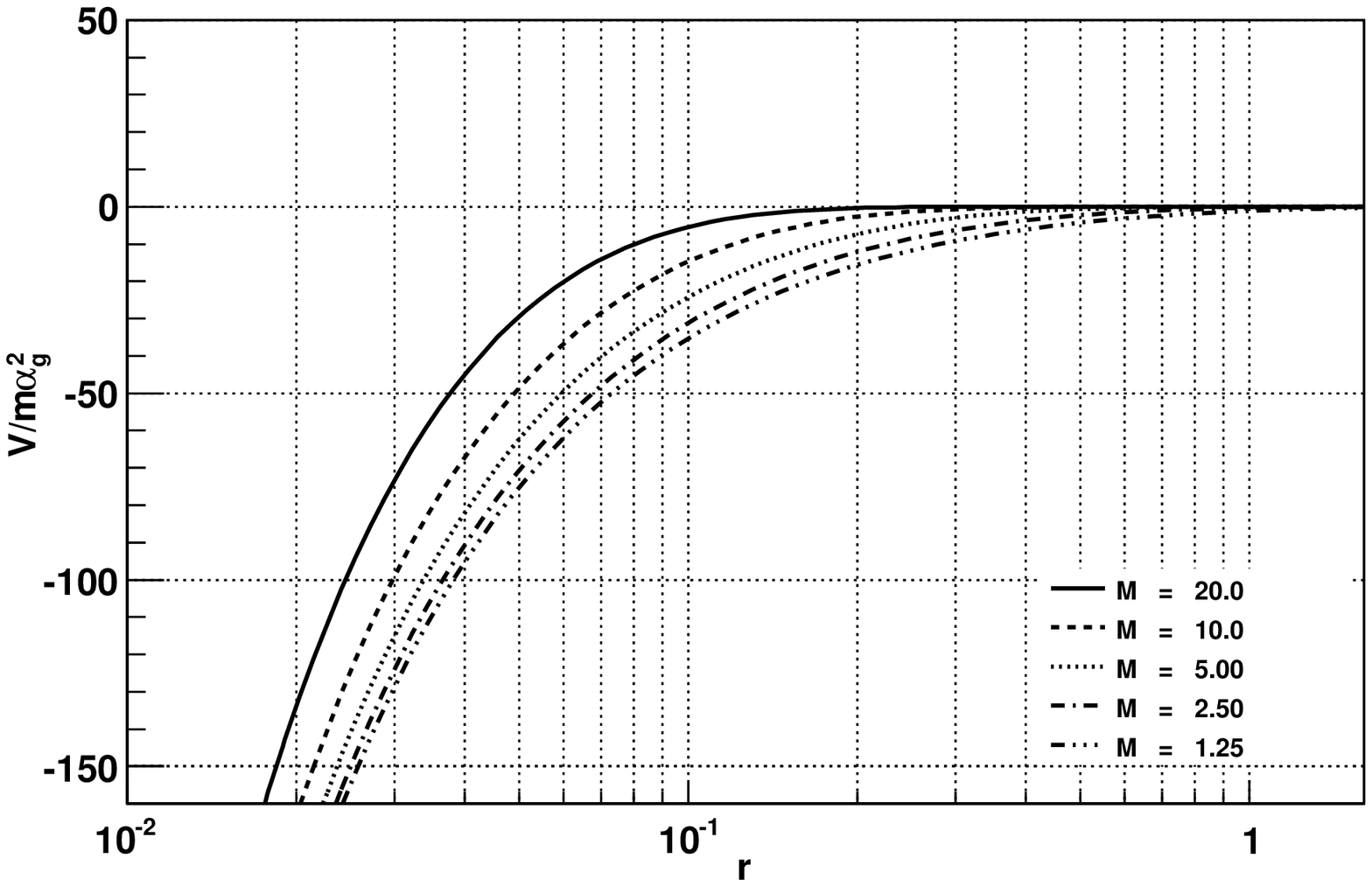}
    \end{array}$
    \spacing{1}
    \caption{The inter-particle potential $V(x_{21}, \mu > 0)$ of equation (\ref{EQ:V_1212MU}) in units of $m\alpha^2_g$ as a function of $r = x_{21} \, m \alpha_g$ for the indicated values of $M$, $\kappa_1 = 0.01$, and $\kappa_2 = 0.1$ (top) and $0.005$ (bottom).}
      \label{FIG:V_1212MU}
  \end{center}
\end{figure}

%
The quartic potential term $V_Q$, equation (\ref{EQ:V_Q1234B}), can be reduced to an evaluation of a single quadrature (refer to section \ref{SEC:scalar_four} of Appendix A for details):
\begin{align}
  V_Q(\x_{1,2,1,2}, \mu \ge 0) =  & \, V_Q(x_{21}, \mu \ge 0) = \nn 
  \\ & \frac{2 \, \pi^5 \, \alpha_\sigma}{x_{21}} \ds\left[\int^{x_{21}}_0 \frac{dv}{v} \, \e^{-2\mu v} \bigg\{\E1(2\mu(x_{21}-v)) - \E1(2\mu(x_{21}+v)) \bigg\}\right.\nn \\
    & \F! + \left.\int_{x_{21}}^\infty \frac{dv}{v} \, \e^{-2\mu v} \bigg\{\E1(2\mu(v-x_{21})) - \E1(2\mu(x_{21}+v)) \bigg\}\right].
  \label{EQ:V_Q1212MU}
\end{align}
The integrals in the expression above are not expressible in terms of common analytic functions, thus one has to evaluate them numerically using, as before, the GNU Scientific Library. 
The total potential energy $V = V_Y + V_C + V_Q$ in the units of $m\alpha_g^2$ as a function of $r = x_{21} \, m\alpha_g$ for the case when $\mu > 0$, suppressing two Yukawa singularities, is
\begin{align}
  V(\x_{1,2,1,2}, \mu > 0) & \; =  V(x_{12}, \mu > 0) = m \, \alpha^2_g \, \Bigg\{  \nn \\
   &  - \, 4 \, \frac{\e^{- r M}}{r} - 2 \, \kappa_1\ds\left(\ln(3) \, \frac{\e^{-rM}}{rM} - \frac{\e^{-rM}}{rM}\E1(rM) + \frac{\e^{rM}}{rM}\E1(3 \, rM)\right) \nn \\
    & \H! + \frac{\kappa_2}{r} \ds\left[\int_0^{rM} \frac{dw}{w} \, \e^{- 2 w} \, \bigg\{ \E1(2(rM - w))- \E1(2(rM + w)) \bigg\} \right.\nn \\
    & \F!\F! \left. + \int_{rM}^\infty \frac{dw}{w} \, \e^{- 2w} \, \bigg\{ \E1(2(w - r M)) - \E1(2(rM + w))\bigg\}\right]\Bigg\} ,
  \label{EQ:V_1212MU}
\end{align}
where $w = v \, \mu $ is a dimensionless integration variable, $M = \ds\frac{\mu}{m \, \alpha_g}$ is the dimensionless mass parameter $\kappa_1 = \ds\frac{4 \, \pi^4 \, \alpha_\eta}{m \, \alpha^2_g}$ and $\kappa_2 = \ds\frac{\alpha_\sigma}{m\, \alpha_g^2}$. Two plots of equation (\ref{EQ:V_1212MU}) are shown in Figure \ref{FIG:V_1212MU} for the parameter values $\kappa_1 = 0.01$ and  $\kappa_2 = 0.1, 0.005$. The choice $\kappa_2 = 0.1$ yields a potential which is repulsive for small separations and possesses a trough. The trough's depth increases with increasing values of the mediating field mass parameter $M$. For large separations and both values of $\kappa_2$, the potential approaches zero without crossing the zero of energy. Both choices of the parameter $\kappa_2$ do not exhibit any confining features in the inter-particle potential; although both can support bound and scattering states. 

\section{Improved Particle-Antiparticle State}
It has been shown in the previous sections that the non-linear terms of the Hamiltonian (\ref{EQ:MODHAMI}) alter the inter-particle potential. The simple particle-antiparticle trial state (\ref{EQ:TRIALTWO}) is incapable of probing these terms. Hence, to observe the effects of the non-linear terms on the particle-antiparticle system, one must consider a multi-component trial state, such as, 
\begin{align}
  | \Psi_{t}\ket = \ds \; C_F \; | \Psi_2 \ket  \; +  \; \frac{C_G}{\sqrt{4}} \; | \Psi_4 \ket,
  \label{EQ:TRIALTWOFOUR}
\end{align}
where the quantities $C_F$ and $C_G$ are linear coefficients specifying the relative size of each basis component, and
\begin{gather}
  | \Psi_2 \ket = \int d\p_{1,2}  \; F(\p_{1,2}) \,  a^{\dagger}({\bf p}_{1}) \, b^{\dagger}({\bf p}_{2}) |0\ket,  \\
  | \Psi_4 \ket = \int d\p_{1..4} \; G(\p_{1...4}) \; a^{\dagger}(\p_1) \, b^{\dagger}(\p_2) \, a^{\dagger}(\p_3) \,b^{\dagger}(\p_4) |0\ket,
\end{gather}
are the adjustable functions containing variational parameters which describe the one-pair and two-pair components of the system respectively. The factor of $1 /\sqrt{4}$ is inserted for convenience to account for the identical particles in the trial state. Note that no such numerical factor is necessary with mono-component trial state. The linear coefficients must be determined variationally, along with other parameters included in $F$ and $G$. The linear coefficients are not independent because of the normalization of $| \Psi_t \ket$. They obey the condition
\begin{equation}
  |C_F|^2 + \frac{|C_G|^2}{4} = 1.
\end{equation}

The improved particle-antiparticle trial state (\ref{EQ:TRIALTWOFOUR}) is flexible enough to describe the transition from a one-pair to a two-pair state, i.e. the so-called string breaking effect. However, it is not the purpose of this research to investigate this phenomenon. 

The matrix element $\bra \Psi_t | \, \hat{H} - E \, | \Psi_t \ket$ consists of many contributions (see section \ref{SEC:scalar_two_four} of Appendix A for details). Be it as is, many of them are irrelevant to the calculation of the inter-particle potential and hence are not included. Varying the matrix element with respect to the functions $F^\ast$ and $G^\ast$ leads to the following system of coupled relativistic equations:
\begin{align}
 F&(\p_{1,2}) \left(\omega_{\p_1} + \omega_{\p_2} - E\right) = \int d\p^{\prime}_{1, 2} \; {\cal Y}_{2,2} (\p^{\prime}_{1,2}, \p_{1,2}) \, F(\p^{\prime}_{1,2}) \nn \\	
  & \F!\F!\F!+ R \int d\p^{\prime}_{1...4} \; \bigg\{{\cal Y}_{2,4}(\p^{\prime}_{1...4}, \p_{1,2}) + {\cal C}_{2,4}(\p^{\prime}_{1...4}, \p_{1,2}) \bigg\}\, G\left(\p^{\prime}_{1...4}\right)
  \label{EQ:EQT_ONE}
  \\
  G& \, (\p_{1..4}) \left(\sum_{i=1}^4 \omega_{\p_i}  - E\right) = R^{-1} \int d\p^{\prime}_{1,2} \; \bigg\{{\cal Y}_{4,2} (\p^{\prime}_{1,2}, \p_{1...4}) + {\cal C}_{4,2}(\p^{\prime}_{1,2}, \p_{1...4}) \bigg\} \, F(\p_{1,2}^{\prime}) \nn \\
  & + \, \int d\p^{\prime}_{1...4} \; \bigg\{{\cal Y}_{4,4}(\p^{\prime}_{1...4},\p_{1...4}) +  {\cal C}_{4,4}(\p^{\prime}_{1...4}, \p_{1...4}) + {\cal Q}_{4,4}(\p^{\prime}_{1...4}, \p_{1...4}) \bigg\}\, G\left(\p^{\prime}_{1...4}\right)
  \label{EQ:EQT_TWO}
\end{align}
where the quantity $R = \ds\frac{C_G}{C_F}$ is the ratio that specifies the relative contribution of each basis state. The quantity $R$ is a complex number and, in general,  its value is determined via a variational calculation of the energy. 

Equations (\ref{EQ:EQT_ONE}) and (\ref{EQ:EQT_TWO}) are relativistic coupled equations, in momentum space, that describe the processes involving one pair and two pair bound state systems but not the processes that involve the emission or absorption of the mediating field quanta. Even though one can, in principle, calculate the energy spectrum from these coupled integral equations, it is more practical to solve them approximately using the variational method which is based on the evaluation of the matrix element of the Hamiltonian. The relativistic kinematics in these equations are described without approximation. On the other hand, the relativistic interactions are approximated by the kernels $\cY_{2,2}$, $\cY_{2,4}$, $\cY_{4,2}$ and $\cY_{4,4}$ arising from the Yukawa (i.e. linear) term of the Hamiltonian density (\ref{EQ:MODHAMY}), $\cC_{2,4}$, $\cC_{4,2}$ and $\cC_{4,4}$ from the cubic term (\ref{EQ:MODHAMC}), and $\cQ_{4,4}$ from the quartic term (\ref{EQ:MODHAMQ}). As explained above, these kernels are adequate approximations when the interaction couplings are weak. The $\cY$ type kernels correspond to one mediating field quanta exchange and virtual-annihilation interactions, diagrammatically illustrated in Figure \ref{FIG:FD1}. The $\cC$ and $\cQ$ type kernels are three- and four-point interactions, illustrated by Figures \ref{FIG:FD2} and \ref{FIG:FD3} respectively.

The primary goal of this dissertation is to study the inter-particle potential, henceforth the emphasis is placed on equation (\ref{EQ:EQT_ONE}). The non-relativistic limit of equation (\ref{EQ:EQT_ONE}) is found and Fourier-transformed to coordinate space. The symmetric definitions of Fourier transform is used:
\begin{align}
  F(\p_{1,2}) = & \, \frac{1}{(2\pi)^3}\int d\x_{1,2} \, \e^{i \, \p_1\cdot\x_1} \, \e^{i \, \p_2\cdot\x_2} \, F(\x_{1,2}), \\
  G(\p_{1..4}) = & \, \frac{1}{(2\pi)^6}\int d\x_{1..4} \, \prod_{i = 1}^{4} \e^{i \, \p_i\cdot\x_i} \, G(\x_{1..4}).
\end{align}

The non-relativistic equation in coordinate space takes on the following form:
\begin{multline}
  -\left(\frac{\nabla^2_{1}}{2m} + \frac{\nabla^2_{2}}{2m} + \epsilon \right) \, F(\x_{1,2}) = \\ \left[ \; Y_{2,2}(\x_{1,2})  
  + R \; \int d\x_{3,4} \, \Big\{ Y_{2,4}(\x_{1..4}) + C_{2,4}(\x_{1..4})\Big\} \, \frac{G(\x_{1..4})}{F(\x_{1,2})} \; \right] F(\x_{1,2}),
  \label{EQ:EQT_POS}
\end{multline}
where $\epsilon = E - 2 \, m$ is non-relativistic energy. The two terms containing the function $G$ are multiplied and divided by the function $F$. In this manner, the coefficients of $F$ in the second line of equation (\ref{EQ:EQT_POS}) can be identified as the correction to the Yukawa potential $Y_{2,2}$. Working entirely in the non-relativistic limit implies keeping only the leading terms in the inverse power of the parameter $m$. The higher order terms correspond to relativistic corrections; which shall not be considered in this non-relativistic approximation. The non-relativistic kernels are
\begin{gather}
  Y_{2,2}(\x_{1,2}) = \, \alpha_g \, \frac{\e^{-\mu x_{21}}}{x_{21}}, \\
  Y_{2,4}(\x_{1..4}) = \, \alpha_g \, \left( \frac{\e^{-\mu x_{14}}}{x_{14}} + \frac{\e^{-\mu x_{24}}}{x_{24}} \right) \, \de(\x_3 - \x_4), \\
  C_{2,4}(\x_{1..4}) =  \, \alpha_\eta \, \int d\q_{1,2} \, \frac{\e^{- i \q_1\cdot(\x_4 - \x_1)} \, \e^{- i \q_2\cdot(\x_4 - \x_2)}}{\left(\mu^2+\left(\q_1 + \q_2\right)^2\right) \left(\mu^2 + \q_1^2\right) \left(\mu^2 + \q_2^2\right)} \; \de(\x_3 - \x_4),
\end{gather}
where $x_{ij} = |\x_i - \x_j|$ are the inter-particle distances, $\alpha_g = \ds\frac{g^2}{16 \, \pi \, m^2}$ is the dimensionless coupling constant, and $\alpha^\prime_\eta = \ds \frac{g^3 \, \eta}{256 \, \pi^6 \, m^3}$ is the coupling constant with dimensions of mass. Note the slightly altered definition, i.e. $\alpha_\eta =  3 \,\alpha^\prime_\eta $. The analytical expression for $Y_{2,2}$ and $Y_{2,4}$ can be determined using the standard technique of integration in the complex plane. The expression for $C_{2,4}$ can not be simplified in terms of the known common functions and is thus left in this form. It should be mentioned that these kernels could have also been obtained by employing only particles, i.e. replacing anti-particles by particles, in the trial state (\ref{EQ:TRIALTWOFOUR}).

In order to proceed with the derivation of the inter-particle potential the functions $F$ and $G$ must be specified. It is difficult to solve equations (\ref{EQ:EQT_ONE}) and (\ref{EQ:EQT_TWO}) for these functions even approximately. Instead, in the spirit of the variational method, one can specify ans\"atze for them. 

The trustworthiness of any variational calculation rests on the choice of ans\"atze. It is best to require for an ansatz to be an eigenstate of the momentum operator, since then, in the centre of mass reference frame, an energy calculation corresponds to the binding energy of the system. This condition can be fulfilled if there is dependence on the coordinate differences only. At sufficiently low energies, the variables $\x_3$ and $\x_4$, describing virtual particles and as such, are integrated out. It is then prudent to consider functions where the centre of mass motion of the particles described by the coordinates $\x_1$ and $\x_2$ factors out:
\begin{align}
  F(\x_{1,2}) = & \,\frac{1}{a^{3/2}} \exp{\left(-i \, {\bf Q}\cdot\ds\frac{(\x_1 + \x_2)}{2}\right)}\, \exp{\left(-\frac{|\x_{12}|}{2 \, a}\right)}, 
  \label{EQ:ANSATZ1}\\
  G(\x_{1..4}) = & \, \frac{1}{a^{9/2}} \exp{\left(-i \, {\bf Q}\cdot\ds\frac{(\x_1 + \x_2)}{2}\right)} \nn \\
  & \F!\F!\F!\times\exp{\left(-\frac{|\x_{12}| + |\x_{13}| + |\x_{14}|+ |\x_{24}| + |\x_{24}| + |\x_{34}|}{2 \, a}\right)}, 
  \label{EQ:ANSATZ2}
\end{align}
where $\x_{ij} = \x_i - \x_j$ are the inter-coordinate vectors, ${\bf Q}$ is the total momentum of particles at $\x_1$ and $\x_2$, and $a$ is a variational parameter. The factor of two and the fractional powers of $a$ are inserted for convenience. By acting with the momentum operator in coordinate space, one can verify that both of these functions, and hence the trial state (\ref{EQ:TRIALTWOFOUR}), are eigenstates of the momentum operator with the eigenvalue of ${\bf Q}$. Note that the exponentials $\e^{-|\x_{ij}|/2 \, a}$ are suitable for the ground state of the system, and this is the state that is considered here.

\begin{figure}
  \begin{center}
      \includegraphics[scale = 0.70]{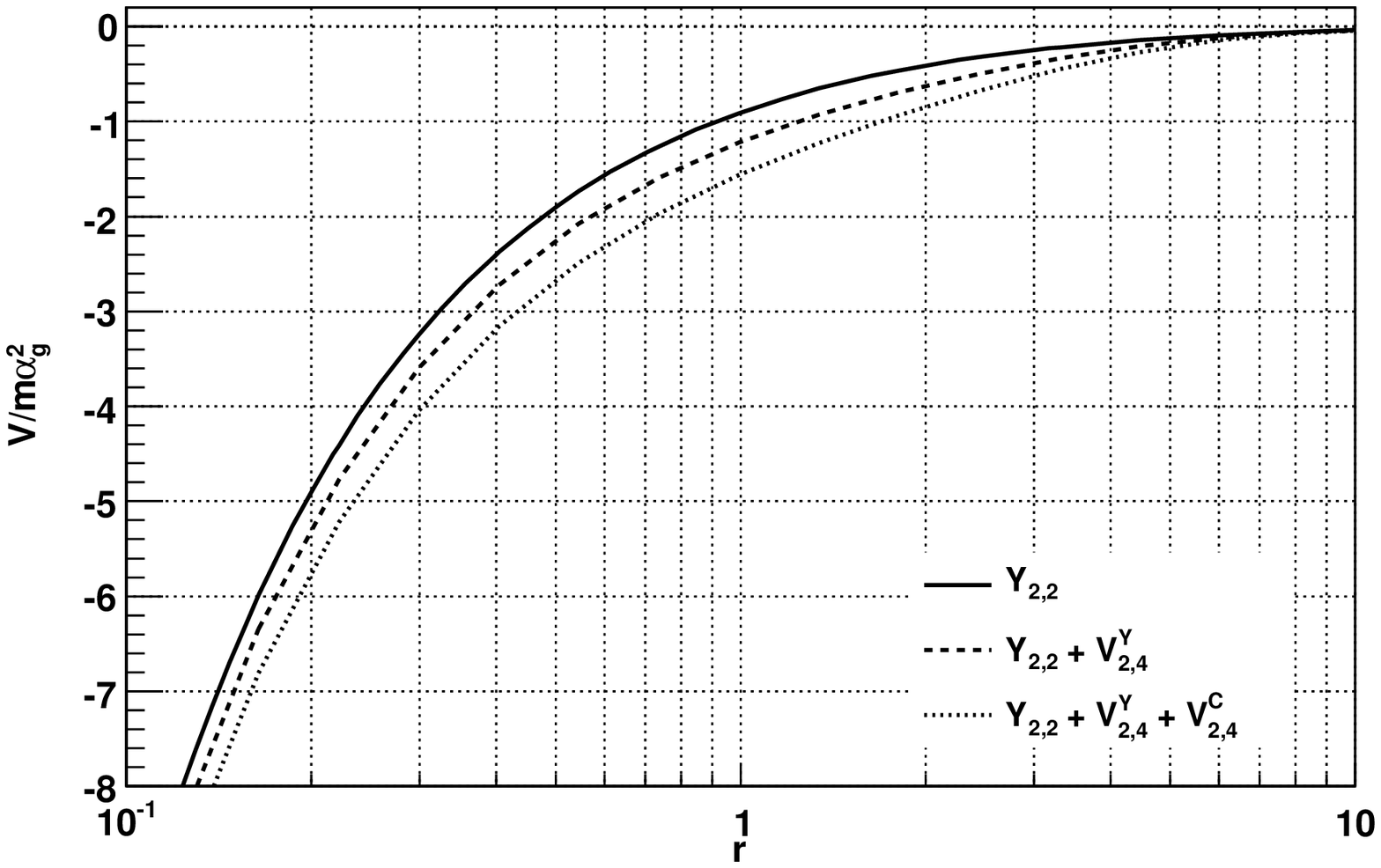} \\
  \end{center}
  \spacing{1}
  \caption{The inter-particle potential $V$, equation (\ref{EQ:V_TOT1}), as a function of distance in units of the Bohr radius with $R = M = 0.1$ and the dimensionless constant $\kappa = 0.1$. The different curves demonstrate how the various contributions alter the potential for the given choice of the parameters. }
  \label{FIG:PLOTS1}
  \begin{center}
    \includegraphics[scale = 0.70]{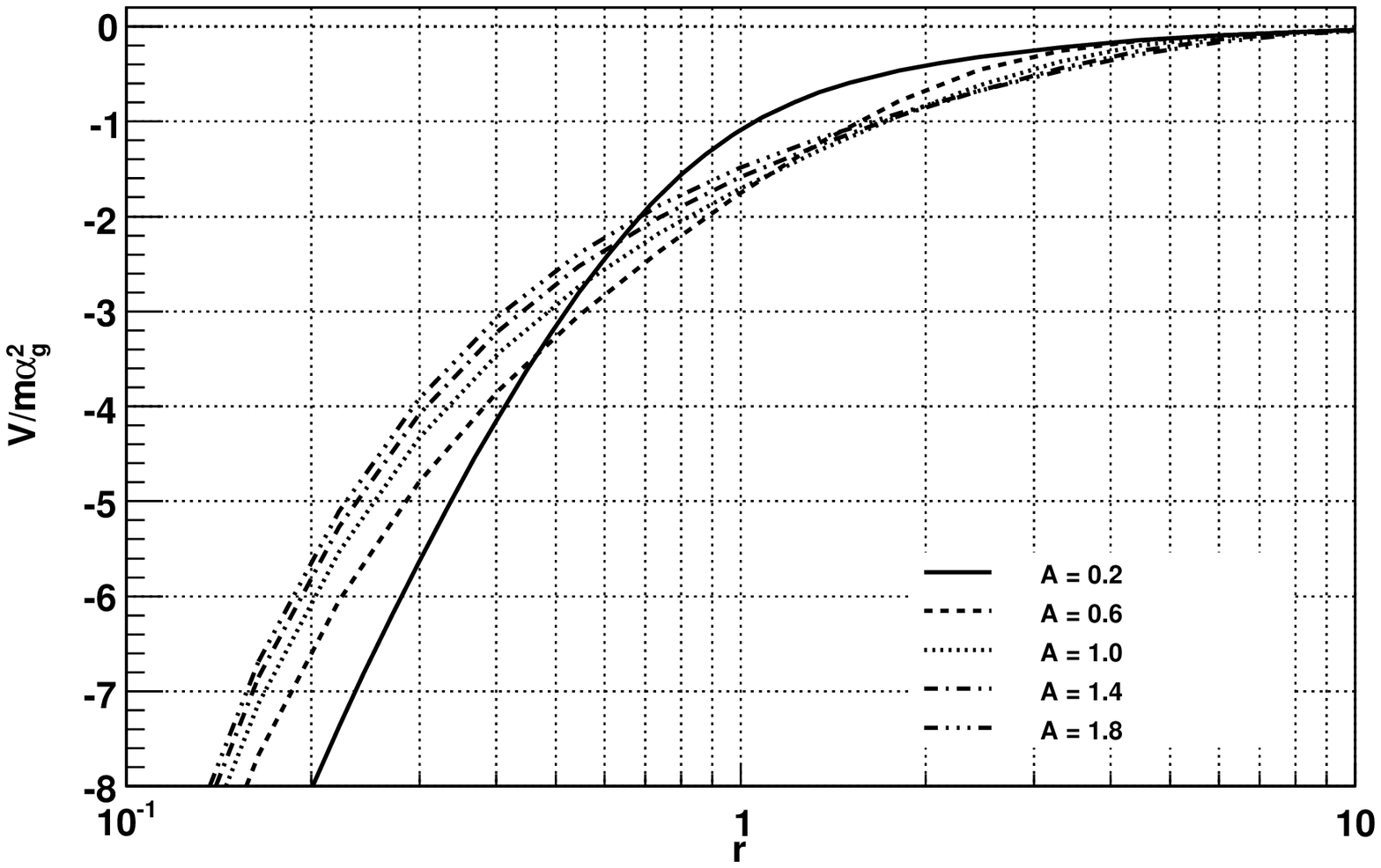} \\
  \end{center}
  \spacing{1}
  \caption{The inter-particle potential $V$, equation (\ref{EQ:V_TOT1}), as a function of distance in units of the Bohr radius for $R = M = 0.1$, $\kappa = 0.1$, and the indicated values of $A$. }
  \label{FIG:PLOTS2}
\end{figure}

Substituting equations (\ref{EQ:ANSATZ1}) and (\ref{EQ:ANSATZ2}) into (\ref{EQ:EQT_POS}), and after some extensive manipulations (see section \ref{SEC:scalar_two_four} of Appendix A for details), the total inter-particle potential in units of $m\, \alpha_g^2$ as a function of length in units of the Bohr radius $ \ds\frac{1}{ m \, \alpha_g}$ becomes
\begin{equation}
  V(x_{12}) = Y_{2,2}(x_{12}) + V^Y_{2,4}(x_{12}) + V^C_{2,4}(x_{12}),
  \label{EQ:V_TOT1}
\end{equation}
where, the contributions to the inter-particle potential emerging from the terms containing the kernels $Y_{2,2}$, $Y_{2,4}$ and $C_{2,4}$ are
\begin{align}
  Y_{2,2}(x_{1,2}) = &\,  -  m \, \alpha^2_g  \; \frac{\e^{- r \, M}}{r}, \\
  V^{Y}_{2,4}(x_{12}) = & \; R \int d\x_{3,4} \, Y_{2,4}(\x_{1..4}) \, \frac{G(\x_{1..4})}{F(\x_{1,2})} \nn \\
   = & \;  - m \, \alpha^2_g \; \frac{8 \, \pi \, R \, \e^{- r \, ( M A + 1 ) / A} }{r \, A^2 \, M^2 \, (M A + 2)^2} \, 
  \, (A \, r \, M^2 \, \e^{r \, M} + 2 \, r \, M \, \e^{r \, M} - 2 \, \e^{r \, M} + 2),
  \label{EQ:V24Y} 
\end{align}
\begin{align}
  V^C_{2,4}(x_{12}) =  & \; R \int d\x_{3,4} \, C_{2,4}(\x_{1..4}) \, \frac{G(\x_{1..4})}{F(\x_{1,2})}  \nn \\
  = & - \frac{2 \, \pi \, R \, \kappa \, m \, \alpha^2_g}{A^3} \, \int^{\infty}_0 w^2 \, dw \int^{+1}_{-1} d\alpha \int_0^{\infty} \frac{d\beta_{1,2,3} }{\beta_{123}^{3/2}} \, \e^{- M^2 \, (\beta_1 + \beta_2 + \beta_3)} \nn \\
  \H!& \times \, \exp{\left(- \frac{\beta_1 \, \r^2 + \beta_2 \, \w^2 + \beta_3 \, (\w + \r)^2}{4 \, \beta_{123}} -\frac{|\w| + |\w + \r|}{A}\right)}.
  \label{EQ:V24C}
\end{align}
%
where $\kappa = \ds\frac{\eta}{g \, \pi}$ and $\beta_{123} = \beta_{1}\beta_{2}+\beta_{1}\beta_{3}+\beta_{2}\beta_{3}$. In the above, the values of $R$ and $A$ must be determined variationally. To do such a calculation is a numerical endeavour which is not worth undertaking in this scalar model. Instead, one can reason out some acceptable values for them. Note that the parameter $R$ appears only as a multiplicative factor in front of the contributions to the potential $V^Y_{2,4}$ and $V^C_{2,4}$. In the non-relativistic regime, and with the perturbative values for the coupling constants, a reasonable value of $R$ should be on the order of less than $0.1$, i.e. the virtual component contributes less than $10\%$ to the bound state system of a particle-antiparticle. The value of the dimensionless scale parameter $A$ is the physical size on the system. In the absence of the non-linear terms of the Hamiltonian, it equals to unity for the ground state. With the perturbative coupling constants one should not expect it to deviate away from unity much.  A small range of values of the parameter $A$ is considered in Figures \ref{FIG:PLOTS1} and \ref{FIG:PLOTS2}, but it should be emphasized that the best value is believed to be in the neighbourhood of unity.

From results as seen in Figures \ref{FIG:PLOTS1} and \ref{FIG:PLOTS2}, one observes that the non-linear interaction terms of the Lagrangian density (\ref{MOD_II_LAGRANGIAN}) provide a slight modification of the Yukawa potential which supports bound states only for $\varepsilon = E - 2 \, m < 0$. However, for $M > 0$ they do not modify the potentials to a confining or a quasi-confining form which permits bound states with energies $\varepsilon > 0$ such as observed in QCD.

\chapter{Quantum Chromodynamics}
\spacing{1}
  \epigraph{In physics, you don't have to go around making trouble for yourself - nature does it for you.}{Frank Wilczek}
\spacing{2}
This chapter commences with a review of the Dirac equation and the quantization of spinors. It is prudent to proceed towards the reformulation of QCD in increments rather than attempting to do everything at once. Once the reformulation is complete the inter-quark potential of QCD bound states is investigated.
\section{QED Quantization}
The starting point of any QFT is the Lagrangian density. The free spinor Lagrangian density is
\begin{equation}
  \cL(x) = - \psib(x) \slashed{\di}  \, \psi(x) - m \, \psib(x) \, \psi(x)
  \label{EQ:LAGRFREEDIRAC}
\end{equation}
where $m$ is the bare mass parameters, the barred notation stands for $\psib = \psi^\dagger \, \gamma_0$, and the Dirac slashed notation stands for $\slashed{\di} = \di_0 \gamma^0 + \nabla \cdot {\bs \gamma}$. The gamma matrices, in the representation used in this dissertation, are defined as
\begin{equation}
  \gamma^0 = \left[\begin{array}{cc}
    I & 0 \\
    0 & - I \end{array}\right], \H!
  \gamma^i = \left[\begin{array}{cc}
    0 & \sigma_i \\
    -\sigma_i & 0 \end{array}\right],
\end{equation}
where $I$ is the $2\times 2$ identify matrix and $\sigma_i$ are the Pauli spin matrices defined by 
\begin{equation}
  \sigma_1 = \left[\begin{array}{cc}
    0 & 1 \\
    1 & 0 \end{array}\right], \H!
  \sigma_2 = \left[\begin{array}{cc}
    0 & -i \\
    i & 0 \end{array}\right], \H!
  \sigma_3 = \left[\begin{array}{cc}
    1 & 0 \\
    0 & -1 \end{array}\right].
  \label{EQ:PAULIMATRICES}
\end{equation}
Note that other representations of the Dirac matrices are connected by a similarity transformation,~\cite{Good:1955em}, 
\begin{equation}
  \gamma_{\mu}^\prime = S \, \gamma_{\mu} \, S^{-1},
\end{equation}
where $S$ is a unitary matrix.

The free Dirac equation, obtained by using the Euler-Lagrange equations, is
\begin{align}
  (i \slashed{\di} - m) \, \psi(x) = 0.
  \label{EQ:DIRAC}
\end{align}
It has four independent four-component solutions denoted by $u_{s}$ and $v_s$, where $s$ can take on two possible values $(+, -)$ designating the up and down spin orientations respectively~\cite{Gross1993}. In momentum representation, the spinors satisfy the following equations: 
\begin{align}
  (\slashed{p} - m) \, u_s & = 0, \\
  (\slashed{u} + m) \, v_s & = 0,  \\
  \bar{u}_s \, (\slashed{p} - m) & = 0, \\
  \bar{v}_s \, (\slashed{p} + m) & = 0, 
\end{align}
where the barred notation stands for $\bar{u} = u^\dagger \, \gamma_0$ and $\bar{v} = v^\dagger \, \gamma_0$ and the slashed notation stands for $\slashed{\p} = E\gamma^0 - \p \cdot {\bf{\gamma}}$. The explicit form of the spinors is
\begin{equation}
  u(\p, s) = N_\p \, \left[\begin{array}{c}
      1 \\
      \ds\frac{{\bs \sigma}\cdot\p}{\omega_\p + m} \end{array} \right] \chi_s, \H!
  v(\p, s) = N_\p \, \left[\begin{array}{c}
      \ds\frac{{\bs \sigma}\cdot\p}{\omega_\p + m} \\
      1 \end{array} \right] \eta_s
  \label{EQ:SPINORS}
\end{equation}
where $\omega_\p = \ds\sqrt{m^2 + \p^2}$, the normalization constant is
\begin{equation}
  N_\p = \ds\sqrt{\frac{\omega_\p + m}{2m}},
  \label{EQ:SPINNORMALIZATION}
\end{equation}
and, the two component spin bases are
\begin{equation}
  \chi_{+} = \left[ \begin{array}{c}
      1 \\ 0 \end{array} \right], \H! 
  \chi_{-} = \left[ \begin{array}{c}
      0 \\ 1 \end{array} \right], \H!
  \eta_{+} = \left[ \begin{array}{c}
      0 \\ 1 \end{array} \right], \H!
  \eta_{-} = \left[ \begin{array}{c}
      -1 \\ 0 \end{array} \right],
\end{equation}
the plus and minus subscripts designate the spin up and down projections of particles and anti-particles respectively. A way to check this assertion is to apply the spin angular momentum operator on the one-particle free states. 

%
%

%
The spinors obey the following orthogonality conditions:
\begin{align}
  u^{\dagger}(\p,s) \, u(\p,\sigma) & = v^{\dagger}(\p, s) \, v(\p,\sigma) = \frac{\omega_\p}{m} \, \de_{s\sigma}, \\
  u^{\dagger}(\p,s) \, v(-\p,\sigma)& = v^{\dagger}(\p,s) \, u(-\p,\sigma) = 0,
\end{align}
which one can easily verify from the explicit equations (\ref{EQ:SPINORS}). The completeness relations, with the normalization given by equation (\ref{EQ:SPINNORMALIZATION}), is 
\begin{align}
  \sum_{s=\pm} u_{\alpha}(\p,s) \ub_{\beta}(\p,s) = \left(\frac{\slashed{p} + m}{2m}\right)_{\alpha\beta} \F!   \sum_{s=\pm} v_{\alpha}(\p,s) \vb_{\beta}(\p,s) = \left(\frac{\slashed{p} - m}{2m}\right)_{\alpha\beta}
  \label{EQ:COMPLETENESS}
\end{align}
which, again, can be verified from equations (\ref{EQ:SPINORS}).

The most general solution of the free Dirac equation is the superposition
\begin{align}
  \psi_a(x) &  = \sum_{s = \pm} \int \frac{d\p}{(2\pi)^{3/2}} \ds\left(\frac{m}{\omega_\p}\right)^{1/2} \, \Bigg\{a_s(\p) \, u_a(\p,s) \, \e^{- i \, p\cdot x} + b^{\dagger}_s(\p) \, v_a(\p,s) \, \e^{i p\cdot x}\Bigg\}, \label{EQ:PSI}\\
  \psi^{\dagger}_a(x) & = \sum_{s = \pm} \int \frac{d\p}{(2\pi)^{3/2}} \ds\left(\frac{m}{\omega_\p}\right)^{1/2} \, \Bigg\{a^{\dagger}_s(\p) \, u^{\dagger}_a(\p,s) \, \e^{i \, p\cdot x} + b_s(\p) \, v^{\dagger}_a(\p,s) \, \e^{-i p\cdot x}\Bigg\} \label{EQ:PSID}.
\end{align}
where it is well known that the choices of the coefficients $a_s(\p)$, $a^\dagger_s(\p)$, $b_s(\p)$ and $b^\dagger_s(\p)$ in equations (\ref{EQ:PSI}) and (\ref{EQ:PSID}) lead to the correct spin-statistics and positive definite energy in the Dirac theory~\cite{Peskin1995}. The spinor index $a$ is explicitly included in these equations.

The canonical quantization is performed by imposing the following equal time anti-commutation relations on the Dirac fields:
\begin{align}
  \big\{\psi_a(\x), \psi^{\dagger}_b({\bf \xp})\big\} & = \de(\x - {\bf \xp}) \, \de_{ab} \\
  \big\{\psi_a(\x), \psi_b({\bf \xp})\big\} = & \; \big\{\psi^{\dagger}_a(\x), \psi^{\dagger}_b({\bf \xp})\big\} = 0
\end{align}
%
The coefficients written in terms of the fields are 
\begin{align}
  a_s(\p)&  = \int \frac{d\x}{(2\pi)^{3/2}} \, \left(\frac{m}{\omega_\p}\right)^{1/2} \, \e^{i p\cdot x} \, u^{\dagger}_a(\p,s) \, \psi_a(x), \label{EQ:A_OP}\\
  a^{\dagger}_s(\p) & = \int \frac{d\x}{(2\pi)^{3/2}} \, \left(\frac{m}{\omega_\p}\right)^{1/2} \,  \e^{- i p\cdot x} \psi^{\dagger}_a(x) \, u_a(\p,s), \\
  b_s(\p)&  = \int \frac{d\x}{(2\pi)^{3/2}} \, \left(\frac{m}{\omega_\p}\right)^{1/2} \, \e^{ - i p\cdot x} \, \psi^{\dagger}_a(x) \, v_a(\p,s), \\
  b^{\dagger}_s(\p) & = \int \frac{d\x}{(2\pi)^{3/2}} \, \left(\frac{m}{\omega_\p}\right)^{1/2} \,  \e^{ i p\cdot x}  v^{\dagger}_a(\p,s) \, \psi_a(x). \label{EQ:B_D_OP} 
\end{align}
Again, the spinor index have been explicitly shown for clarity. From equations (\ref{EQ:A_OP})-(\ref{EQ:B_D_OP}), it follows that the anti-commutation relations for the coefficients are:
\begin{align}
  \big\{a_s(\p), a^{\dagger}_\sigma(\q)\big\} =   \big\{b_s(\p), b^{\dagger}_\sigma(\q)\big\} = \de(\p - \q) \, \de_{s\sigma},
\end{align}
and all other anti-commutators vanish. As operators the coefficients $a^{\dagger}_s$ and $b^{\dagger}_s$ are identified as the particle/anti-particle creation operators respectively, whereas $a_s$ and $b_s$ are the particle/anti-particle annihilation operators obeying $a_s| 0 \ket = b_s | 0\ket = 0$, where $|0 \ket$ is the vacuum state. 


The free classical Dirac Hamiltonian follows from the Lagrangian density (\ref{EQ:LAGRFREEDIRAC}). It turns out to be 
\begin{align}
  H = \int d\x \, \cH(x) = \int d\x \; \psib(x) \,( - i \, \nabla \cdot {\bs \gamma} + m) \, \psi(x).
\end{align}
The quantized version of this classical expression is obtained with the use of equations (\ref{EQ:PSI}) and (\ref{EQ:PSID}). After some extensive algebra one arrives at
\begin{align}
  H_\psi = \sum_{s = \pm} \int d\p \; \omega_\p \, \bigg\{ a^{\dagger}_s(\p) \, a_s(\p) + b^{\dagger}_s(\p) \, b_s(\p) \bigg\},
  \label{EQ:QEDHPSI}
\end{align}
where the Hamiltonian has been normal-ordered to remove the infinite energy of the vacuum. 
%

\section{QCD Reformulation}
The QCD Lagrangian density (units: $\hbar = c = 1$)~\cite{PDBook, Peskin1995, book_Srednicki}, suppressing the spinor and flavour indices, is 
\begin{align}
  \cL = \, - \frac{1}{4} \, (F_{\mu\nu}^a)^2 - \frac{1}{2 \, \xi} (\di^\mu A_\mu)^2 + \psib_i(i \slashed{D}_{ij} - m) \, \psi_j 
  \label{EQ:LQCD1}
\end{align}
where, the shorthand definitions are 
\begin{gather}
  F_{\mu\nu}^a = \di_\mu \, A_\nu^a - \di_\nu \, A_\mu^a + g \, f^{abc} \, A_\mu^b \, A_\nu^c, \\
  (D_\mu)_{ij}  = \di_\mu \de_{ij} - i \, g A_\mu^a \, T^a_{ij}. 
\end{gather}
The quark field $\psi_i$ is a Dirac spinor where the index $i = 1,2,3$ is a colour index. The vector boson field $A^a_\mu$ represents gluons carrying a Lorentz index $\mu = 0,..., 3$ and a colour index of the adjoint representation $a = 1,...,8$.  The dimensionless coupling constant $g$ characterizes the strength of the strong interaction. $F_{\mu\nu}^a$ is the non-Abelian field strength tensor, while $D_\mu$ is the covariant derivative which is the first ingredient in building a local non-Abelian gauge symmetry. The eight $SU(3)$ group generators $T^a_{ij}$ and the structure constants $f^{abc}$ obey the commutation relation:
\begin{align}
  [T^a, T^b] = i \, f^{abc} \, T^c,
  \label{EQ:SUNGENERATORS}
\end{align}
where $f^{abc}$ are completely antisymmetric. The term containing $\xi$ is known as the gauge fixing term. Physical observables do not, in principle, depend on the value of $\xi$. However, the canonical quantization of the gauge field is impossible in its absence since the conjugate momentum is undefined, i.e. $\pi^0 = \ds\frac{\partial \cL}{\partial \dot{A}_0} = 0$ if $\xi = \infty$. 

In the perturbative S-matrix formalism one has to include ghost fields. The ghost fields are non-physical auxiliary fields required to ensure the unitarity of the S-matrix and appear in the beyond-leading-order Feynman diagrams. They are irrelevant for the calculations presented in this dissertation, therefore they have been omitted.

The QCD Lagrangian density is invariant under the simultaneous transformations
\begin{gather}
  \psi_i \rightarrow  \psi^\prime_i = U_i^j \, \psi_j, \\
  A^a_\mu \, T_{ai}^{j}\rightarrow A^{\prime a}_{\mu} \, T_{ai}^j = U^j_k \, A^a_\mu \, T^{kl}_a \; \left[U^{-1}\right]_{li} + \frac{i}{g} \, \left[\di_{\mu}U\right]^{j}_{k} \, \left[U^{-1}\right]^{k}_{i}, \\
  F^{a}_{\mu\nu} \, T^{j}_{ai} \rightarrow F^{\prime a}_{\mu\nu} \, T^{j}_{ai} = U^{j}_{k} \, F^{a}_{\mu\nu} \, T^{kl}_{a} \, \left[U^{-1}\right]_{li}.
\end{gather}
where all the indices are explicitly exhibited. The gauge transformation matrix is
\begin{equation}
  U = \e^{- i \, g \, \frac{1}{2} \, T_i \alpha_i(x)},
\end{equation}
where the colour indices have been suppressed. The $\alpha_i(x)$ are the eight independent rotation ``angles''. 

It is convenient to express the Lagrangian density (\ref{EQ:LQCD1}) in the following form
\begin{align}
  \cL = \cL_A + \cL_\psi + \cL_{\psi A} + \cL_{3A} + \cL_{4A},
  \label{EQ:LQCD2}
\end{align}
where 
\begin{gather}
  \cL_A = - \frac{1}{4} (\di_\mu A_\nu^a - \di_\nu A_\mu^a) (\di^\mu A^{\nu a} - \di^\nu A^{\mu a}) - \frac{1}{2 \, \xi} (\di^\mu \, A^a_\mu)^2 \label{EQ:LA}, \\
  \cL_\psi =  \psib (i\, \slashed{\di} - m) \psi,  \label{EQ:LPSI} \\
  \cL_{\psi A} = g \, \psib_i \, \slashed{A}^a T^a_{ij} \psi_j, \label{EQ:LPA}\\
  \cL_{3A} = - g\, f^{abc} (\di_\mu A^a_\nu) \, A^{\mu \, b} \, A^{\nu \, c}, \label{EQ:L3A} \\
  \cL_{4A} = - \frac{1}{4} \, g^2 (f^{abc} A_\mu^b A_\nu^c) (f^{ade} A^{\mu d} A^{\nu e}) \label{EQ:L4A}.
\end{gather}
In this form, each term of the Lagrangian density is identified by the type of the interaction it represents. In the above contributions to the QCD Lagrangian density, the spinor index has been suppressed. 

The first step in reformulating the Lagrangian density (\ref{EQ:LQCD2}) is to obtain the solution to the classical equation of motion for the gauge field $A_\mu^a$. Having the solution for the non-interacting (i.e. with $g = 0$) theory, one can form the ``formal solution'' for the gauge field $A_\mu^a$ in the interacting case (i.e. with $g \neq 0$) similar to how it was done in the Higgs-like scalar theory. Integrating $\cL_A$ by parts and discarding the total derivative term, since its contributes nothing to the action, one obtains
\begin{align}
  \cL_{A} = \frac{1}{2}A_{\mu}^a\left(\di^2 g^{\mu\nu} - \di^{\mu} \di^{\nu}\right) A_{\nu}^a + \frac{1}{2\xi} A_{\mu}^a\di^{\mu}\di^{\nu}A_{\nu}^a,
  \label{FFL2}
\end{align}
where  $\di^2 = \ds\frac{\di^2}{\di t^2} - \nabla^2$ is the d'Alembert operator and $g^{\mu\nu}$ is the Minkowski metric with the signature $\left[1, -1, -1, -1\right]$. Consequently, the least action principle, i.e. $\de \ds\int \cL \, dx = 0$, leads to the following equation of motion 
\begin{align}
  \left(\di^2 g^{\mu\nu} - (1 - \frac{1}{\xi}) \, \di^{\mu}\di^{\nu}\right) A_{\nu}^a = 0.
\end{align}
This equation has a Green function solution. The Green function (or the free propagator) satisfies the identical equation but with a delta function source on the right hand side:
\begin{align}
  \left(\di^2 g^{\mu\nu} - (1 - \frac{1}{\xi}) \, \di^{\mu}\di^{\nu}\right) \Delta_{\nu\rho}^{ab}(x-y) = \de^{ab}\de(x-y) \de^\mu_{\, \rho},
\end{align}
or, in momentum space, 
\begin{align}
  \left(- k^2 g^{\mu\nu} + (1 - \frac{1}{\xi}) \, \frac{k^\mu \, k^\nu}{k^2}\right) \widetilde{\Delta}_{\nu\rho}^{ab}(k) =  \de^{ab}\de^\mu_{\, \rho}.
  \label{EQ:GREEN_MOM}
\end{align}
It is customary to solve for the momentum space propagator and then relate it to coordinate space. The momentum space propagator is given by
\begin{align}
  \widetilde{\Delta}_{\mu\nu}^{ab}(k) = -\frac{\de^{ab}}{k^2 + i\epsilon}\ds\left(g_{\mu\nu} - (1 - \xi)\frac{k_\mu k_\nu}{k^2}\right),
\end{align}
where the inclusion of $\epsilon$ instructs to use the Feynman prescription to handle singularities. One can easily check that 
\begin{align}
  \left[\widetilde{\Delta}^{ab}_{\mu\nu}(k)\right]^{-1} \widetilde{\Delta}^{\nu\rho \, bc}(k) = \de^{ac}\de_\mu^\rho,
\end{align}
where $\widetilde{\Delta}^{-1}$ is the operator acting on $\widetilde{\Delta}$ in equation (\ref{EQ:GREEN_MOM}). 
The momentum and coordinate space propagators are related by a Fourier transform:
\begin{align}
  \Delta_{\mu\nu}^{ab}(x-y) = \int\frac{d^4k}{(2\pi)^4} \, \e^{- i k\cdot(x-y)} \, \widetilde{\Delta}_{\mu\nu}^{ab}(k), \;\;\;   \widetilde\Delta_{\mu\nu}^{ab}(k) = \int d^4x \, \e^{+ i k\cdot(x-y)} \, \Delta_{\mu\nu}^{ab}(x-y).
  \label{EQ:GREEN_C_M}
\end{align}
Hence, the coordinate space Green function is  
\begin{align}
  \Delta_{\mu\nu}^{ab}(x-y) = \int\frac{d^4k}{(2\pi)^4} \, \e^{-i k\cdot(x-y)} \, \Bigg\{-\frac{\de^{ab}}{k^2 + i \epsilon}\ds\left(g_{\mu\nu} - (1 - \xi)\frac{k_\mu k_\nu}{k^2}\right)\Bigg\}.
\end{align}

Returning to the full Lagrangian density (\ref{EQ:LQCD2}), one finds the complete classical equation of motion for the gauge field $A_\mu^a$ to be  
\begin{equation}
  \left(\di^2 g^{\mu\nu} - (1 - \frac{1}{\xi}) \, \di^{\mu}\di^{\nu}\right) A_{\nu}^a = \rho^{\mu \, a} (x)
  \label{EQ:EM_A}
\end{equation}
where the ``source'' of this inhomogeneous equation is 
\begin{align}
  \rho^{\mu \, a}(x)  = & - g \, \psib_i(x) \, \gamma^\mu \,  T^a_{ij} \, \psi_j(x) + g \, f^{abc} \di^\nu \, \Big(A^b_\nu(x) \,  A^{\mu c}(x) \Big) \nn \\
  & - g \, f^{abc} \, \Big((\di^\mu A^{\nu b}(x) - \di^\nu A^{\mu b}(x)\Big) \,  A^c_\nu(x) + g^2 \, f^{abc} \, f^{cde} A^b_\nu(x) \,  A^{\mu d}(x) \, A^{\nu e}(x). 
\end{align}
Equation (\ref{EQ:EM_A}) can be written in integral from 
\begin{align}
  A^a_\nu(x) = \ds\big(A_0\big)^a_\mu + \int d\xp \, \Delta^{ab}_{\mu\nu}(x - \xp) \, \rho^{\nu b}(\xp).
  \label{EQ:FORMAL_SOL}
\end{align}
This expression is dubbed the ``formal solution'' to equation (\ref{EQ:EM_A}). In actuality, it is the integral form of equation (\ref{EQ:EM_A}). The homogeneous solution $A_0$ is irrelevant since no systems with external gluons will be considered. Therefore, $A_0$ is simply omitted. Everything that has been performed so far in reformulating the QCD Lagrangian density is in analogy to that of the Higgs-like scalar model.

To proceed further one must specify a gauge. It is convenient to choose the Feynman gauge where $\xi = 1$ since the Green function takes on the following simple appearance:
\begin{align}
  \Delta_{\mu\nu}^{ab}(x-y) = - \int\frac{d^4k}{(2\pi)^4} \, \frac{\de^{ab} \, g_{\mu\nu} \, \e^{-i k\cdot(x-y)}}{k^2 + i \epsilon}.
  \label{EQ:G_PROP}
\end{align}
Substituting the formal solution (\ref{EQ:FORMAL_SOL}) into the Lagrangian density (\ref{EQ:LQCD2}), and after some extensive algebra, one arrives at the reformulated Lagrangian density
\begin{align}
  \cL = \cL_\psi + \cL_{\psi A}^R + \cL_{3A}^R + \cL_{4A}^R, 
  \label{EQ:LQCD3}
\end{align}
where, the reformulated contributions are
\begin{align}
  \cL_A + \cL_{\psi A} & \rightarrow \cL_{\psi A}^{R} = - \frac{1}{2} \, g^2 \, \psib \, \gamma^\mu T^a \psi \int d\xp \Delta^{ab}_{\mu\nu} (x - \xp) \, \rho^{\nu b}(\xp), \\
  \cL_{3A} & \rightarrow \cL_{3A}^R =  \, - \, g \, f^{abc}  \int d\xp \, \di_\mu \, \Delta^{ad}_{\nu\sigma}(x - \xp) \, \rho^{\sigma d}(\xp) \nn \\
  & \F!\F! \times \int d\xpp \Delta^{\mu\alpha \, be}(x - \xpp) \, \rho^{e}_{\alpha}(\xpp) \int d\xppp \Delta^{\nu\beta \, cf}(x - \xppp) \, \rho^{f}_{\beta}(\xppp), \\
  \cL_{4A} & \rightarrow \cL_{4A}^R = - \frac{1}{4} \, g^2 \left(f^{abc} \int d\xp \Delta^{bi}_{\mu\sigma}(x - \xp) \, \rho^{\sigma i}(\xp) \int d\xpp \Delta^{cj}_{\nu\tau}(x - \xpp) \, \rho^{\tau j}(\xpp)\right) \nn \\ 
  & \F!\F! \times \left(f^{ade} \int d\zp \Delta^{\mu\alpha \, dk}(x - \zp) \, \rho^{k}_{\alpha}(\zp) \int d\zpp \Delta^{\nu\beta \, el}(x - \zpp) \, \rho^{l}_{\beta}(\zpp)\right). 
\end{align}
The arrow represents the process of reformulation, the superscript $R$ stands for reformulated quantity and the summation of the colour indices is implied, i.e. $\psib \gamma^\mu T^a \psi \equiv \psib_i \gamma^\mu T^a_{ij} \psi_j$. The letters $a,b ...$ stand for the colour indices of the adjoint representation and the Greek letters are Lorentz indices. To obtain this result, one has to work directly with the action $S = \ds\int dx \, \cL$ rather than with just the Lagrangian density. 

Expression (\ref{EQ:LQCD3}) is a non-local Lagrangian density, i.e. it involves integration over more than one space-time coordinate. Note that the free quark term $\cL_{\psi}$ does not get reformulated since it contains no gluon fields.  Recall that the purpose of reformulation is to eliminate the gluon field from the Lagrangian density while preserving its effects through its propagator. However, the ``source" term in (\ref{EQ:LQCD3}) implicitly contains the gauge field $A_\mu^a$. Therefore, infinitely many iterations of the substitution of the formal solution (\ref{EQ:FORMAL_SOL}) must be performed in order to completely eliminate $A_\mu^a$.  Realistically, this is impossible. Instead, one must truncate the ``source" term $\rho_\mu^a$ and work to a given order of iteration. In lowest iterative order, the ``source'' term is truncated to
\begin{align}
  \rho_\mu^a(x) = - g \, \psib_i(x) \gamma^\mu T^a_{ij} \psi_j(x).
  \label{EQ:SOURCE_T}
\end{align}
where the spinor indices are suppressed. Using this lowest-order truncation in the terms of the Lagrangian density (\ref{EQ:LQCD3}) leads to
\begin{align}
  \cL_{\psi A}^R = & \, + \frac{1}{2} \, g^2 \, \psib(x) \, \gamma^\mu T^a \psi(x) \int \frac{d\xp \, dk}{(2\pi)^4} \, \frac{\e^{ -i k\cdot(x - \xp)}}{k^2} \, \psib(\xp) \, \gamma_\mu T^a \psi(\xp), 
  \label{EQ:INTQCD1} \\
  \cL_{3A}^R = & \, - i \, g^4 \, f^{abc} \int d\xp \, d\xpp \, d \xppp \, \frac{dk \, dq \, dp}{(2\pi)^{12}} \, \frac{\e^{- i k\cdot(x - \xp)}}{k^2} \, \frac{\e^{- i q\cdot(x - \xpp)}}{q^2} \, \frac{\e^{- i p\cdot(x - \xppp)}}{p^2} \nn \\
  & \F!\H! \times \bigg\{\psib(\xp) \, \gamma_\nu T^a \psi(\xp) \bigg\} \,\bigg\{k_\mu \, \psib(\xpp) \, \gamma^\mu T^b \psi(\xpp) \bigg\}\, \bigg\{\psib(\xppp) \, \gamma^\nu T^c \psi(\xppp)\bigg\}, \\
  \cL_{4A}^R =  & - \frac{1}{4} \, g^6 \, f^{abc} f^{ade} \int d\xp \, d\xpp \, d\zp \, d\zpp \, \frac{dk \, dq \, dp \, dl}{(2\pi)^{16}} \; \frac{\e^{- i k\cdot(x - \xp)}}{k^2} \, \frac{\e^{- i q\cdot(x - \xpp)}}{q^2} \nn \\
  & \F!\H! \times\frac{\e^{- i p\cdot(x - \zp)}}{p^2} \, \frac{\e^{- i l\cdot(x - \zpp)}}{l^2} \; \bigg\{ \psib(\xp) \, \gamma_\mu T^b \psi(\xp)\bigg\} \, \bigg\{\psib(\xpp) \, \gamma_\nu T^c \psi(\xpp) \bigg\} \nn \\
  & \F!\H! \times\bigg\{\psib(\zp) \, \gamma^\mu T^d \psi(\zp) \bigg\} \, \bigg\{\psib(\zpp) \, \gamma^\nu T^e \psi(\zpp) \bigg\}.
  \label{EQ:REF_LQCD}
\end{align}
The reformulated Lagrangian density contains only quark fields; the interactions involving the gluon field are represented by the gluon propagator. Notice, also, that $\cL^R$ contains terms up-to order $g^6$, hence an energy calculation is expected to be accurate to this order in the coupling constant. As in the Higgs-like scalar model, one can examine the effects of the interaction terms, including the non-linear terms, in the context of bound states of quarks by considering trial states with quark quanta only. 

Finally, the reformulated Hamiltonian density corresponding to the Lagrangian density (\ref{EQ:LQCD3}) follows from the usual consideration
\begin{align}
  \cH_R = \Pi_{\psi} \dot{\psi} + \dot{\psib} \, \Pi_{\psib} - \cL_R,
  \label{EQ:HAM}
\end{align}
where, the conjugate momenta are defined in the usual way:
\begin{align}
  \Pi_{\psi} = \frac{\di \cL}{\di \dot{\psi}} = i \, \psib \, \gamma^0 \F!   \Pi_{\psib} = \frac{\di \cL}{\di \dot{\psib}} = - i \, \gamma^0 \, \psi.
\end{align}
Expressing the fields in terms of the operators, integrating out the spatial coordinates and normal-ordering the ladder operators to remove the infinite vacuum energy, one obtains the Hamiltonian 
\begin{gather}
  H_R = \int d\x \, : \cH(x) : \,  =  H_\psi - \int d\x \, : \Big(\cL_{\psi \, A}^R + \cL_{3A}^R + \cL_{4A}^R \Big) : \;  \equiv H_\psi + H_{\psi \, A}^R + H_{3A}^R + H_{4A}^R
  \label{EQ:QCDHAMREFORM}
\end{gather}
where, the free Hamiltonian equals
\begin{align}
  H_\psi = \sum_{i}^{3} \sum_{s = \pm} \int d\p \; \omega_\p \, \bigg\{ a^{\dagger}_{s, i}(\p) \, a_{s,i}(\p) + b^{\dagger}_{s,i}(\p) \, b_{s,i}(\p) \bigg\},  
\end{align}
Notice that it is practically identical to the one given in equation (\ref{EQ:QEDHPSI}); there is an extra summation of the colour index. The quantization of the quark field follows the same procedure as for colourless Dirac spinors described in the previous section. The principal difference is exactly in this index. Due to the inclusion of this index, the anti-commutation relations have to be slightly modified by supplementing a Kronecker delta function in the colour index:
\begin{align}
  \big\{a_{s,i}(\p), a^{\dagger}_{\sigma,j}(\q)\big\} =   \big\{b_{s,i}(\p), b^{\dagger}_{\sigma,j}(\q)\big\} = \de(\p - \q) \, \de_{s\sigma} \, \de_{ij}.
\end{align}
This generalization is straightforward and there should be not obscureness if this discussion is skipped.

Unfortunately, there are no simple expressions for the interactions terms $H_{\psi \, A}^R$, $H_{3A}^R$ and $H_{4A}^R$ when they are written in terms of the creation and annihilation operators (\ref{EQ:A_OP})-(\ref{EQ:B_D_OP}) (now containing an extra colour index). They are non-trivial and will not be written out explicitly. Instead, the matrix elements in the context of a definite trial state will be provided. 

As before, to obtain relativistic equations for stationary states (i.e. time-independent) of bound state systems of quark one must switch from the interaction picture to the Schr\"odinger picture by means of equation (\ref{EQ:PICTURES}). Henceforth, all matrix elements are written in the Schr\"odinger picture.

\section{Particle-Antiparticle State in QED}
As a preamble, the potential of a fermion-antifermion state in QED is investigated to get acquainted with the derivation method in the case of Dirac spinor matter fields. The amount of algebra in QCD increases drastically, therefore it is wise to demonstrate the method and intermediate steps on an almost ``trivial'' example 

The simplest particle-antiparticle trial state in QED is
\begin{align}
  | \Psi_2 \ket = \sum_{s,\sigma = \pm} \int d\p_{1,2} \, \cF_{s,\sigma}(\p_{1,2}) \, a^{\dagger}_s(\p_1) \, b^{\dagger}_\sigma(\p_2) | 0 \ket,
  \label{EQ:TRIALQEDTWO}
\end{align}
where $\cF_{s, \sigma}(\p_{1,2})$ are coefficient functions containing adjustable parameters and $s$ and $\sigma$ are the spin indices of the particle and anti-particle respectively. There is no colour index here; it will be added in the next section. The QED interaction is given by the term $\cL_{\psi A}^R$ (\ref{EQ:INTQCD1}) except that there is no colour related features in it. 

The relativistic coefficient functions $\cF_{s, \sigma}(\p_{1,2})$, in principle, should be selected a priori to be the eigenstate of the quantum numbers $J^{PC}$ ($J$ - total angular momentum, $P$ - parity, $C$ -charge conjugation) as one desires; see for example the works~\cite{Dykshoorn:1990zv} and~\cite{Zhang1990}. However, for the purpose of deriving the non-relativistic inter-particle potential, it is not essential to specify these quantum numbers right away. In the end, only spin (index) function must be specified in the non-relativistic limit which, as is illustrated below, decouples from the momentum (and hence angular momentum) dependence. In some sense, the approach adapted by Terekidi~\cite{Terekidi_thesis} is followed in this dissertation.

To implement the variational method the matrix element corresponding to the trial state (\ref{EQ:TRIALQEDTWO}) is calculated and then varied with respect to the coefficient function $\cF^\ast$. Carrying out these two operations leads to the following relativistic equation in momentum space:
\begin{align}
  \cF_{s_1, \sigma_1}(\p_{1,2}) \, (\omega_{\p_1} + \omega_{\p_2} - E) = \sum_{s_2, \sigma_2} \int d\p^\prime_{1,2} \, Y_{s_1, \sigma_1}^{s_2, \sigma_2}(\p_{1,2}, \p^{\prime}_{1,2}) \, \cF_{s_2, \sigma_2} (\p^\prime_{1,2}),
\end{align}
where, the QED interaction kernel is
\begin{align}
  & Y_{s_1, \sigma_1}^{s_2, \sigma_2} (\p_{1,2}, \p^{\prime}_{1,2}) = \, \frac{g^2 \, m^2}{2(2\pi)^3} \, \frac{\de(\p_1 + \p_2 - \p^\prime_1 - \p^\prime_2)}{\ds\sqrt{\omega_{\p_1} \omega_{\p_2} \omega_{\p^\prime_1} \omega_{\p^\prime_2}}} \nn \\
  & \times \Bigg\{ \, \frac{\vb(\p^\prime_2, \sigma_2) \gamma^\mu v(\p_2, \sigma_1) \, \ub(\p_1, s_1) \gamma_\mu u(\p^\prime_1, s_2)} {(p^\prime_1 - p_1)^2}+ \frac{\ub(\p_1, s_1) \gamma^\mu u(\p^\prime_1, s_2) \, \vb(\p^\prime_2, \sigma_2) \gamma_\mu v(\p_2, \sigma_1)}{(p^\prime_2 - p_2)^2} \nn \\
  & \H!- \frac{\ub(\p_1, s_1) \gamma^\mu v(\p_2, \sigma_1) \, \vb(\p^\prime_2, \sigma_2) \gamma_\mu u(\p^\prime_1, s_2)}{(p^\prime_1 + p^\prime_2)^2} - \frac{\vb(\p^\prime_2, \sigma_2) \gamma^\mu u(\p^\prime_1, s_2) \, \ub(\p_1, s_1) \gamma_\mu v(\p_2, \sigma_1)}{(p_1 + p_2)^2} \Bigg\}.
\end{align}
As in the Higgs-like scalar model, the first two terms correspond to a mediating field quanta (i.e. photon) exchange and the last two correspond to a virtual annihilation interaction. This expression is nearly identical to equation (\ref{EQ:Y_22}) with $\mu = 0$; where the exception is the spinor products in the numerators.

%
The derivation of the inter-particle potential is continued in the centre of mass frame such that the wavefunction takes on the form $\cF_{s, \sigma}(\p_{1,2}) = F_{s, \sigma}(\p_1) \, \de(\p_1 + \p_2)$. The equation simplifies to 
\begin{equation}
  F_{s_1, \sigma_1}(\p_1) \, (2 \, \omega_\p - E) = \sum_{s_2, \sigma_2} \int d\p^\prime \, Y_{s_1, \sigma_1}^{s_2, \sigma_2}(\p, \p^{\prime}) \, F_{s_2, \sigma_2} (\p^\prime),
  \label{EQ:QEDTWOCOFM}
\end{equation}
where, the interaction kernel becomes 
\begin{align}
  Y_{s_1, \sigma_1}^{s_2, \sigma_2} (\p, \p^{\prime}) = \, & \frac{g \, m^2}{(2\pi)^3} \frac{1}{\ds{\omega_{\p} \omega_{\p^\prime}}} \, \Bigg\{ \, \frac{\ub(\p, s_1) \gamma^\mu u(\p^\prime, s_2) \, \vb(-\p^\prime, \sigma_2) \gamma_\mu v(-\p, \sigma_1) }{(p^\prime - p)^2} \nn \\
  & \H! - \ub(\p, s_1) \gamma^\mu v(-\p, \sigma_1) \, \vb(-\p^\prime, \sigma_2) \gamma_\mu u(\p^\prime, s_2) \, \bigg(\frac{1}{4 \, \omega_{\p^\prime}^2} - \frac{1}{4 \, \omega_{\p}^2}\bigg) \Bigg\}.
  \label{EQ:Y22QEDREL}
\end{align}
where like terms have been collected, Once again, this is basically a repetition of the Higgs-like scalar model with the exception of the spinor products. Notice that there are spinor arguments with negative momentum; they come as a result of the delta function.

Unfortunately, and contrary to the customary practice, the products of the Dirac spinors and gamma matrices can not be worked out using the completeness relations (\ref{EQ:COMPLETENESS}) and the trace theorems of the gamma matrices. Therefore, the only option to proceed with the derivation of the inter-particle potential from the kernel is by using the brute method of matrix multiplication.  

In the non-relativistic limit (see Appendix B section \ref{SEC:QED_two} for the identities involving spinors and gamma matrices), the Dirac spinors and gamma matrices multiply out to yield the simplified kernel 
\begin{align}
  Y_{s_1, \sigma_1}^{s_2, \sigma_2}(\p, \p^{\prime}) = \frac{g^2}{(2\pi)^3} \Bigg\{ \frac{\de_{s_1 \, s_2} \, \de_{\sigma_1 \, \sigma_2}}{(\pp - \p)^2} - \frac{\chi^\dagger_{s_1} \sigma_i \, \eta_{\sigma_1}  \eta^\dagger_{\sigma_2} \sigma_i \, \chi_{s_2}}{4 m^2}\Bigg\}. 
  \label{EQ:Y22QED}
\end{align}
Notice how the virtual annihilation term still carries a non-trivial dependence on spin. In addition, in the non-relativistic limit, the spin and momentum dependencies of a coefficient function factorize, i.e. $F_{s_1, \sigma_1}(\p) = \Theta_{s_1, \sigma_1} \, f(\p)$. This implies that there are four different equations for $f(\p)$ in each spin configuration. Recall that spin states transform irreducibly under the symmetry group $SU(2)$ in the fundamental representation. The four spin index functions $\Theta_{s_1, \sigma_1}$ for a particle-antiparticle pair are obtained by working out the product of two fundamental $SU(2)$ representation. This is widely known~\cite{Griffiths1987} and the result is
\begin{align}
  \Theta^{\tx{singlet}}_{s_1,\sigma_1} = \frac{1}{\sqrt{2}} \epsilon_{s_1 \sigma_1}, \F! 
  \Theta^{\tx{triplet}}_{s_1, \sigma_1} 
  = \left\{ 
  \begin{array}{l l}
    \Theta_{1,1} = 1 &  m_z = +1 \\
    \Theta_{1,2} =  \Theta_{2,1} = \ds\frac{1}{\sqrt{2}} & m_z = 0 \\
    \Theta_{2,2} = 1 & m_z = -1 \\
  \end{array} \right.
  \label{EQ:TWOSPIN}
\end{align}
where $\epsilon_{s_1 \sigma_1}$ is the two index antisymmetric symbol, $m_z$ is the spin projection and all the unspecified values of $\Theta_{s_1, \sigma_1}$ in the singlet and triplet configurations are zero. 

Having the results of equations (\ref{EQ:Y22QED}) and (\ref{EQ:TWOSPIN}) in command, one can perform a Fourier transform of equation (\ref{EQ:QEDTWOCOFM}) to coordinate representation. Simultaneously, multiplying by $\Theta_{s_1, \sigma_1}$ and carrying out all index multiplications, including the index $i$ in equation (\ref{EQ:Y22QED}), leads to a Schr\"odinger-like equation for the particle-antiparticle system corresponding to the trial state (\ref{EQ:TRIALQEDTWO})~\footnote{It is evident that a trial state containing more Fock-space components leads to coupled equations.}:
\begin{equation}
  -\frac{\nabla}{m} \, \psi(x) + V(x) \, \psi(x) = E \, \psi(x),
\end{equation}
where, the potential energy in the singlet and triplet configurations of spin are
\begin{gather}
  V^{\tx{singlet}}(x) = \,- \alpha_g \, \frac{1}{x},  \\
  V^{\tx{triplet}}(x) = \, - \alpha_g \left( \frac{1}{x} - \frac{\pi}{m^2} \, \de(\x)\right).
  \label{EQ:V_2}
\end{gather}
Here $\alpha_g = \ds\frac{g^2}{4\pi}$ is the dimensionless coupling constant of QED and $x$ is the particle-antiparticle separation. Notice that the delta function arise naturally in this formalism, and as seen, is only present when the spins are in the triplet configuration. It is, actually, a relativistic correction as is evident from the $\ds\frac{1}{m^2}$ factor. If, while taking the non-relativistic limit of equation (\ref{EQ:Y22QEDREL}), the terms in the leading powers of the inverse mass have been kept, then one could have obtained spin-orbit, spin-spin and other delta function lowest order relativistic corrections to the potential energy. 

\section{Three-Quark (Baryon) Trial State}
It has been learned from the Higgs-like scalar model that in order to probe the effects of the non-linear terms of the Lagrangian density it is necessary for a trial state to have a component with at least three particle quanta. It is anticipated that the same should concur in QCD. As it is simpler to work with mono-component trial states, a three quark trial state which models a baryon with quarks of different masses, is considered next.

The simplest three quark trial state with each quark of a different flavour (i.e. mass) is given by
\begin{align}
  |\Psi_3 \ket = \sum_{i,j,k} \, \Omega_{i,j,k} \sum_{s,r,t} \int d\p_{1..3} \, F_{s,r,t}(\p_{1..3}) \, a^{\dagger}_{s,i,a}(\p_1) \,  a^{\dagger}_{r,j,b}(\p_1) a^{\dagger}_{t,k,c}(\p_3)| 0 \ket,
  \label{EQ:TRIALQCDTHREE}
\end{align}
where the indices $i,j, k = 1..3$ refer to colour, $s,r,t = \pm$ refer to spin, and $a,b,c$ are the different flavours. The colour index function $\Omega$ holds the information about the combined colour configuration of the quarks. It has been brought in front to accent that the colour dependence of the coefficient function $F$ does not couple to the spin and momentum dependencies even relativistically (i.e. the colour dynamics is decoupled from angular momentum as can be seen from the Lagrangian density). Recall, that the colour index function of a quark transforms in the fundamental representation of $SU(3)$. Therefore, the combined colour configuration $\Omega$ has to be obtained by finding the products of three fundamental $SU(3)$ representations. The result is widely known~\cite{Griffiths1987} and its derivation also appears in section \ref{SEC:YOUNG} of Appendix B.

The dimensions of the product of three fundamental $SU(3)$ representations breaks down as follows:
\begin{equation}
3 \otimes 3 \otimes 3 = 1 \oplus 8 \oplus 8 \oplus 10,
\end{equation}
where the singlet representation corresponds to the physically allowable colour configuration. The index function of the singlet is a three index, totally anti-symmetric object with the appropriate normalization:
\begin{equation}
  \Omega_{i,j,k} = \frac{1}{\sqrt{6}} \epsilon_{ijk}.
\end{equation}
Note that if the trial state (\ref{EQ:TRIALQCDTHREE}) contained quarks of the same flavour only, then the overall trial state would have to be anti-symmetric under interchanges of quark momenta and indices if it is to model the ground state. Since the colour index function is always anti-symmetric, consequently the remaining (spin and momentum) parts would be symmetric. However, this is not the case here. Recall that the coefficient function $F$ contains adjustable parameters which must be calculated variationally via an energy calculation.

The matrix element of the Hamiltonian (\ref{EQ:QCDHAMREFORM}) with the trial state (\ref{EQ:TRIALQCDTHREE}) is given in section \ref{SEC:QCD_three} of Appendix B. Variationally differentiating it with respect to the coefficient function $F^\ast$ leads to the following relativistic equation in momentum space:
\begin{align}
  F_{s,r,t}&(\p_{1,2,3}) \, \left(\omega^A_{\p_1} + \omega^B_{\p_2} + \omega^C_{\p_3} - E\right) = \nn \\
  & \F! \sum_{s^\prime, r^\prime, t^\prime} \int d\pp_{1,2,3} \, \bigg\{ \cY_{s,r,t}^{s^\prime, r^\prime, t^\prime}(\pp_{1,2,3}, \p_{1,2,3}) + \cC_{s,r,t}^{s^\prime, r^\prime, t^\prime}(\pp_{1,2,3}, \p_{1,2,3}) \bigg\} \, F_{s^\prime, r^\prime, t^\prime}(\pp_{1,2,3}),
  \label{EQ:EQQCDTHREE}
\end{align}
where the slightly generalized definition $\omega^i_\p = \ds\sqrt{m_i^2 + \p^2}$ accounts for different flavours. 

The relativistic interaction kernel $\cY$ in equation (\ref{EQ:EQQCDTHREE}) is
\begin{align}
  &\cY_{s,r,t}^{s^\prime, r^\prime, t^\prime}(\pp_{1,2,3}, \p_{1,2,3}) = \frac{1}{2} \frac{g^2}{(2\pi)^3} \left(\frac{4}{3} \right) \nn \\
  & \times \left[\frac{m_A m_B \, \de(\p_1 + \p_2 - \pp_1 - \pp_2) \, \de(\p_3 - \pp_3) \, \de_{t_1 t_2}}{(\omega^A_{\p_1}\omega^A_{\pp_1}\omega^B_{\p_2}\omega^B_{\pp_2})^{1/2}} \, \frac{\ub(\p_1, s) \gamma_{\mu} u(\pp_1, \ss) \ub(\p_2, r) \gamma^{\mu} u(\pp_2, \rr)}{(p_2^B - p^{\prime B}_2)^2} \right. \nn \\
    & \H! + \frac{m_A m_B \, \de(\p_1 + \p_2 - \pp_1 - \pp_2) \, \de(\p_3 - \pp_3) \, \de_{t_1 t_2}}{(\omega^A_{\p_1}\omega^A_{\pp_1}\omega^B_{\p_2}\omega^B_{\pp_2})^{1/2}} \, \frac{\ub(\p_2, r) \gamma_{\mu} u(\pp_2, \rr) \ub(\p_1, s) \gamma^{\mu} u(\pp_1, \ss)}{(p_1^A - p^{\prime A}_1)^2} \nn \\
   & \H! + \frac{m_A m_C \, \de(\p_1 + \p_3 - \pp_1 - \pp_3) \, \de(\p_2 - \pp_2) \, \de_{r_1 r_2}}{(\omega^A_{\p_1}\omega^A_{\pp_1}\omega^C_{\p_3}\omega^C_{\pp_3})^{1/2}} \, \frac{\ub(\p_1, s) \gamma_{\mu} u(\pp_1, \ss) \ub(\p_3, t) \gamma^{\mu} u(\pp_3, \tt)}{(p_3^C - p^{\prime C}_3)^2} \nn \\
    & \H! + \frac{m_A m_C \, \de(\p_1 + \p_3 - \pp_1 - \pp_3) \, \de(\p_2 - \pp_2) \, \de_{r_1 r_2}}{(\omega^A_{\p_1}\omega^A_{\pp_1}\omega^C_{\p_3}\omega^C_{\pp_3})^{1/2}} \, \frac{\ub(\p_3, t) \gamma_{\mu} u(\pp_3, \tt) \ub(\p_1, s) \gamma^{\mu} u(\pp_1, \ss)}{(p_1^A - p^{\prime A}_1)^2} \nn \\
    & \H! + \frac{m_B m_C \, \de(\p_2 + \p_3 - \pp_2 - \pp_3) \, \de(\p_1 - \pp_1) \, \de_{s_1 s_2}}{(\omega^B_{\p_2}\omega^B_{\pp_2}\omega^C_{\p_3}\omega^C_{\pp_3})^{1/2}} \, \frac{\ub(\p_2, r) \gamma_{\mu} u(\pp_2, \rr) \ub(\p_3, t) \gamma^{\mu} u(\pp_3, \tt)}{(p_3^C - p^{\prime C}_3)^2} \nn \\
    & \H! \left. + \frac{m_B m_C \, \de(\p_2 + \p_3 - \pp_2 - \pp_3) \, \de(\p_1 - \pp_1) \, \de_{s_1 s_2}}{(\omega^B_{\p_2}\omega^B_{\pp_2}\omega^C_{\p_3}\omega^C_{\pp_3})^{1/2}} \, \frac{\ub(\p_3, t_1) \gamma_{\mu} u(\pp_3, \tt) \ub(\p_2, t) \gamma^{\mu} u(\pp_2, \tt)}{(p_2^B - p^{\prime B}_2)^2} \right].
\end{align}
This kernel gives rise to pairwise Coulomb interactions emerging from one-gluon exchange; there are no virtual annihilation terms in the absence of anti-quark quanta in the trial state (\ref{EQ:TRIALQCDTHREE}).

The cubic interaction kernel $\cC$ in equation (\ref{EQ:EQQCDTHREE}) is 
\begin{align}
  & \cC_{s,r,t}^{s^\prime, r^\prime, t^\prime} (\pp_{1,2,3}, \p_{1,2,3}) =  \, \frac{i\, g^4}{(2\pi)^6} \, F_C \, m_A \, m_B \, m_C \; \frac{\de(\p_1 + \p_2 + \p_3 - \pp_1 -\pp_2 - \pp_3)}{(\omega^A_{\p_1}\omega^A_{\pp_1}\omega^B_{\p_2}\omega^B_{\pp_2}\omega^C_{\p_3}\omega^C_{\pp_3})^{1/2}} \nn \\
  & \times \left[\ub(\p_1, s) \gamma_\nu u(\pp_1, \ss) \ub(\p_2, r) \gamma^\mu u(\pp_2, \rr) \ub(\p_3, t) \gamma^\nu u(\pp_3, \tt) \frac{(p_1^A - p^{\prime A}_1)_\mu}{(p_1^A - p_1^{\prime A})^2 (p_2^B - p_2^{\prime B})^2 (p_3^C - p_3^{\prime C})^2} \right. \nn \\
    & \H! + \ub(\p_1, s) \gamma_\nu u(\pp_1, \ss) \ub(\p_3, t) \gamma^\mu u(\pp_3, \tt) \ub(\p_2, r) \gamma^\nu u(\pp_2, \rr) \frac{(p_1^A - p^{\prime A}_1)_\mu}{(p_1^A - p_1^{\prime A})^2 (p_2^B - p_2^{\prime B})^2 (p_3^C - p_3^{\prime C})^2} \nn \\
    & \H! + \ub(\p_2, r) \gamma_\nu u(\pp_2, \rr) \ub(\p_1, s) \gamma^\mu u(\pp_1, \ss) \ub(\p_3, t) \gamma^\nu u(\pp_3, \tt) \frac{(p_2^B - p^{\prime B}_2)_\mu}{(p_1^A - p_1^{\prime A})^2 (p_2^B - p_2^{\prime B})^2 (p_3^C - p_3^{\prime C})^2} \nn \\
    & \H! + \ub(\p_2, r) \gamma_\nu u(\pp_2, \rr) \ub(\p_3, t) \gamma^\mu u(\pp_3, \tt) \ub(\p_1, s) \gamma^\nu u(\pp_1, \ss) \frac{(p_2^B - p^{\prime B}_2)_\mu}{(p_1^A - p_1^{\prime A})^2 (p_2^B - p_2^{\prime B})^2 (p_3^C - p_3^{\prime C})^2} \nn \\
    & \H! + \ub(\p_3, t) \gamma_\nu u(\pp_3, \tt) \ub(\p_1, s) \gamma^\mu u(\pp_1, \ss) \ub(\p_2, r) \gamma^\nu u(\pp_2, \rr) \frac{(p_3^C - p^{\prime C}_3)_\mu}{(p_1^A - p_1^{\prime A})^2 (p_2^B - p_2^{\prime B})^2 (p_3^C - p_3^{\prime C})^2} \nn \\
    & \H! + \left. \ub(\p_3, t) \gamma_\nu u(\pp_3, \tt) \ub(\p_2, r) \gamma^\mu u(\pp_2, \rr) \ub(\p_1, s) \gamma^\nu u(\pp_1, \ss) \frac{(p_3^C - p^{\prime C}_3)_\mu}{(p_1^A - p_1^{\prime A})^2 (p_2^B - p_2^{\prime B})^2 (p_3^C - p_3^{\prime C})^2} \right]. 
\end{align}
This is a three-way interaction which, in addition to spin and momentum dynamics, involves colour dynamics. The colour dynamics reside in the factor $F_C$ which surprisingly turns out to to be vanishing. There are six separate colour factors in each term but all lead to the same result (refer to section \ref{SEC:QCD_three} of Appendix B for details). For example,
\begin{align}
  F_C & \, =  f^{abc} \, \Omega_{i_1, j_1, k_1} \, \Omega_{i_2, j_2, k_2} \; T^a_{i_1 i_2} \, T^b_{j_1 j_2} \,  T^c_{k_1 k_2}  =  \frac{1}{6} \, f^{abc} \, \epsilon_{i_1 j_1 k_1} \, \epsilon_{i_2 j_2 k_2} \; T^a_{i_1 i_2} \, T^b_{j_1 j_2} \,  T^c_{k_1 k_2} \nn \\
  & = \, \frac{1}{6} \, f^{abc}  \; T^a_{i_1 i_2} \, T^b_{j_1 j_2} \,  T^c_{k_1 k_2} \, \left(\de_{i_1 j_2} \de_{j_1 k_2} \de_{k_1 i_2} + \de_{i_1 k_2} \de_{j_1 i_2} \de_{k_1 j_2}\right) \nn \\
  & = \, \frac{1}{6} \, f^{abc}  \; \left( T^a_{j_2 i_2} \, T^b_{k_2 j_2} \, T^c_{i_2 k_2} + T^a_{k_2 i_2} \, T^b_{i_2 j_2} \,  T^c_{j_2 k_2}\right) = \frac{1}{6} \, f^{abc}  \; \tx{Tr}\left( T^a \, T^c \, T^b + T^a \, T^b \,  T^c\right) \nn \\
  & = \, \frac{1}{6} \, f^{abc}  \; \tx{Tr}\left( - T^a \, T^b \, T^c + T^a \, T^b \,  T^c\right) = 0,
  \label{EQ:QCDTHREEC33}
\end{align}
where the fact that the generating matrices $T$ are traceless, the structure constants $f^{abc}$ are anti-symmetric under index interchanges and the following identity involving epsilon tensors have been used:
\begin{align}
  \epsilon_{i_1 j_1 k_1} \, \epsilon_{i_2 j_2 k_2} = \tx{det} 
  \left[\begin{array}{ccc} 
    \de_{i_1 i_2} & \de_{i_1 j_2} & \de_{i_1 k_2} \\
    \de_{j_1 i_2} & \de_{j_1 j_2} & \de_{j_1 k_2} \\
    \de_{k_1 i_2} & \de_{k_1 j_2} & \de_{k_1 k_2}
    \end{array} \right].
\end{align}
%

This vanishing contribution to the energy from the cubic term of the Hamiltonian $H^R_{3A}$ seems to be somewhat mysterious. The quartic term $H^R_{4A}$ does not contribute anything either; it is a four way interaction hence a trial state, such as in equation (\ref{EQ:TRIALQCDTHREE}), with only three particle/antiparticle operators does not to probe it. It is improbable that the non-linear terms do not participate at all, especially when the effects of confinement and string breaking are ascribed to them. It might be that this surprising and counter-intuitive result only occurs in this leading order and for the simplified trial state (\ref{EQ:TRIALQCDTHREE}). Indeed, reference~\cite{Lucha:1991vn} illustrates the same vanishing result for a system of three quarks, yet using a different method. In the next section though, it is shown that a multi-component trial state does produce non-vanishing non-linear contributions to the interaction and a reason for this is suggested. 

\section{Multi-Component Quark-Antiquark (Meson) State in QCD}
The following multi-component trial state, consisting of a heavy quark-antiquark pair and a light quark-antiquark pair, is treated in this section: 
\begin{align}
  | \Psi_{t} \ket = C_F \, | \Psi_{2} \ket + C_G \, | \Psi_{4} \ket,
  \label{EQ:QCDTRIAL24}
\end{align}
where the two Fock-space components are
\begin{align}
  | \Psi_2 \ket = & \sum_{i,j} \, \Omega_{ij} \sum_{\kappa, \lambda} \int d\p_{1,2} \, \cF_{\kappa, \lambda}(\p_{1,2}) \, a^{\dagger}_{\kappa,i, a}(\p_1) \, b^{\,\dagger}_{\lambda,j, a}(\p_2) | 0 \ket,  
  \label{EQ:QCDTRIAL2} \\
  | \Psi_4 \ket = & \sum_{i,j,k,l} \, \Lambda_{ijkl} \sum_{\kappa, \lambda, \mu, \nu} \int d\p_{1..4} \, \cG_{\kappa, \lambda, \mu, \nu}(\p_{1..4}) \, a^{\dagger}_{\kappa,i,a}(\p_1) \, b^{\,\dagger}_{\lambda,j,a}(\p_2) \, a^{\dagger}_{\mu,k,b}(\p_3) \, b^{\,\dagger}_{\nu,l,b}(\p_4) | 0 \ket,
  \label{EQ:QCDTRIAL4}
\end{align}
where the unsummed indices $a, b$ label the two flavours of quarks, $\kappa, \lambda, \mu, \nu$ are the spin indices, and $i, j, k, l$ are the colours indices along with the colour wavefunctions $\Omega$ and $\Lambda$ for the two and four components respectively. As previously noted, the colour index functions factor out in front even relativistically. The relativistic coefficient functions $\cF$ and $\cG$ are, in principle, such that $| \Psi_t \ket$ is an eigenstate of the quantum numbers $J^{PC}$ one desires. The coefficients $C_F$ and $C_G$ are parameters which, modulo normalization, must be determined variationally together with all the parameters contained in $\cF$ and $\cG$. 

The colour index of a quark (anti-quark) transforms in the fundamental (conjugate) representation of $SU(3)$. To find the colour index functions $\Omega$ and $\Lambda$ one must consider group products of $SU(3)$ representations. The dimensions of the group products of $SU(3)$ representation can be determined using the Young tableaux method as is explained in Appendix B section \ref{SEC:YOUNG}.

The dimensions of the two required decompositions are
\begin{gather}
  3 \, \otimes \, \bar{3} = 1 \, \oplus \, 8,
  \label{EQ:ONE_SING} \\
  3 \, \otimes \, \bar{3} \otimes 3 \, \otimes \, \bar{3} = 1 \, \oplus \, 1 \, \oplus \,  8 \, \oplus \, 8 \, \oplus \,  8 \, \oplus \,  8 \, \oplus \, 10 \, \oplus \, 10 \, \oplus \,  27.
  \label{EQ:TWO_SING}
\end{gather}
It is known experimentally that the physically allowed representations are colourless (i.e. zero in each of the eight colour charges; see below) corresponding to the one dimensional representations. The singlet in equation (\ref{EQ:ONE_SING}) provides the colour wavefunction for the two component (\ref{EQ:QCDTRIAL2}); it is nothing but the delta function with the appropriate normalization.  Similarly, the two singlets in (\ref{EQ:TWO_SING}) provide the colour index function for the four component (\ref{EQ:QCDTRIAL4}). The index wavefunction is, again, comprised of delta functions and, in this case, has to be equally weighted between the two singlets. The properly normalized index functions are 
\begin{gather}
  \Omega_{ij} =  \ds\frac{1}{\sqrt{3}} \, \delta_{ij},
  \label{EQ:OMEGAIJ}\\
  \Lambda_{ijkl} = \ds\frac{1}{2\sqrt{6}}\left(\de_{ij} \, \de_{kl} + \de_{il} \, \de_{kj}\right) 
  \label{EQ:LAMBDAIJKL}.
\end{gather}
Recall, that the total colour charge operator can be obtained from the colour charge current density as follows:
\begin{align}
  Q^a \equiv & \; \int d\x \; \psib_i(x) \, \gamma^0 \, T^a_{ij} \, \psi_j(x) \nn\\ 
   = & \; \sum_{i,j=1..3} \sum_{s=\pm} \int d\p \; \bigg\{ a^\dagger_{s,i}(\p) \, T^a_{ij} \, a_{s,j}(\p) - b^\dagger_{s,i}(\p) \, T^a_{ij} \, b_{s,j}(\p)\bigg\}
\end{align}
One can verify that the trial state (\ref{EQ:QCDTRIAL24}) with the colour index functions given by (\ref{EQ:OMEGAIJ}) and (\ref{EQ:LAMBDAIJKL}) is an eigenstate of $Q^a = 0$ for $a=1, ..., 8$.

The procedure of extracting the quark-antiquark potential is the same as discussed in the previous sections. The relevant contributions to the matrix element $\bra \Psi_{t} | H - E | \Psi_{t} \ket$ are given in section \ref{SEC:QQ} of Appendix B. The variation of the matrix element with respect to both coefficient functions $\cF^\ast$ and $\cG^\ast$ leads to the coupled equations that describe a bound state of a meson:
\begin{align}
  & \cF_{\kappa_1, \lambda_1}(\p_{1,2}) \left( \omega_{\p_1}^A + \omega_{\p_2}^A - E \right) = \; \sum_{\kappa_2, \lambda_2} \int d\p_{3,4} \, \left(\cY_{2,2}\right)^{\kappa_2, \lambda_2}_{\kappa_1, \lambda_1}(\p_{1..4}) \; \cF_{\kappa_2, \lambda_2}(\p_{3,4}) \label{EQ:EQRELQCD2} \\
  & \; + \; R \sum_{\kappa_2, \lambda_2, \mu_2, \nu_2} \int d\p_{3..6} \; \bigg\{ \left(\cY_{2,4}\right)^{\kappa_2, \lambda_2, \mu_2, \nu_2}_{\kappa_1, \lambda_1}(\p_{1..6}) + \left(\cC_{2,4}\right)^{\kappa_2, \lambda_2, \mu_2, \nu_2}_{\kappa_1, \lambda_1}(\p_{1..6}) \bigg\} \; \cG_{\kappa_2, \lambda_2, \mu_2, \nu_2}(\p_{3..6}), \nn \\
  & \cG_{\kappa_1, \lambda_1, \mu_1, \nu_1}(\p_{1..4}) \left(\omega^A_{\p_1} + \omega^A_{\p_2} + \omega^B_{\p_3} + \omega^B_{\p_4} - E\right) = \label{EQ:EQRELQCD4} \\ 
  & \; + \; \frac{1}{R} \sum_{\mu_2, \nu_2} \int d\p_{5,6} \; \bigg\{ \left(\cY_{4,2}\right)^{\mu_2, \nu_2}_{\kappa_1, \lambda_1, \mu_1, \nu_1}(\p_{1..6}) + \left(\cC_{4,2}\right)^{\mu_2, \nu_2}_{\kappa_1, \lambda_1, \mu_1, \nu_1}(\p_{1..6}) \bigg\} \; \cF_{\mu_2, \nu_2}(\p_{5,6}) \nn \\
  & \; + \sum_{\kappa_2, \lambda_2, \mu_2, \nu_2} \int d\p_{4..8} \; \bigg\{ \left(\cY_{4,4}\right)^{\kappa_2, \lambda_2, \mu_2, \nu_2}_{\kappa_1, \lambda_1, \mu_1, \nu_1}(\p_{1..8}) + \left(\cQ_{4,4}\right)^{\kappa_2, \lambda_2, \mu_2, \nu_2}_{\kappa_1, \lambda_1, , \mu_1, \nu_1}(\p_{1..8}) \bigg\} \; \cG_{\kappa_2, \lambda_2, \mu_2, \nu_2}(\p_{4..8}), \nn
\end{align}
where the colour indices have been summed over and the resulting colour factors included in the kernels, and $R = \ds\frac{C_G}{C_F}$ is the ratio that specifies relative contributions of each Fock-space component. The kernel $\cC$ have been deliberately omitted because, just as in (\ref{EQ:QCDTHREEC33}), the colour summation, despite the presence of the ``spectator'' particle which just produces a delta function in the colour index, sets this kernel to zero. 

The relativistic kernels pertaining to the equation for the function $F$ are 
\begin{align}
  & \left(\cY_{2,2}\right)^{\kappa_2, \lambda_2}_{\kappa_1, \lambda_1}(\p_{1..4}) = - \frac{ 2 \, g^2 \, m_A^2}{3(2\pi)^3} \, \frac{\de(\p_1 + \p_2 - \p_3 - \p_4)}{\left(\omega^A_{\p_1} \omega^A_{\p_2} \omega^A_{\p_3} \omega^A_{\p_4} \right)^{1/2}} \nn \\
  & \F!\times \Bigg\{\ub(\p_1, \kappa_1) \, \gamma_\mu \, u(\p_3, \kappa_2) \; \vb(\p_4, \lambda_2) \, \gamma^\mu \, v(\p_2, \lambda_1) \left(\frac{1}{(p_4 - p_2)^2} + \frac{1}{(p_3 - p_1)^2}\right)\Bigg\}, 
  \label{EQ:RELY22}\\
  &\left(\cY_{2,4}\right)^{\kappa_2, \lambda_2, \mu_2, \nu_2}_{\kappa_1, \lambda_1}(\p_{1..6}) = \,- \frac{2 \, g^2 \, m_A \, m_B}{3(2\pi)^3} \nn \\
  & \F! \times \Bigg\{\frac{\de(\p_1 - \p_3 - \p_5 - \p_6) \, \de(\p_2 - \p_4) \, \de_{\lambda_1 \lambda_2}}{\left(\omega^A_{\p_1} \omega^A_{\p_3} \omega^B_{\p_5} \omega^B_{\p_6} \right)^{1/2}} \, \nn \\
  & \F!\F! \times \ub(\p_1, \kappa_1) \, \gamma_\nu \, u(\p_3, \kappa_2) \; \vb(\p_6, \nu_2) \, \gamma^\nu \, u(\p_5, \mu_2) \left(\frac{1}{(p_1 - p_3)^2} - \frac{1}{(p_5 + p_6)^2}\right) \nn \\
& \F!\H! + \frac{\de(\p_2 - \p_4 - \p_5 - \p_6) \, \de(\p_1 - \p_3) \, \de_{\kappa_1 \kappa_2}}{\left(\omega^A_{\p_2} \omega^A_{\p_4} \omega^B_{\p_5} \omega^B_{\p_6} \right)^{1/2}} \, \nn \\
  & \F!\F! \times \vb(\p_4, \lambda_2) \, \gamma_\nu \, v(\p_2, \lambda_1) \; \vb(\p_6, \nu_2) \, \gamma^\nu \, u(\p_5, \mu_2) \left(\frac{1}{(p_2 - p_4)^2} - \frac{1}{(p_5 + p_6)^2}\right) \Bigg\},
\end{align}
\begin{align}
  & \left(\cC_{2,4}\right)^{\kappa_2, \lambda_2, \mu_2, \nu_2}_{\kappa_1, \lambda_1}(\p_{1..6}) = \nn \\
  &  \F!\, \frac{ f^{abc}f^{abc} \, g^4 \, m_A^2 \, m_B}{(2\pi)^6}  \frac{\de(\p_1 + \p_2 - \p_3 - \p_4 - \p_5 - \p_6)}{\left(\omega_{\p_1}^A \omega_{\p_2}^A \omega_{\p_3}^A \omega_{\p_4}^A \omega_{\p_5}^B \omega_{\p_6}^B\right)^{1/2}} \; \frac{1}{(p_1 - p_3)^2} \frac{1}{(p_2 - p_4)^2} \frac{1}{(p_5 + p_6)^2} \nn \\
  &  \F! \times \bigg\{(p_3 - p_1)_\mu \, \ub(\p_1, \kappa_1) \, \gamma_\nu \,  u(\p_3, \kappa_2) \; \vb(\p_4, \lambda_2) \, \gamma^\mu \, v(\p_2, \lambda_1) \; \vb(\p_6, \nu_2) \, \gamma^\nu \, u(\p_5, \mu_2) \nn \\
  & \F!\H! - \, (p_3 - p_1)_\mu \, \ub(\p_1, \kappa_1) \, \gamma_\nu \, u(\p_3, \kappa_2) \; \vb(\p_6, \nu_2) \, \gamma^\mu \, u(\p_5, \mu_2) \; \vb(\p_4, \lambda_2) \, \gamma^\nu \, v(\p_2, \lambda_1) \nn \\
  & \F!\H! + \,  (p_4 - p_2)_\mu \, \vb(\p_4, \lambda_2) \, \gamma_\nu \, v(\p_2, \lambda_1) \; \vb(\p_6, \nu_2) \, \gamma^\mu \, u(\p_5, \mu_2) \;\ub(\p_1, \kappa_1) \, \gamma^\nu \, u(\p_3, \kappa_2) \nn \\
  & \F!\H! - \, (p_4 - p_2)_\mu \, \vb(\p_4, \lambda_2) \, \gamma_\nu \, v(\p_2, \lambda_1) \; \ub(\p_1, \kappa_1) \, \gamma^\mu \, u(\p_3, \kappa_2) \; \vb(\p_6, \nu_2) \, \gamma^\nu \, u(\p_5, \mu_2) \nn \\
  & \F!\H! + \, (p_5 + p_6)_\mu \, \vb(\p_6, \nu_2) \, \gamma_\nu \, u(\p_5, \mu_2) \; \ub(\p_1, \kappa_1) \, \gamma^\mu \,  u(\p_3, \kappa_2) \; \vb(\p_4, \lambda_2) \, \gamma^\nu \, v(\p_2, \lambda_1) \nn \\
  & \F!\H! - \, (p_5 + p_6)_\mu \, \vb(\p_6, \nu_2) \, \gamma_\nu \, u(\p_5, \mu_2) \; \vb(\p_4, \lambda_2) \, \gamma^\mu \, v(\p_2, \lambda_1) \; \ub(\p_1, \kappa_1) \,\gamma^\nu \, u(\p_3, \kappa_2) \bigg\}. 
  \label{EQ:RELC24}
\end{align}
Equations (\ref{EQ:EQRELQCD2}) and (\ref{EQ:EQRELQCD4}) are coupled relativistic integral equations for the wavefunctions $\cF$ and $\cG$ which are difficult, if not impossible, to solve. The function $\cF$ describes a quark-antiquark pair of mass $m_A$, while $\cG$ describes a two quark-antiquark pair state with each quark flavour of mass $m_A$ and $m_B$ respectively. These equations are, in principle, capable of describing the process of string breaking in QCD, i.e. the transition from a single meson state $q_A \, \bar{q}_A$ to a two meson state $q_A \, \bar{q}_A \, q_B \, \bar{q}_B$. The relativistic kinematics are described fully and without approximations. On the other hand, the dynamics are described approximately by the kernels $\cY_{2,2}$, $\cY_{2,4}$, $\cY_{4,2}$, $\cY_{4,4}$, $\cC_{2,4}$, $\cC_{4,2}$ and $\cQ_{4,4}$ because of the approximate nature of the trial state (\ref{EQ:QCDTRIAL24}) and the first-order iterative approximation in the reformulation procedure. The emphasis in this dissertation is on the quark-antiquark bound state described by the function $\cF$, therefore it is required to calculate the kernels $\cY_{4,4}$, $\cC_{4,2}$ and $\cQ_{4,4}$ for this purpose. Henceforth, equation (\ref{EQ:EQRELQCD4}) will not be investigated. The colour indices have been summed in $\cY_{2,2}$ and $\cY_{2,4}$ and the results appear as multiplicative factors in front. In $\cC_{2,4}$, the colour factor is expressed by the contraction of the structure constants $f^{abc}f^{abc}$. Note that the spin and momentum dependencies are still coupled in these relativistic equations whereas the colour dependence is separate and calculated fully.

In the non-relativistic limit, the spin and momentum dependencies of the wavefunctions decouple (i.e. separate into factors). That is, the functions $\cF$ and $\cG$ can be written as the products
\begin{gather}
  \cF_{\kappa, \lambda}(\p_{1,2}) = \Theta_{\kappa, \lambda} \, F(\p_{1,2}), \label{EQ:SPINFUNC1}\\  
  \cG_{\kappa, \lambda, \mu, \nu}(\p_{1..4}) = \Xi_{\kappa, \lambda, \mu, \nu} \, G(\p_{1..4})  \label{EQ:SPINFUNC2}.
\end{gather}
The spin indices $\kappa$, $\lambda$, $\mu$ and $\nu$ transform in the $SU(2)$ representation. Therefore, the spin indices in $\Theta$ and $\Xi$ transform according to a particular product of $SU(2)$ representation. The dimensions of these products of representations have been worked out using the Young tableaux method in Appendix B section \ref{SEC:YOUNG}: 
\begin{gather}
  2 \otimes \bar{2} = 1 \oplus 3, \\
  2 \otimes \bar{2} \otimes 2 \otimes \bar{2} = 1 \oplus 1 \oplus 3 \oplus 3 \oplus 3 \oplus 5.
\end{gather}
The index functions corresponding to these representations, such as $\Theta$ and $\Xi$ in equations (\ref{EQ:SPINFUNC1}) and (\ref{EQ:SPINFUNC2}), are eigenstates of the spin operator such that, schematically (spin indices must be included to be precise), 
\begin{gather}
  {\bf S}^2 \, \Lambda = \, s(s+1) \, \Lambda  
  \label{EQ:S2} \\ 
  S_3 \, \Lambda = \, m_3 \, \Lambda 
  \label{EQ:S3}
\end{gather}
where $\Lambda = \Theta, \,  \Xi$ is either of the spin index functions (i.e. equation (\ref{EQ:SPINFUNC1}) and (\ref{EQ:SPINFUNC2})), $s$ is the total spin quantum number and $m_3$ is the projection of the spin (i.e. the eigenvalue of the Pauli matrix $\sigma_3$ in the representation given by the equation (\ref{EQ:PAULIMATRICES})).

According to equation (\ref{EQ:V_2}), the triplet spin configuration produces a delta function contribution to the potential energy. Hence, it is expected that an intricate delta function dependence might arise if the spin index function is not in the singlet representation. The simplest spin configuration to consider, then, is the singlet where the spin eigenvalues are ${\bf S}^2 = S_3 = 0$. The corresponding properly normalized spin index functions are
\begin{gather}
  \Theta_{\kappa, \lambda} = \ds\frac{1}{\sqrt{2}}\, \epsilon_{\kappa, \lambda}, 
  \label{EQ:SPINTHETA}\\
  \Xi_{\kappa, \lambda, \mu, \nu} = \ds\frac{1}{3\sqrt{2}} \left(\epsilon_{\kappa \lambda} \, \epsilon_{\mu \nu} + \epsilon_{\kappa \nu} \, \epsilon_{\mu \lambda}\right)
  \label{EQ:SPINXI},
\end{gather}
where $\epsilon$ is the totally-antisymmetric tensor. One can verify, by explicitly writing out all spin indices, that spin index functions (\ref{EQ:SPINTHETA}) and (\ref{EQ:SPINXI}) satisfy the spin eigenvalue equations (\ref{EQ:S2}) and (\ref{EQ:S3}).
%

%
The non-relativistic equation for the function $F$ in the spin singlet configuration can be obtained by multiplying the relativistically reduced equation (\ref{EQ:EQRELQCD2}) by $\Theta$, of equation (\ref{EQ:SPINTHETA}), and summing over all the spin indices (see section \ref{SEC:QQ} of Appendix B for some details). Simultaneously, the functions $F$ and $G$ are expressed in the centre of mass frame, such that the total energy corresponds to the rest mass of the system:
\begin{gather}
  F(\p_{1,2}) = f(\p_1) \; \de(\p_1 + \p_2), \\
  G(\p_{1..4}) = g(\p_{1,2,3}) \; \de(\p_1 + \p_2 + \p_3 + \p_4). 
\end{gather}
Consequently, one can integrate out one momentum dependence using the overall momentum conserving delta functions and thus reduce the number of variables in the equations. Such an integration leads to slightly altered interaction kernels which exhibit a degree of skewness (loss of symmetry) in the dependence on the momentum variables. 

The non-relativistic equation for the quark-antiquark system in the singlet configuration of spin and in the centre of mass frame is
%
\begin{align}
  f(\p_1) \left( \frac{\p^2_1}{m_A} - {\cal E} \right) = \int d\p_3 \, \cY_{2,2}(\p_{1,3}) \, f(\p_3) + R \int d\p_{3,4,5} \; \cC_{2,4}(\p_{1,3,4,5}) \; g(\p_{3,4,5}).
  \label{EQ:EQFORSMALLF}
\end{align}
where ${\cal E} = E - 2 \, m_A$ is the non-relativistic energy and all spin indices have been summed. The non-relativistic kernel $Y_{2,4}$ has been omitted since its leading contribution is of order ${\cal O}(m^{-3})$. The remaining non-relativistic kernels $Y_{2,2}$ and $C_{2,4}$ have the leading contribution in order ${\cal O}(m^{-2})$: 
\begin{align}
  Y_{2,2}(\p_{1,3}) = & \, \frac{4}{3} \frac{g^2}{(2 \, \pi)^3} \, \Biggl(\frac{1}{(\p_3 - \p_1)^2}\Biggr) \; \Biggl(1 + \frac{(\p_3 - \p_1)^2}{4 \, m^2_A} + \frac{(\p_1^2 + \p_3^2)}{2 \, m^2_A}\,  \Biggr), \\
  C_{2,4}(\p_{1,3,4,5}) = &
  \, i \, \frac{ f^{abc}f^{abc} \, g^4}{4 \, (2\pi)^6 \, m^2_B} \, \frac{1}{(\p_3 - \p_1)^2 \, (\p_4 + \p_1)^2}  \; \Biggl(\frac{13}{2 \, m_A^2} \; \p_1\cdot\p_3\times\p_4 \nn \\
  & \F! - \frac{3 \, (m_B + m_A)}{m_B^2 \, m_A} \; \Bigl(\p_1\cdot \p_5 \times \p_3 \; + \; \p_1\cdot\p_5\times\p_4 \; + \; \p_5\cdot\p_3\times\p_4\Bigr) \Biggr),
\end{align}
where an extensive simplification has been performed to bring the kernels to the present form. The appearance of the factor $i$ in $C_{2,4}$ looks troublesome since the inter-particle potential can not be imaginary. However, the quantity $R$ is not restricted to be real and one can make a choice such that it is purely imaginary to enforce the overall term to be real. One can see from the spin and spinor indices summation in Appendix B section \ref{SEC:QQ} that at least there is a degree of consistency in this approach since both cross terms $Y_{2,4}$ and $C_{2,4}$ have an $i$ in front. Even so, there still remains the choice of the phase which leaves the sign in front undetermined. The variational method, if implemented, would select the sign corresponding to the lower energy (more on this matter is written below). Also, notice that the discernible momentum dependence implies that the Fourier transform of equation (\ref{EQ:EQFORSMALLF}) to coordinate space would not separate the inter-particle potential and the wavefunctions into separate factors. 

To perform a Fourier transform of equation (\ref{EQ:EQFORSMALLF}) one multiplies by $\ds\frac{\e^{-i\, \p_1\cdot\x}}{(2 \, \pi)^{3/2}}$ and integrates over $\p_1$. As already mentioned, such an operation does not lead to an equation where the wavefunction and inter-particle potential stand as separate factors in all terms. In the terms where the decoupling does not occur, one is forced to multiply and divide by $f(\x)$. Subsequently, an ansatz for $f$ must be provided, such as the following choice
%
\begin{align}
  f(x) = \sqrt{\frac{1}{\pi \, a^3}} \, \exp{\Bigl(-\frac{x}{a}\Bigr)},\F! f(p) = \frac{\sqrt{8} \, a^{3/2}}{\pi \, (\p^2 \, a^2 + 1)^2},
  \label{EQ:HYDROGEN}
\end{align}
which is the properly normalized wavefunction for the ground state of hydrogen (or, more precisely, positronium), with $a$ being the characteristic size.

Similarly, an ansatz for $g$ has to be provided. In the coordinate representation, the function $g$ has to be brought to a suitable frame where the particles described by the coordinates $\x_1$ and $\x_2$ must be in the centre of mass frame. The dependence on the coordinates $\x_3$ and $\x_4$ can be arbitrary since it is integrated out. There is no unique choice for such a choice of coordinates. However, the following transformation has been found useful:
\begin{align}
    \left[\begin{array}{c}
      \x_3 \\ \x_4 \\ \x_5 \\ \x_6 
    \end{array}\right] =
  \left[\begin{array}{cccc}
      1 & 1/2 & 0 & 0 \\
      1 & -1/2 & 0 & 0 \\
      1 & 1/2 & -1 & 0 \\
      1 & 1/2 & 0 & -1 \\
      \end{array}\right] \,
  \left[\begin{array}{c}
      {\bf X } \\ \x_{34} \\ \x_{45} \\ \x_{56} 
    \end{array}\right].
  \label{EQ:COORDINATES}
\end{align}
One can carry out a schematic Fourier transform of $G$ using the coordinate transformation (\ref{EQ:COORDINATES}) to obtain the following dependence in momentum space:
\begin{align}
  \int d\x_{3..6} \, & \prod_{i=3}^{6} \, \e^{i \p_i \cdot \x_i} \, G(\x_{3..6}) = \int d{\bf X} \, d\x_{34} \, d\x_{45} \, d\x_{56} \; 
  \e^{i \, {\bf X}\cdot(\p_3 + \p_4 + \p_5 + \p_6)} \, \nn \\
  & \F!\F!\F!\F! \times  \e^{ - i \, 1/2 \, \x_{34}\cdot(\p_3 + \p_5 + \p_6 - \p_4)} \; \e^{-i \, \x_{45}\cdot\p_5} \;   \e^{-i \, \x_{56}\cdot\p_6} G(\x_{34}, \x_{45},\x_{56})\nn \\
  & = (2 \, \pi)^3 \de(\p_3 + \p_4 + \p_5 + \p_6) \; g(\, \p_4, \p_5, \p_6 \,)
  \label{EQ:FTG}
\end{align}
The delta function refers to the centre of mass motion which is a non-normalizable factor. For the remaining function $g$, in equation (\ref{EQ:FTG}), one can make a variational ansatz such as 
\begin{align}
  g(\p_4, \p_5, \p_6) \sim \frac{a^{9/2}}{f^{abc}f^{abc}} \, \frac{(2 \, \pi)^4 \, 2^{3/2} \, \pi}{(\p_4^2 \, a^2 + 1)^2} \, \frac{1}{(\p_5^2 \, a^2 + 1)^2} \, \frac{1}{(\p_6^2 \, a^2 + 1)^4}, 
  \label{EQ:SMALLG}
\end{align}
where the factors in front are included for convenience and the normalization of this wavefunction is absorbed into the definition of $R$ (in fact, all constants in front can be absorbed into the definition of $R$). Equation (\ref{EQ:SMALLG}) reflects, as in equation (\ref{EQ:HYDROGEN}), a factorized hydrogen-like dependence for each three-momentum variable and is only appropriate for the ground state since it contains no angular dependence. A similar ansatz for the four-component coefficient function $G$ has been used in paper~\cite{EmamiRazavi:2006yx}.

Upon calculating the Fourier transform of equation (\ref{EQ:EQFORSMALLF}) using the ans\"atze (\ref{EQ:HYDROGEN}) and (\ref{EQ:SMALLG}), the equation in coordinate representation becomes:
\begin{align}
  -\, \frac{\nabla^2}{m^A} \,  f(\x) +  \Bigl(V_1(x)  + V_2(x)\Bigr) \,  f(\x) = \, {\cal E} \, f(\x).
\end{align}
The $V_1$ contribution to the inter-particle potential is obtained from
\begin{multline}
  -  V_1(x) \, f(\x) = \frac{1}{(2 \, \pi)^{3/2}} \, \int d\p_{1,3} \, \e^{-i\, \p_1\cdot\x} \;  Y_{2,2}(\p_{1,3}) \, f(\p_3)  \\
  = \; \frac{4}{3} \, \frac{\alpha_s}{2 \pi^2}\int d\p_{1,3} \, \e^{- i \, \p_1\cdot\x}\Bigg\{\frac{F(\p_3)}{(\p_3 - \p_1)^2}  + \frac{f(\p_3)}{4 \, m^2_A} 
    \\ + \frac{a^3}{2 \pi^2 m_A^2} \, \e^{x/A} \, \frac{(\p_1^2 + \p_3^2)}{(\p_3 - \p_1)^2} \, \frac{f(\x)}{(\p_3^2 \, a^2 + 1)^2}\,  \Bigg\},
  \label{EQ:PSIV1}
\end{multline}
where $\alpha_s = \ds\frac{g^2}{4 \, \pi}$ is the dimensionless coupling constant of the strong interaction. The last term in the second line of equation (\ref{EQ:PSIV1}) has been multiplied and divided by $f(x)$, and (\ref{EQ:HYDROGEN}) used to approximate $f$ in coordinate and momentum spaces. It is manifest that the contribution to the potential energy from the last term is non-local and the procedure of multiplying and dividing by $f$ makes a local approximation for it. It follows from (\ref{EQ:PSIV1}) that 
\begin{align}
  V_1(x) = - \frac{4}{3} \, \alpha_s \Bigg\{ \frac{1}{x} +  \frac{\pi}{m^2_A} \, \delta(\x) \, + \frac{a^3}{4 \,  \pi^4 \, m^2_A } \, \exp\Bigl(\frac{x}{a}\Bigr) \int d\p_{1,3} \, \frac{\exp(-i \, \p_1 \cdot\x)(\p^2_1 + \p^2_3)}{(\p_3 - \p_1)^2 \, (\p^2_3 \, a^2 + 1)^2} \Bigg\},
\end{align}
where the $\ds\frac{1}{x}$ term has been obtained by the usual integration in spherical coordinates followed by the radial integration in the complex plane. With little effort, one can reduce the integral in the last line to a double quadrature:
\begin{multline}
  V_1(x) = - \frac{4}{3} \, m_A \, \alpha^2_s \; \Bigg\{ \frac{1}{r} + \alpha^2_s \, \pi \, \delta({\bf r}) \\
  + \alpha_s^2 \, \frac{2}{\pi^2 \,A^2 \, r} \, \exp\Bigl(\frac{r}{A}\Bigr)\int_0^{\infty} d p_1 \, \sin\Bigg( p_1 \frac{r}{A}\Bigg) \int_0^{\infty} d p_3 \; p_3 \; \frac{p_1^2 + p_3^2}{(p^2_3 + 1)^2} \, \ln\bigg|\frac{p_1 + p_3}{p_1 - p_3}\bigg| \Bigg\}
  \label{EQ:V1}
\end{multline}
where $r = |{\bf r}| = |\x| \, m_A \, \alpha_s$ is the dimensionless quark-antiquark separation variable, $A = a \, m_A \, \alpha_s$ is the dimensionless variational parameter and the remaining integrals are expressed in terms of dimensionless momentum variables. In light of the fact that the second and third terms in equation (\ref{EQ:V1}) are of the same order in $\alpha_s$, it is conceivable that the result of the intricate double integral of the third term also leads to a delta function dependence as in the second term. 
The $V_2$ contribution to the inter-particle potential comes from
\begin{align}
  - & V_2(x) \; f(\x) =  \frac{R}{(2 \, \pi)^{3/2}} \, \int d\p_{1,3,4,5} \, \e^{-i\, \p_1\cdot\x} \;  C_{2,4}(\p_{1,3,4,5}) \, g(\p_{3,4,5}) \nn \\
  = & \; R \, i \,\frac{\alpha^2_s \, a^6}{m_B^2} \, \e^{x /a}\;  f(\x) \; \int d\p_{1,3,4,5} \, \e^{ - i \, \p_1\cdot\x} \nn \\
  & \times \, \Biggl( \frac{1}{(\p_3 - \p_1)^2} \, \frac{1}{(\p_4 + \p_1)^2} \, \frac{1}{(\p_3^2 \, a^2 + 1)^2} \, \frac{1}{(\p_4^2 \, a^2 + 1)^2} \, \frac{1}{(\p_5^2 \, a^2 + 1)^4} \Biggr) \nn \\ 
  & \times \, \Biggl(\frac{13}{2 \, m_A^2} \; \p_1\cdot\p_3\times\p_4 - \frac{3 \,  (m_B + m_A)}{m_B^2 \, m_A} \; \Bigl(\p_1\cdot \p_5 \times \p_3 \; + \; \p_1\cdot\p_5\times\p_4 \; + \; \p_5\cdot\p_3\times\p_4\Bigr)\Biggr) 
  \label{EQ:PSIV2}
\end{align}
where, again, it has been multiplied and divided by $f$ and the ans\"atze (\ref{EQ:HYDROGEN}) and (\ref{EQ:SMALLG}) have been substituted. Then, it follows that the $V_2$ contribution to the potential is
\begin{align}
  V_2(\x) & = \; \mp  \, |R^\prime| \frac{m_A \, \alpha^7_s}{\xi^2 \, A^5} \; \e^{r /A}\; \int d\p_{1,3,4,5} \; \e^{ - i \, \p_1\cdot{\bf r} /A} \nn \\
  & \times \, \Biggl(\frac{1}{(\p_3 - \p_1)^2} \, \frac{1}{(\p_4 + \p_1)^2} \, \frac{1}{(\p_3^2 + 1)^2} \, \frac{1}{(\p_4^2 + 1)^2} \, \frac{1}{(\p_5^2 + 1)^4} \Biggr) \nn \\
  & \times \, \Biggl(\frac{13}{2} \; \p_1\cdot\p_3\times\p_4 - \frac{3 \,  (\xi + 1)}{\xi^2} \; \Bigl(\p_1\cdot \p_5 \times \p_3 \; + \; \p_1\cdot\p_5\times\p_4 \; + \; \p_5\cdot\p_3\times\p_4 \Bigr) \Biggr).
   \label{EQ:V2}
\end{align}
where the notation $m_B = \xi \, m_A$ has been introduced and the necessary substitutions have been made as before to make the integration variables dimensionless. The redefinition $R^\prime = \e^{i \, \pi / 2} \, R = i \, R$ does not provide the overall sign as there are two choices of the phase which yield a real number. The correct sign will be determined below. 

In the non-relativistic limit, the heavier quark flavour is suited to be either the charm with the mass of $1.25\pm0.09 \; \tx{GeV}$ or the bottom quark with the mass of $4.70\pm0.07 \; \tx{GeV}$~\cite{PDBook}. On the other hand, to model a realistic string breaking effect, the lighter quark flavour could be the up or the down quark bearing approximately identical but ill-defined masses in the neighbourhood of $1.5 - 7.0 \; \tx{MeV}$ ~\cite{PDBook}. Hence, the value of $\xi = 0.001$ is a decent benchmark and approximately corresponds to the mass difference of the proposed flavours. With $\xi = 0.001$ the first term in equation (\ref{EQ:V2}) can be ignored. 

The numerical value of the coupling constant $\alpha_s$ must be obtained from equation (\ref{EQ:ALPHA_S}). It should correspond to the bound state energy of, say, a bottom quark-antiquark system in the singlet spin configuration which is $9.86 \; \tx{GeV}$ in energy~\cite{PDBook}. This is the QCD bound state whose potential will be modelled.
 As an input to equation (\ref{EQ:ALPHA_S}), one can use the experimentally measured value $\alpha_s(m_Z) = 0.117$ where $m_{Z} = 91.19 \; \tx{GeV}$~\cite{PDBook}.  Upon substitution, one retrieves $\alpha_s(9.86) = 0.167$. When expressing the potential in units of $m \,\alpha^2_s$, the contribution $V_2$ is of order $\alpha_s^5$ which justifies the usage of such a large value for the coupling constant. The bottom quark-antiquark system, one unit of distance in Bohr radii corresponds to approximately $2.51\times10^{-16}\; \tx{m}$ of actual length.

Equation $(\ref{EQ:V2})$, as it stands, contains two, yet undetermined, variational parameters $A$ and $R$ (or $R^\prime$). The numerical values of these parameters should be obtained from a variational calculation of the matrix element $\bra \Psi_t | \, H - E \, | \Psi_t \ket$. Such a calculation requires great effort which is not undertaken in this work. Instead, the approach shall be to multiply equation $(\ref{EQ:EQFORSMALLF})$ by $f^\ast(\p_1)$ and, thereafter, integrate out the free variable $\p_1$:
\begin{multline}
  \int d\p_1 \, f^\ast(\p_1) \, f(\p_1) \; \left( \frac{\p^2_1}{m_A} - {\cal E} \right) = \int d\p_{1,3} \, f^\ast(\p_1) \, \cY_{2,2}(\p_{1,3}) \, f(\p_3) \; + \\ R \int d\p_{1,3,4,5} \; f^\ast(\p_1) \; \cC_{2,4}(\p_{1,3,4,5}) \; g(\p_{3,4,5}).  
  \label{EQ:ENERGYEXPRESSION}
\end{multline}
This is an expression for the total non-relativistic energy of a meson in the singlet spin configuration and it could be used to estimate one of the two variational parameters. A variational calculation of energy, according to a theorem~\cite{Ballentine1989}, always yields a result which is greater than the true value. As QCD bound states are confining (i.e. ${\cal E} > 0$), one can obtain the following expression containing the parameters $A$ and $R$ (see Appendix B section \ref{SEC:PARAMETER}):
\begin{align}
  {\cal E} = E - 2 \, m_A = & \; m_A \, \alpha_s^2 \, \Biggl(\frac{1}{A^2} - \frac{4}{3}\bigg( \frac{1}{A} + \frac{\alpha_s^2(1 + 16 \, \pi^{-2} \, C_1)}{A^3} \bigg)\Biggr) \pm | R^\prime| \, \frac{m_A \, \alpha_s^7}{A^5} \, C_2,
   \label{EQ:ENERGYA}
\end{align}
where the constant $C_1 \approx1.85044$ emerges from the calculation of the first term on the left hand side of equation $(\ref{EQ:ENERGYEXPRESSION})$ (this calculation appears in Appendix B section \ref{SEC:PARAMETER}). The constant $C_2$:
\begin{align}
  C_2 & = \; \frac{A^5}{m_A \, \alpha_s^7 \, i} \, \int d\p_{1,3,4,5} \; f(\p_1) \; C_{2,4}(\p_{1,3,4,5}) \; g(\p_{3,4,5}) \nn \\
  \, & =  \, - \frac{8}{\xi^2} \int d\p_{1,3,4,5} \nn \\
  \, & \times \, \Biggl(\frac{1}{(\p_1^2 +1)^2} \, \frac{1}{(\p_3^2 +1)^2} \, \frac{1}{(\p_4^2 +1)^2} \, \frac{1}{(\p_5^2 +1)^2} \, \frac{1}{(\p_3 -\p_1)^2} \, \frac{1}{(\p_4 + \p_1)^2} \Biggr) \nn \\
  \, & \times  \, \Biggl(\frac{13}{2} \; \p_1\cdot\p_3\times\p_4 - \frac{3 \, (\xi + 1)}{\xi^2} \; \Bigl(\p_1\cdot \p_5 \times \p_3 \; + \; \p_1\cdot\p_5\times\p_4 \; + \; \p_5\cdot\p_3\times\p_4 \Bigr) \Biggr)
  \label{EQ:C2}
\end{align}
%
is a multidimensional integral expression which one has to solve numerically using, in practice, the Monte Carlo method.

Unfortunately, it turns out that Monte Carlo integration of equation (\ref{EQ:C2}) does not produce reliable results. The troublesome pieces of the integrand are the triple scalar vector products in the numerator. To circumvent this difficulty, one can place an upper bound on the integral by considering upper bounds on the scalar vector products:
\begin{align}
  \p\cdot\q\times{\bf k} = |\p |\, |\q| \, |{\bf k}| \, \sin{\theta} \, \cos{\phi} \leq |\p |\, |\q| \, |{\bf k}|,
\end{align}
where $\phi$ is the angle between the vectors $\q$ and ${\bf k}$ and $\theta$ is the angle between the vectors $\p$ and $\q\times{\bf k}$. 

When all scalar vector products are replaced by their upper bound estimates, equation (\ref{EQ:C2}) can be reduced to a triple quadrature in the radial coordinates of the momentum variables:
\begin{align}
   C_2 & \leq \; \frac{24 \, (\xi + 1)}{\xi^4} \, \frac{4 \, \pi^4}{3}\int dp_1 \, dp_3 \, dp_4 \; p_3 \, p_4 \left( p_1 \, p_3 + p_1 \, p_4 + p_3 \, p_4 \right)\nn \\
  \, & \times \, \Biggl(\frac{1}{(p_1^2 + 1)^2} \, \frac{1}{(p_3^2 +1)^2} \, \frac{1}{(p_4^2 +1)^2}  \Biggr) \, 
  \ln\Bigg(\frac{(p_3 + p_1)^2 + \omega^2}{(p_3 - p_1)^2 + \omega^2}\Bigg) \;   \ln\Bigg(\frac{(p_4 + p_1)^2 + \omega^2}{(p_4 - p_1)^2 + \omega^2}\Bigg),
\end{align}
where only the leading terms in the parameter $\xi$ have been kept. To obtain this result, angular integrations resembling those in the calculation of the constant $C_1$ have been employed. The parameter $\omega$ is a regulator which has been inserted to ensure that a numerical evaluation of this triple quadrature converges. In general, one must perform a handful of numerical integrations where the value of $\omega$ is reduced in each successive trial. If the integral converges then its numerical evaluation should approach a constant value as $\omega$ tends zero. In this case, the triple quadrature is found to converge to the value
\begin{align}
  C_2 \leq 1.005\times 10^{16}.
  \label{EQ:NUMERICC2}
\end{align}
%

%

\begin{figure}[t]
  \center{
    \includegraphics[scale=0.6]{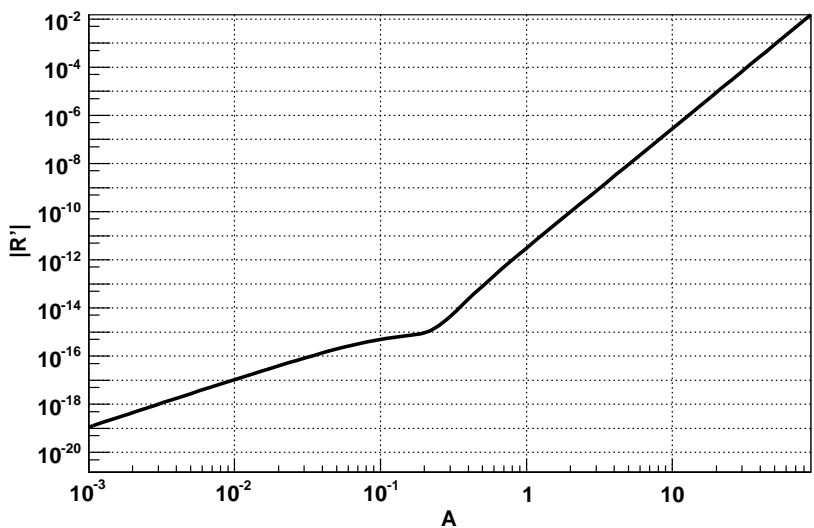}
  }
  \spacing{1}
  \caption{The allowed values for the parameter $A$ and $R^{\prime}$ according to equation (\ref{EQ:ENERGYA}) with the plus sign.}
  \label{FIG:PLOTRA}
\end{figure}

Turning attention back to equation (\ref{EQ:ENERGYA}) itself, the energy of the bottom quark-antiquark state ${\cal E}$, expressed in units of $m \, \alpha_s^2$, is 3.55. Accordingly, one can solve the equation to determine the values of $A$ and $R$ which could produce such energy. Doing so, one finds that only the plus sign in equation (\ref{EQ:ENERGYA}) (correspondingly the negative sign in equation (\ref{EQ:V2})) is realizable for this energy. Figure (\ref{FIG:PLOTRA}) shows a plot of the allowed values of the parameters $A$ and $R^\prime$. Recall, the optimal value of $A$ is believed not to deviate much from unity. 

\begin{figure}[t]
  \center{
    \includegraphics[scale=0.64]{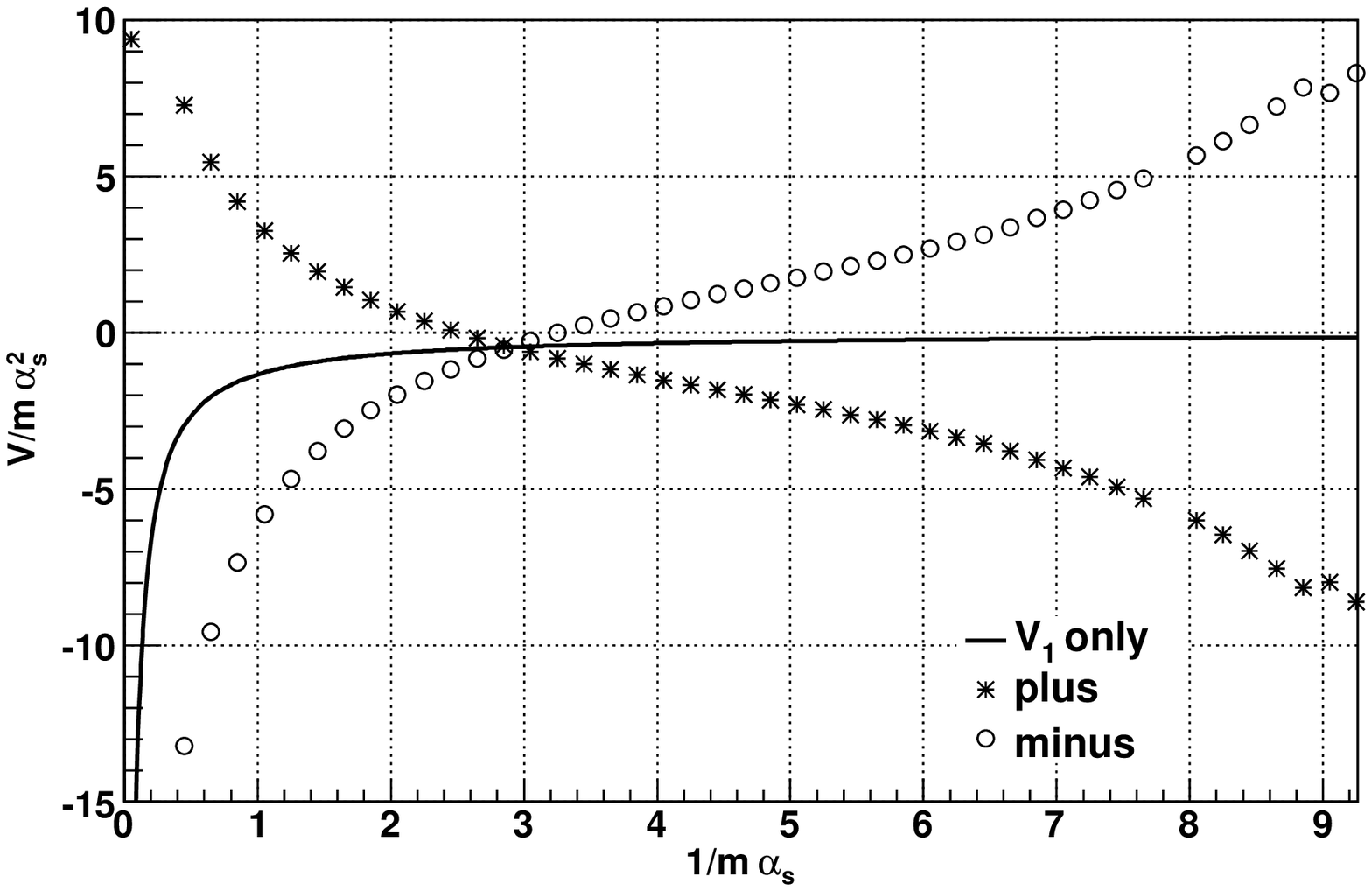}
  }
  \spacing{1}
  \caption{The quark-antiquark potential $V_1 + V_2$ for $A = 1.0$ and $|R| = 3.0\times10^{-12}$. The solid line shows the Coulombic $V_1$ contribution only. The two sets of plotted points corresponds to the two choices of the overall sign in $V_2$ as indicated. The lower curve (labelled ``plus'') is clearly unphysical.}
  \label{FIG:QQPOT1}
  \center{
    \includegraphics[scale=0.64]{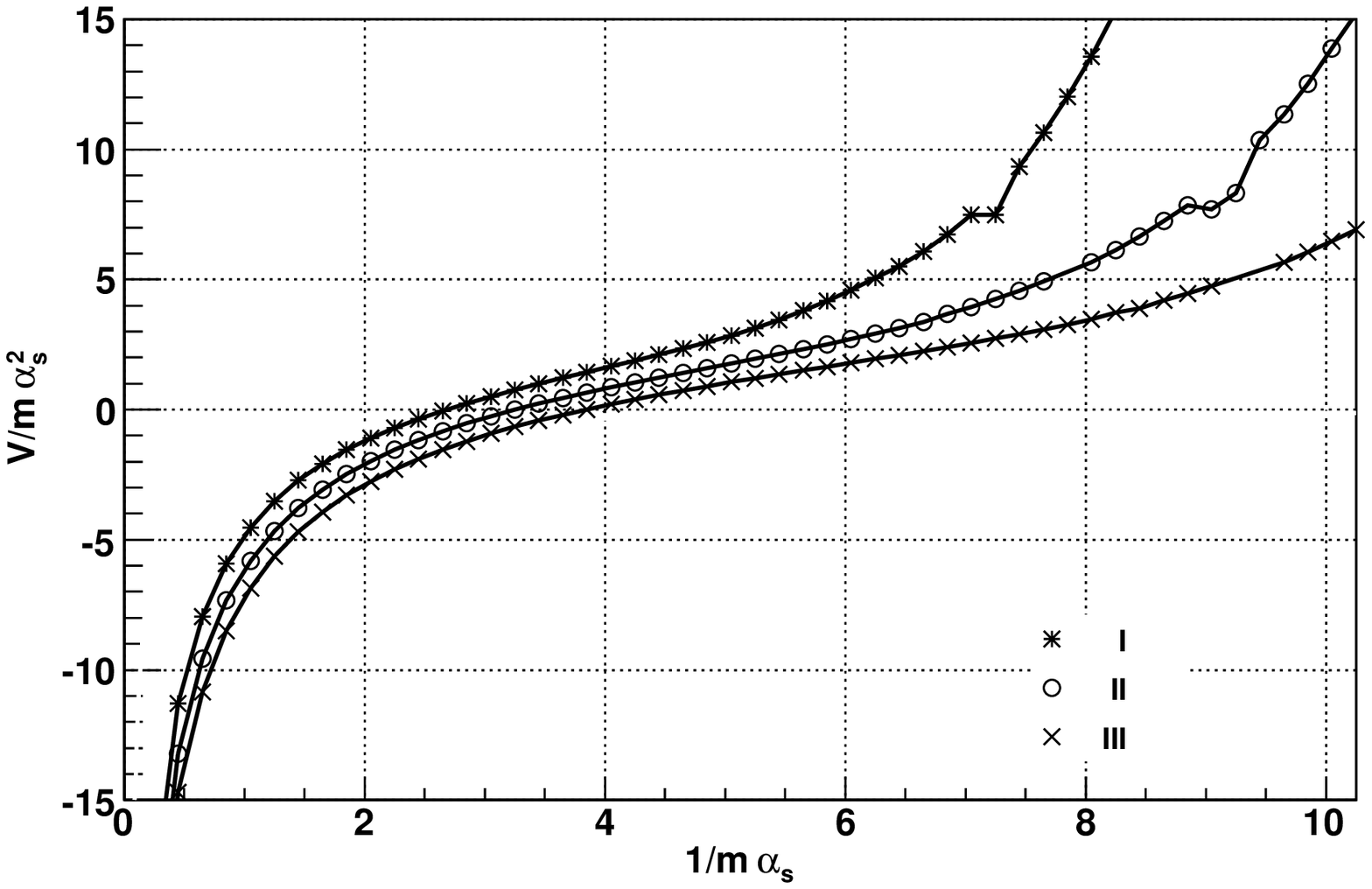}
  }
  \spacing{1}
  \caption{The quark-antiquark potential $V_1 + V_2$ for three choices of the parameters $(A, |R|)$ with the appropriate (-) sign in $V_2$. The choices are I: $(0.8, 9.9\times10^{-13})$, II: $(1.0, 3.1\times10^{-12})$ and III: $(1.2, 7.7\times10^{-12})$. }
  \label{FIG:QQPOT2}
\end{figure}

Having figured out all the values for the inputs, it is time to return to the expression for $V_2$, equation (\ref{EQ:V2}). The multidimensional expression can be reduced to the following triple quadrature:
\begin{align}
  V_2(\x) & = \; \mp |R^\prime| \, \frac{m_A \, \alpha^7_s}{A^4} \; \frac{4 \pi^4 \, (\xi + 1)}{\xi^4} \; \frac{\e^{r /A}}{r} \nn \\
  & \H! \times \int dp_1 \, dp_3 \, dp_4 \; \frac{p_3 \, p_4}{p_1} \, \sin\bigg( p_1 \frac{r}{A} \bigg) \, ( p_1 \, p_3 + p_1 \, p_4 + p_3 \, p_4 )\nn \\
  & \H! \times \, \Biggl(\frac{1}{(p_3^2 + 1)^2} \, \frac{1}{(p_4^2 + 1)^2} \, \ln\Bigg(\frac{(p_3 + p_1)^2 + \omega^2}{(p_3 - p_1)^2 + \omega^2}\Bigg) \, \ln\Bigg(\frac{(p_4 + p_1)^2 + \omega^2}{(p_4 - p_1)^2 + \omega^2}\Bigg) \Biggr) 
  \label{EQ:V2A}
\end{align}
where similar steps have been taking as those in the calculation of the constant $C_1$. The parameter $\omega$ is a regulator which is, ultimately, to be taken to zero and, as before, $R^\prime = i \, R$ is a redefinition with the now-determined sign. The correct sign, according to equation (\ref{EQ:V2A}), must be negative. For an instructive purpose, inter-particle potential curves corresponding to both signs in equation (\ref{EQ:V2A}) are plotted in Figure (\ref{FIG:QQPOT1}). The correct sign (-) exhibits a confining potential at larger separation distances, whereas the wrong sign (+) yields a potential which does not support stable bound states. 

It is useful to examine the behaviour of the potential for somewhat different values of the parameters $A$ and $R$. To this end, Figure \ref{FIG:QQPOT2} illustrates the quark-antiquark potential for the indicated values of the parameters $A$ and $R$ which are obtained from the curve in Figure \ref{FIG:PLOTRA}. The quark-antiquark potential is substantially altered from its Coulombic behaviour by the $H_{3A}^R$ term of the QCD Hamiltonian (\ref{EQ:QCDHAMREFORM}). The apparent linear segment is characteristic of all three curves (at least in the domain $r \lesssim 7$ beyond which the numerical results become unreliable). The higher values of $A$ seems to render the linear segment longer. Beyond the separation distances shown on the graph, the points become increasingly scattered. The likely cause of this scattering is in the difficulty of evaluating accurately the product of the exponential factor $\e^{r/A}$ and the numerical calculation of the triple quadrature in $V_2$, equation (\ref{EQ:V2A}). This scattering can be diminished by requiring greater numerical accuracy in the evaluation of the triple quadrature but, of course, at the expanse of computational time. In any case, the validity of the curves is questionable beyond the linear segment and this is likely due to the limited accuracy in evaluating the triple quadrature. This is further discussed in the final section. 

It is of interest to extract the string tension $\sigma$ (i.e. the slope of the linear segment) of the quark-antiquark potential and compare it with the value known from lattice gauge calculations. In Greensite's book~\cite{greensite2011an}, the value obtained from LQCD is quoted to be $\sigma \approx 0.18 \, \tx{GeV}^2$. In Figure \ref{FIG:QQPOT2}, the approximate values of the slopes are $0.16$, $0.13$, $0.11 \; \tx{GeV}^2$ respectively. These derived values are in reasonable agreement with the LQCD calculations. Thus, the present results are gratifying given the approximate knowledge of the parameters $A$ and $R$.

Note that in the approach of this dissertation, the effect of the non-Abelian terms on the inter-particle interactions is to modify the shape of the potential but leave the coupling constant $\alpha_s$ unchanged. Within a perturbative calculation, the inclusion of virtual quark-antiquark pairs would give rise to vacuum-polarization effects. Ultimately, these would lead to a distance-dependent coupling constant $\alpha_s$ that would modify the strength of the Coulomb potential. 

%
This is quite different from the perturbative S-matrix formalism where the coupling constant $\alpha_s$ becomes energy dependent but the Coulombic shape of the potential remains unchanged. 
Nevertheless, it is gratifying to see that the inclusion of a virtual quark-antiquark pair in the trial state (\ref{EQ:QCDTRIAL4}) changes the potential between the heavy valence quark-antiquark pair from purely attractive Coulombic one to one which exhibits linear confinement.

\chapter{Concluding Remarks}
\spacing{1}
\epigraph{... until non-abelian gauge theories are solved
  analytically, there is likely to be disagreement about ... and the
  origin of the confining force.}{Jeff Greensite ({\sl An Introduction to
  the Confinement Problem })}
\spacing{2}

The research presented in this dissertation is concerned with the derivation of the inter-particle potentials in few particle systems in QFTs with non-linear mediating fields. The derivation of the potential is implemented in the Hamiltonian formalism of QFT. The method employs a reformulation of the original Hamiltonian to enable the usage of simpler trial states. The reformulation follows from the classical equations of motion and the Greens function for the mediating field. The equation of motion for the mediating field is non-linear and has no known analytical solutions. In the absence of other readily available alternatives, an iterative technique is used to approximate the solution to the leading order in the coupling constants of the underlying theory. 

The iterative technique is only valid in the perturbative regime. Therefore, the developed method of deriving inter-particle potentials is not universal. To extend its applicability to the non-perturbative regime would require an ingenious method of solving equations (\ref{EQ:CHI_FORMAL_SOL}) and (\ref{EQ:FORMAL_SOL}) possibly using a variational technique that would reformulate the Lagrangian densities (\ref{EQ:SCLLAGR}) and (\ref{EQ:LQCD1}) but differently. At this point, it is only a speculation and suggests a possible direction of continuing this research.

To apply the described method, in its current form, requires the calculation of matrix elements of the Hamiltonian in the context of few particle systems. With the reformulation, it is not necessary to include mediating field quanta in the trial state, unless one wishes to include processes that involve their physical (as opposed to virtual) emission and absorption. Application of the variational method leads to relativistic equations where the interactions among the particles are represented by kernels which are accurate to the leading order in the coupling constants. However, in order to capture as much relevant physics as possible one is compelled to employ multi-component trial states such as in equations (\ref{EQ:TRIALTWOFOUR}) and (\ref{EQ:QCDTRIAL24}). The multi-component trial states lead to coupled relativistic equations which are more difficult to analyze. The amount of algebra increases drastically. This is unavoidable and a substantial complication of the method.

Once the relativistic equations are established, they are then reduced to the non-relativistic limit and Fourier-transformed to the coordinate representation. In coordinate representation, the interaction kernels become inter-particle potentials which, due to the complexity of the non-linear interactions, are expressed in the form of multi-dimensional integrals. These integrals are solvable using numerical methods. For the systems described by multi-component trial states, one requires the complete knowledge (i.e. the exact wave function), in general, of all but one component. Thus, one is forced to postulate ans\"atze which contain variational parameters to provide a reasonable representation the system. This complication is another drawback of this method since it requires a tedious calculation to determine the optimal values of the variational parameters. 

The result of the chapter devoted to the scalar model with a Higgs-like mediating field is that there is no confinement for few particle systems described by the trial states (\ref{EQ:TRIALTHREE}), (\ref{EQ:TRIALFOUR}) and (\ref{EQ:TRIALTWOFOUR}) in the case when $\mu > 0$. The non-linear interaction terms (\ref{EQ:MODHAMC}) and (\ref{EQ:MODHAMQ}) do not raise the inter-particle potential to the region where energy is positive. On the contrary, the inclusion of these terms seems to lower the inter-particle potential as can be seen in Figures (\ref{FIG:V_122SEPARATE}) and  (\ref{FIG:PLOTS1}). For the $\mu = 0 $ case, the inter-particle potential must be regularized to be finite. This regularization introduces an infinite constant in the shifted energy of equation (\ref{EQ:REC2}) and hence is considered unphysical. It is uncertain whether the appearance of this infinity is due to the inadequate choices of ansatz or the possibility that the theory given by the Lagrangian density (\ref{EQ:SCLLAGR}) is ill-defined when $\mu = 0$.


The reformulation of the QCD Lagrangian is readily generalizable from the scalar Higgs-like model. Actually, the reformulation is more appealing in QCD than in the scalar Higgs-like model since it does not involve an intermediate Lagrangian as that of equation (\ref{EQ:MODLAGR}). On the other hand, the reformulation of QCD requires a choice of gauge. Ideally, the choice of gauge should not alter the final result. Notwithstanding, a computational technique usually loses its attractiveness when the gauge invariance becomes broken. The gluon propagator in the Feynman gauge, equation (\ref{EQ:G_PROP}), leads to the reformulated Lagrangian terms given by equations (\ref{EQ:INTQCD1})-(\ref{EQ:REF_LQCD}). It would be quite interesting to see what these terms look like with no gauge fixing and how the results would be affected by different gauge choices.

The vanishing of the colour index in the calculation of the matrix element of the Hamiltonian for the three quark system shows that the mono-component trial state (\ref{EQ:TRIALQCDTHREE}) is insufficient to describe this bound state properly. To observe the effects of the non-linear terms on a three quark system would require constructing a three plus five particle trial state. The reason why a simple three quark trial state fails is probably due to the fact that it does not contain any virtual pairs and cannot accommodate string breaking. It is the inclusion of a virtual pair, as seen in the analysis of the two-component trial state (\ref{EQ:QCDTRIAL24}), which invokes the extra interaction that induces confinement. Recall that there are no gluon bound states observed in nature despite the fact that there are LQCD predictions of them~\cite{PDBook}. Therefore, this suggests that it is primarily the virtual quarks, and not gluons, which provide the means for the strong interaction to trap valence quarks in strongly bound states. In fact, without explicitly using any gluon quanta in the trial state (\ref{EQ:QCDTRIAL24}) (though virtual gluons are still represented by their propagators), the derived quark-antiquark potential shown in Figure \ref{FIG:QQPOT2} shows evidence of linear confinement.  

Perhaps, the most important results of this dissertation are reflected in Figures \ref{FIG:QQPOT1} and \ref{FIG:QQPOT2}. These figures imply the possibility of linear confinement in quark-antiquark systems. The deviation from a linear form of the potential for large separations, in addition to inaccuracy of the numerical evaluation of the triple quadrature in $V_2$ equation (\ref{EQ:V2A}), is from the limitations of the variational ansatz given by equation (\ref{EQ:QCDTRIAL24}) (although, this is suspected to be a small effect). 
To improve upon this, one might consider more flexible forms for the functions $F$ and $G$ containing more non-linear parameters. This should translate into more accurate and realistic potential curves for larger separation distances (i.e. perhaps a longer linear segment). The drawback of more flexible forms for the functions $F$ and $G$ is that it would require more variational parameters which have to be thoroughly calculated, and not estimated, if an improvement on what has been done here is sought.  


%
The derivation of the quark-antiquark potential for a heavy meson produces an estimate of the string tension $\sigma$. The string tension is frequently written as $\sigma = k \, \Lambda_{\tx{QCD}}$, where $k$ is a constant to be determined and $\Lambda_{\tx{QCD}}$ is the renormalization-scheme dependent QCD scale. In perturbation theory, the determination of $\Lambda_{\tx{QCD}}$ requires QCD theory and experimental input. In LQCD, the determination of the string tension $\sigma$ does not require knowledge of $\Lambda_{\tx{QCD}}$. In this dissertation, the string tension is similarly obtained directly from the linear form of the confining potential with the use of the coupling constant $\alpha_s$ and the current quark masses as inputs. 


From the usual perturbative treatment of QFT, one expects loop effects to dominate the mechanism of confinement. In the present formulation, the Feynman-like diagrams for the cubic and quartic interactions, shown in Figures \ref{FIG:FD2} and \ref{FIG:FD3}, and consequently the matrix elements do not posses these properties. However, the relevant loop effects are believed to be incorporated by the reformulation of the Lagrangian density. One should also note that in the context of LQCD, the exclusion of the fermion propagator in the calculation (the quenched approximation) does not alter the potential drastically~\cite{Knechtli:2011pz}. 

In this dissertation, the light quarks in the two plus four QCD calculation have been treated non-relativistically. This is acceptable since in the heavy quark-antiquark equation (\ref{EQ:EQFORSMALLF}) the kinetic energies of the light quarks do not appear and the non-relativistic approximation is sufficient in the potential energy. Recall that in the non-relativistic limit, the potential terms have been expanded to the leading order in inverse powers of the mass $m_B$.  

A main feature of the reformulation of the Lagrangian density is the non-inclusion of the gluon field quanta explicitly in the trial state (though they are included implicitly through the propagators). It is then prudent to wonder whether the inclusion of the gluon field explicitly is required at all to describe bound states of QCD. Certainly, the non-linear gluon interaction terms in QCD are necessary to explain confinement. However, is it possible to devise new non-linear Lagrangian densities without the gluon field, based on new symmetries and principles, which would copy their effects and describe the phenomenon of quark confinement? 

One possible direction of research to achieve this might have to do with the gauge indices of QCD. It is simple to notice that the terms of the QCD Lagrangian (\ref{EQ:LQCD1}) are completely factorisable into products of the colour index quadratic forms and the quark and gluon fields. Such factorization between the momentum and spin dependencies does not occur in the QED and QCD Lagrangian densities. One might then consider exploring the idea of uniting the colour, spin and momentum dependencies by modifying the Lagrangian density in unconventional ways. Of course, the guiding principle, in such an approach, should be that in the appropriate limit the standard QCD Lagrangian is retrieved.

In this dissertation, the results of the QCD calculations show evidence of linear confinement in quark-antiquark pairs and this is consistent with LQCD. Besides the potential itself, it is desirable to calculate the bound state energy of quark-antiquark pairs using the trial state (\ref{EQ:QCDTRIAL24}).  This would require the calculation of all remaining interaction kernels in equations (\ref{EQ:EQRELQCD2}) and (\ref{EQ:EQRELQCD4}) and an extensive numerical toil. Even more so, it is prudent to perform a relativistic calculation such that, more common, light quark-antiquark states could be examined. The research was conducted using several approximations and simplification which could be improved upon in the future, as discussed in this section. The author hopes that the work recounted in this dissertation can be useful to other researchers. 

\chapter{Appendix A: Higgs-like Scalar Model}
\section{Three-Particle State}
\label{SEC:scalar_three}
Some intermediate steps of the derivations with the three-particle trial state (\ref{EQ:TRIALTHREE}) are presented in this section. All matrix elements are calculated using Maple\cite{maple}. 
\subsection*{Derivation of the kernels}
The matrix element in the Schr\"odinger picture for the three identical particle trial state (\ref{EQ:TRIALTHREE}) is comprised of
\begin{equation}
  \bra \Psi_3 | \, \hat{H} - E \, | \Psi_3 \ket =   \bra \Psi_3 | \, \hH_{\phi} - E \, | \Psi_3 \ket  + \bra \Psi_3 | \, \hH_{I_1} \, | \Psi_3 \ket 
  + \bra \Psi_3 | \, \hH_{I_2}  \, | \Psi_3 \ket 
\end{equation}
where the contributions are
\begin{align}
  \bra \Psi_3 | \, \hat{H}_{\phi} -E \, | \Psi_3 \ket = & \int d\p_{1,2,3} \big(\omega_{\p_1} + \omega_{\p_2} + \omega_{\p_3} - E \big) \; F^{\ast}(\p^{\prime}_{1,2,3}) \, F_S(\p_{1,2,3}), \\
  \bra \Psi_3 | \, \hat{H}_{I_{1}} \, | \Psi_3 \ket =  & -\frac{g^2}{8\,(2\pi)^{3}} \int \frac{d\p^{\prime}_{1,2,3} \, d\p_{1,2,3}}{\ds\sqrt{\omega_{\p^{\prime}_{1}}\omega_{\p^{\prime}_{2}}\omega_{\p_1}\omega_{\p_2}}} \; F^{\ast}_S(\p^{\prime}_{1,2,3}) \, F_S(\p_{1,2,3}) \nonumber \\
  & \times \delta(\p^{\prime}_1 + \p^{\prime}_2 - \p_1 - \p_2) \, \delta(\p^{\prime}_3 - \p_3) \, \left[\frac{1}{\mu^2-(p^{\prime}_1 - p_1)^2}\right], \\
  \bra \Psi_3 | \, \hat{H}_{I_{2}} \, | \Psi_3 \ket = & - \frac{g^{3}\eta}{8(2\pi)^{6}} \int \frac{d\p^{\prime}_{1,2,3} \, d\p_{1,2,3}}{\ds\sqrt{\omega_{\p^{\prime}_{1}}\omega_{\p^{\prime}_{2}}\omega_{\p^{\prime}_{3}}\omega_{\p_1}\omega_{\p_2}\omega_{\p_3}}} \; F^{\ast}_S(\p^{\prime}_{1,2,3}) \, F_S(\p_{1,2,3}) \nonumber \\
  & \times \delta(\p^{\prime}_1 + \p^{\prime}_2 +\p^{\prime}_3 - \p_1 - \p_2 - \p_3) \nonumber \\
  & \times \left[\frac{1}{\mu^2-(p^{\prime}_1 + p^{\prime}_2 - p_1 - p_2)^2}\frac{1}{\mu^2-(p^{\prime}_1 - p_1)^2}\frac{1}{\mu^2-(p^{\prime}_2 - p_2)^2}\right]. 
\end{align}
In working out the variational derivative the following identity is used:
\begin{equation}
  \frac{\delta F_{S}(\p_{1,2,3})}{\delta F(\q_{1,2,3})} = \sum_{i_1, i_2, i_3}^6 \delta(\p_{i_1} - \q_1) \, \delta(\p_{i_2} - \q_2) \, \delta(\p_{i_3} - \q_3),
  \label{EQ:VARIATIONAL1}
\end{equation}
where the summation is on the six permutation of the indices $i_1, i_2$ and $i_3$. 

The relativistic kernels for the three-particle trial state are given in equations (\ref{EQ:Y_33}) and (\ref{EQ:C_33}) in the body of the dissertation. In the non-relativistic limit the kernels become
\begin{align}
  Y_{3,3} & \, (\p^{\prime}_{1,2,3}, \p_{1,2,3}) = -\frac{g^2}{8\,(2\pi)^{3}m^2} \; \left[\frac{\delta(\p^{\prime}_1 + \p^{\prime}_2 - \p_1 - \p_2) \, \delta(\p^{\prime}_3 - \p_3)}{\mu^2+(\p^{\prime}_1 - \p_1)^2}\right. \nn \\
    & + \left. \frac{\delta(\p^{\prime}_1 + \p^{\prime}_3 - \p_1 - \p_3) \, \delta(\p^{\prime}_2 - \p_2)}{\mu^2+(\p^{\prime}_3 - \p_3)^2} +  \frac{\delta(\p^{\prime}_2 + \p^{\prime}_3 - \p_2 - \p_3) \, \delta(\p^{\prime}_1 - \p_1)}{\mu^2+(\p^{\prime}_2 - \p_2)^2}\right] \\
  C_{3,3} & \,(\p^{\prime}_{1,2,3}, \p_{1,2,3}) = - \frac{3 \, g^3 \, \eta}{4(2\pi)^6m^3} \nn \\
  & \times \frac{\delta(\p^{\prime}_1 + \p^{\prime}_2 +\p^{\prime}_3 - \p_1 - \p_2 - \p_3)}{(\mu^2+(\p^{\prime}_1 - \p_1)^2)(\mu^2+(\p^{\prime}_2 - \p_2)^2)(\mu^2+(\p^{\prime}_3 - \p_3)^2)}
\end{align}
where the symmetry property of $F_S$ is used to simplify the expression. 

To obtain equation (\ref{EQ:V_C123}), the kernel $\cC_{3,3}$ is multiplied by $\ds\prod_i^3\, \e^{i \, \p_i\cdot\x_i}$ and the variable of integration is shifted by $\p^{\prime}_i - \p_i = \q_i$. Integrating over the vectors $\q_i$ leads to equation (\ref{EQ:V_C123A}). This integral is found using the standard technique where the radial integral is evaluated in the complex plane using Cauchy's integration formula. 
%

\section*{Gaussian parametrization}
An alternative method for evaluating the cubic contribution $V_C$ for the three identical particle system is presented in this section. The invariance of $V_C$ under rotations and translations of the coordinates becomes explicit in this method. 

The denominators of equation (\ref{EQ:V_C123}) can be written in a way which enables one to perform Gaussian integration over the momentum variables. The following identity and the Gaussian integration formula are used:
\begin{align}
  \int_{0}^{\infty}d\beta \, \e^{-A\beta} = \, & \frac{1}{A} \H! \tx{for} \H! A > 0, 
  \label {EQ:WILSON} \\
  \int_{-\infty}^{\infty}\prod_{i} \, dx_{i} \, \exp\left[-\frac{1}{2}K_{ij}x_{i}x_{j}-L_{j}x_{j}-W\right] = \, & \textrm{Det} \, \left[\frac{K}{2\pi}\right]^{-\frac{1}{2}}\exp\left[\frac{1}{2}K^{-1}_{ij}L_{i}L_{j}-W\right],
  \label{EQ:GAUSS}
\end{align}
where $K$ is an invertible and symmetric matrix, $L$ is a vector and $W$ is a constant. 

The three-particle trial state (\ref{EQ:TRIALTHREE}) yields the cubic interaction kernel (\ref{EQ:V_C123}). After integrating over the delta function in (\ref{EQ:V_C123}) there remain only two momentum integrations. Using the identity of equation (\ref{EQ:WILSON}) on each factor separately, the the following expression is obtained:
\begin{multline}
  V_C(\x_{ij}, \mu > 0) =  - \alpha_{\eta} \, \pi^3 \int_{0}^{\infty} d\beta_{1,2,3} \int d\q_{1,2} \\ 
  \e^{ -(\mu^2 + \q^2_1) \, \beta_1 - (\mu^2 + \q^2_2) \, \beta_2 - (\mu^2 + (\q_1 +\q_2)^2) \, \beta_3} \e^{- i\q_1\cdot\x_{21}} \, \e^{- i\q_2\cdot\x_{31}},
  \label{EQ:V_C123M2A}
\end{multline}
where $\x_{ij} = \x_i - \x_j$ are the inter-particle vectors. Upon expanding the squares in the exponentials and defining $\q=\q_1+\q_1$, a six dimensional Gaussian integral is identified. The matrix $K$ and its inverse $K^{-1}$, the vectors $L_{i}$ and $W$ for the Gaussian integral are given in block diagonal form
\begin{gather}
  K = \left[
    \begin{array}{cc}
      2(\beta_{1}+\beta_{3}) & 2\beta_{3}  \\
      2\beta_{3} & 2(\beta_{2}+\beta_{3})  \\ \end{array}
    \right], 
  K^{-1} = \frac{1}{4 \, \beta_{123}}\left[
    \begin{array}{cc}
      2(\beta_{2}+\beta_{3}) & -2\beta_{3}  \\
      -2\beta_{3} & 2(\beta_{1}+\beta_{3})  \\ \end{array}
    \right],  
  \label{EQ:MK3} \\
  L= i \, \left[
    \begin{array}{cc}
      \x_{21}, & \x_{31} \end{array}
    \right],\F!
  W = \left(\beta_{1}+\beta_{2}+\beta_{3}\right)\mu^{2}.
  \label{EQ:VLM}
\end{gather}
where $\beta_{123} = \beta_{1}\beta_{2}+\beta_{1}\beta_{3}+\beta_{2}\beta_{3}$. Applying equation (\ref{EQ:GAUSS}), one arrives at
\begin{multline}
  V_C(\x_{ij}, \mu > 0) = - \alpha_{\eta} \, \pi^3 \int_{0}^{\infty}d\beta_{1,2,3} \; \frac{\e^{-\mu^{2}(\beta_{1}+\beta_{2}+\beta_{3})}}{\beta_{123}^{3/2}} \exp\ds\left(-\frac{\beta_{1}\x_{21}^{2}+\beta_{2}\x_{31}^{2}+\beta_{3}\x_{32}^{2}}{4 \, \beta_{123}}\right). 
\end{multline}
This is the expression given by equation (\ref{EQ:V_C123C}). It shows explicitly that the cubic potential term $V_C$ depends only on the inter-particle distances. Numerical integration over the parameters $\beta_i$ is required to complete the calculation.
%
\subsection*{Derivation of the cubic potential energy $V_C$ for $\x_2 = \x_3$}
The cubic inter-particle potential for the three-particle trial state (\ref{EQ:TRIALTHREE}) in the case when $\mu > 0$ and $\x_2 = \x_3$ is determined from the following expression:  
\begin{equation}
  V_C(\x_{1,2,2}, \mu) = V_C(x_{21}, \mu)= - \alpha_{\eta}\pi^3 \, \int d\v \, \frac{\e^{-\mu v}}{v} \, \frac{\e^{-2\mu|\v+\x_{21}|}}{|\v+\x_{21}|^2},
  \label{EQ:STEP1}
\end{equation}
where $\x_{21} = \x_2 - \x_1$ and the non-vector notation means the magnitude of the vector. After the trivial azimuthal integration one obtains: 
\begin{equation}
  V_C(x_{21}, \mu) = - 2 \, \pi^4\alpha_{\eta} \, \int_0^\infty dv \, v^2 \int_{-1}^{+1} \, dw \; \frac{\e^{-2 \, \mu v}}{v^2}\frac{\e^{-\mu|\v+\x_{21}|}}{|\v+x_{21}|}, 
  \label{EQ:STEP1A}
\end{equation}
where $w$ = $\cos\theta$ with $\theta$ being the polar angle of the vector $\v$. The polar integration is performed using the substitution $\varrho^2=x_{21}^2+v^2+2 \,x_{21} \, v \, w$, with $\varrho_{1}=(x_{21}+v)$ and $\varrho_{2}=|x_{21}-v|$ as the upper and the lower limits of integration. Thereupon, the integral of equation (\ref{EQ:STEP1A}) can be written as
\begin{equation}
  V_C(x_{21}, \mu) = -\frac{2 \, \pi^4\alpha_{\eta}}{x_{21}} \, \int_0^{\infty} \frac{dv}{v} \, \e^{-2\mu v} \int_{\vr_2}^{\vr_1} d\vr \, \e^{-\mu \vr}.
  \label{EQ:STEP2} 
\end{equation}
To integrate over the variable $\vr$, one splits the interval of integration accordingly and ends up with the following radial integral: 
\begin{multline}
  V_C(x_{21}, \mu) = - \frac{2 \, \pi^4\alpha_{\eta}}{\mu \, x_{21}} \\
  \times \Bigg\{\e^{-\mu x_{21}} \int_0^{x_{21}} \frac{dv}{v} \, \left(\e^{-\mu v}- \e^{-3 \mu v}\right) + \left(\e^{\mu x_{21}} - \e^{-\mu x_{21}} \right) \int_{x_{21}}^{\infty} \frac{dv}{v} \, \e^{-3\mu v} \Bigg\}.
\end{multline}
This integral is expressible in terms of the exponential integral defined by equation (\ref{EQ:EXPINT}). Thus, the cubic potential term for the three-particle trial state in the case when $\mu > 0$ and $\x_2 = \x_3$ evaluates to
\begin{equation}
  V_C(x_{21}, \mu) = -\frac{2 \, \pi^4\alpha_{\eta}}{\mu \, x_{21} } \Bigg\{\e^{-x_{21} \mu}\ds\left[\ln\left(3\right) - \E1\left(x_{21} \mu\right)\right] + \e^{x_{21} \mu} \, \E1\left(3x_{21} \mu\right) \Bigg\},
\end{equation}
which is equation (\ref{EQ:V_C122MU}).

\subsection*{Derivation of the cubic potential $V_C$ for the equidistant case with $\mu =0$}
It is instructive to consider the inter-particle potential for the three-particle system described by the trial state (\ref{EQ:TRIALTHREE}) in the case when $\mu = 0$ and the coordinates are at the vertices of an equilateral triangle.  Even though the $\mu =0$ limit leads to an unphysical result, the technique of integrating out the angular coordinates is interesting in itself. 

The cubic contribution $V_C$ to the potential is determined from the following expression:
\begin{equation}
  V_C(\x_{1,2,3}, \mu = 0) = - \alpha_{\eta} \, \pi^3 \int \, d\v \, \frac{1}{v} \, \frac{1}{|\v + \x_{21}|}\frac{1}{|\v + \x_{31}|}.
\end{equation}
A cut off on the upper limit in the radial integral is inserted in anticipation of an infinity. This yields
\begin{equation}
  V_C(\x_{1,2,3}, \mu = 0) = -\lim_{R\rightarrow \infty} \, \alpha_{\eta}\, \pi^3 \int^R_0 v \,dv \, \int d\Omega \, \frac{1}{|\x_{21}+\v|}\frac{1}{|\x_{31}+\v|}  
\end{equation}
where $d\v = v^2\, dv \, d\Omega$ is the volume element in spherical coordinates. The angular integration is performed using a number of steps. First, the $|\r_1-\r_2|^{-1}$ expansion identity is applied on the remaining denominators 
\begin{equation}
  \frac{1}{|\r_1-\r_2|} = \sum_{l=0}^{\infty} \, \frac{r^l_<}{r^{l+1}_{>}} \, P_{l}(\alpha) \\
\end{equation}
where $r_{<}$($r_{>}$) is the smaller(larger) of the lengths of $\r_1$ and $\r_2$, and $P_{l}$ are the Legendre polynomials. This leads to
\begin{equation}
  V_C(\x_{1,2,3}, \mu = 0) = - \lim_{R\rightarrow \infty} \, \alpha_{\eta} \, \pi^3 \sum^{\infty}_{l_1=0} \, \sum^{\infty}_{l_2=0} \; \int_0^R dv \, v \int d\Omega  \; \frac{r_<^{l_1}}{r_>^{l_1+1}} \, \frac{s_<^{l_2}}{s_>^{l_2+1}} \; P_{l_1}(\alpha_2) \, P_{l_2}(\alpha_3),
  \label{EQ:V_TA}
\end{equation}
where $r_{</>}$ is the smaller/larger of $x_{21}$ and $v$, $s_{</>}$ is the smaller/larger of $x_{31}$ and $v$, $\alpha_{2}$ is the angle between the vectors $\x_{21}$ and $\v$, and $\alpha_3$ is the angle between the vectors $\x_{31}$ and $\v$.

Next, the Legendre polynomials are rewritten in terms of the spherical harmonics and integrated using the orthogonality relation of the spherical harmonics. The relevant expressions are
\begin{equation}
  P_{l}({\alpha}) = \frac{4 \, \pi}{2l+1} \, \sum_{m=-l}^{l} \; Y^{m\,\ast}_{l}(\theta_{1}, \phi_{1}) \, Y^{m}_{l}(\theta_{2}, \phi_{2}),
\end{equation}
where $(\theta_1, \phi_1)$ and $(\theta_2, \phi_2)$ specify the directions of two unit vectors which are separated by the cosine of the angle $\alpha$, and
\begin{equation}
  \delta_{m m^{\prime}} \, \delta_{l l^{\prime}} = \int d\Omega \; Y^{m\ast}_{l}(\theta, \phi) \, Y^{m^{\prime}}_{l^{\prime}}(\theta, \phi).     
\end{equation}
Using these two expressions in equation (\ref{EQ:V_TA}) enables one to perform the angular integration. Then, resumming over the index $m$ leads to the Legendre polynomials with the argument being $\hat{\x}_{21}\cdot\hat{\x}_{31}$ where $\hat{\x}_{ij} = \x_{ij} / |\x_{ij}|$. The resulting expression is
\begin{equation}
  V_C(\x_{1,2,3}, \mu = 0) = - \lim_{R\rightarrow \infty} \, \alpha_{\eta} \, \pi^3 \, \sum^{\infty}_{l=0} \frac{4\pi}{2l+1} \; P_{l}(\hat\x_{21}\cdot\hat\x_{31}) \, \int_0^R dv \, v \; \frac{r_<^l}{r_>^{l+1}} \, \frac{s_<^l}{s_>^{l+1}}.
\end{equation}
One can set $\x_{21} \leq \x_{31}$ without loss of generality and break the intervals of integration into three segments. The subsequent expression holds for arbitrary coordinates and $\mu = 0$:
\begin{multline}
  V_C(\x_{1,2,3}, \mu = 0) = -  \lim_{R\rightarrow \infty} \, 4 \,\pi^4 \, \alpha_{\eta} \, \sum_{l=0}^{\infty}  \frac{P_{l}(\hat{\x_{21}}\cdot\hat{\x_{31}})}{2l+1} \, \Bigg\{\int^{x_{21}}_0 v \,dv \; \frac{v^{2l}}{x_{21}^{l+1}x_{31}^{l+1}} \\
  + \int_{x_{21}}^{x_{31}} v \, dv \; \frac{x_{21}^l}{v^{l+1}} \, \frac{v^l}{x^{l+1}_{31}} + \int_{x_{31}}^R v \, dv \; \frac{x_{21}^l x_{31}^l}{v^{2l+2}} \Bigg\}.
\end{multline}
The free variables in this expression are the coordinate separations $x_{21}$, $x_{31}$ and the quantity $\hat{\x}_{21}\cdot\hat{\x}_{31}$. This last quantity can be written in terms of $x_{21}$, $x_{31}$ and $x_{32}$, thus showing that the inter-particle potential depends on the three distances $x_{ij}$ only. Specializing to the case where the coordinates are at the vertices of an equilateral triangle leads to a simpler expression to evaluate:
\begin{align}
    V_C(x_{21} = x_{31} = x_{23}, \mu = 0) = & V_C(\Delta, \mu = 0) = \nn \\
    & - 4 \, \pi^4 \, \alpha_{\eta} \, \sum^{\infty}_{l = 0} \, \frac{P_l(0.5)}{2l+1} \, \Bigg\{\int_0^\Delta dv \frac{v^{2l+1}}{\Delta^{2l+2}} + \int_\Delta^R dv \frac{\Delta^{2l}}{v^{2l+1}}\Bigg\},
    \label{EQ:V_89}
\end{align}
where $\Delta = |\x_2 - \x_1| = |\x_3 - \x_1| = |\x_3 - \x_2|$ and $\hat{\x}_{21}\cdot\hat{\x}_{31} = 0.5$ for the present equidistant case. The integrals in equation (\ref{EQ:V_89}) are elementary. The $l=0$ term in the summation has to be dealt with separately so that
\begin{equation}
  V_C(\Delta, \mu = 0) = - 4 \, \pi^4 \, \alpha_{\eta}\left[\frac{1}{2} + \ln\left(\frac{R}{\Delta}\right)\right] - 4 \, \pi^4 \, \alpha_{\eta} \, \sum^{\infty}_{l = 1} \, \frac{P_l(0.5)}{2} \left(\frac{1}{l} - \frac{1}{l+1}\right).
\end{equation}
The summation is expressible in analytical form using identities involving the Legendre polynomials~\cite{Gradsh1980}. The expression can be simplified to
\begin{equation}
  V_C(\Delta, \mu =0) =  -4 \, \pi^4 \, \alpha_{\eta} \left[1+ \ln\left(\frac{2}{3}\right) - \ln\left(\frac{\Delta}{R}\right)\right].
\end{equation}
%

\section{Four-Particle State}
\label{SEC:scalar_four}
Some intermediate steps of the derivations with the four-particle trial state (\ref{EQ:TRIALFOUR}) are presented in this section. All matrix elements were calculated using Maple. 
\subsection*{Derivation of the kernels}
The matrix element in the Schr\"odinger picture for the four identical particle trial state (\ref{EQ:TRIALFOUR}) is comprised of
\begin{equation}
  \bra \Psi_4 | \, \hat{H} - E \, | \Psi_4 \ket =   \bra \Psi_4 | \, \hH_{\phi} - E \, | \Psi_4 \ket + 
  \bra \Psi_4 | \, \hH_{I_1} \, | \Psi_4 \ket + \bra \Psi_4 | \, \hH_{I_2} \, | \Psi_4 \ket + \bra \Psi_4 | \, \hH_{I_3} \, | \Psi_4 \ket
\end{equation}
where the contributions are
\begin{align}
  \bra\Psi_4| \, \hat{H}_{\phi} - E \, |\Psi_4 \ket  = &  \int d\p_{1..4} \; F^{\ast}(\p_{1..4}) \, F_{S}(\p_{1..4}) \; \ds\left(\omega_{\p_1} + \omega_{\p_2} + \omega_{\p_3} + \omega_{\p_4} - E \right), \nn \\
  \bra \Psi_4| \, \hat{H}_{I_1} \, | \Psi_4 \ket = & - \frac{g^2}{16(2\pi)^3} \; \int \frac{d\p^{\prime}_{1..4} \, d\p_{1..4}}{\ds\sqrt{\omega_{\p^{\prime}_1}\omega_{\p^{\prime}_2}\omega_{\p_1}\omega_{\p_2}}} \, F^{\ast}_{S}(\p^{\prime}_{1..4}) \, F_{S}(\p_{1..4}) \nn \\
  & \times \, \de(\p^{\prime}_1 + \p^{\prime}_2 - \p_1 - \p_2)\, \de(\p^{\prime}_3 - \p_3) \, \de(\p^{\prime}_4 - \p_4) \, \left[\frac{1}{\mu^2-(p^{\prime}_{1}-p_{1})^2}\right], \\
  \bra\Psi_4 | \,\hat{H}_{I_2} \, | \Psi_4 \ket = & - \frac{g^3\eta}{8(2\pi)^6} \, \int  \frac{d\p^{\prime}_{1..4} \, d\p_{1..4}}{\ds\sqrt{\omega_{\p^{\prime}_1}\omega_{\p^{\prime}_{2}}\omega_{\p^{\prime}_3}\omega_{\p_1}\omega_{\p_2}\omega_{\p_3}}} \, F^{\ast}_S(\p^{\prime}_{1..4}) \, F_S(\p_{1..4})   \nonumber \\
  & \times \, \de(\p^{\prime}_1 + \p^{\prime}_2 + \p^{\prime}_3 - \p_1 - \p_2 - \p_3)\,  \de(\p^{\prime}_4 - \p_4) \nonumber \\
  & \times \, \left[\frac{1}{\mu^2-(p^{\prime}_1 + p^{\prime}_2 - p_1 - p_2)^2}\frac{1}{\mu^2-(p^{\prime}_1 - p_1)^2}\frac{1}{\mu^2-(p^{\prime}_2 - p_2)^2}\right],
\end{align}
and
\begin{align}
  \bra \Psi_4 | \, \hat{H}_{I_{3}} \, | \Psi_4 \ket = \, & \frac{g^4\sigma}{16(2\pi)^9} \; \int \frac{d\p^{\prime}_{1..4} \, d\p_{1..4}}{\ds\sqrt{\omega_{\p^{\prime}_1}\omega_{\p^{\prime}_2}\omega_{\p^{\prime}_3}\omega_{\p^{\prime}_4}\omega_{\p_1}\omega_{\p_2}\omega_{\p_3}\omega_{\p_4}}} \, F^{\ast}_{S}(\p^{\prime}_{1..4}) \, F_{S}(\p_{1..4}) \nn \\
  & \times \, \de(\p^{\prime}_1 + \p^{\prime}_2 + \p^{\prime}_3 + \p^{\prime}_4 - \p_1 - \p_2 - \p_3 - \p_4) \nn \\ 
  & \times \, \left[\frac{1}{\mu^2-(p^{\prime}_1 + p^{\prime}_2 + p^{\prime}_3 - p_1 - p_2 - p_3)^2}\right]  \nn \\
  & \times \, \left[\frac{1}{\mu^2-(p^{\prime}_1 - p_1)}\frac{1}{\mu^2-(p^{\prime}_2 - p_2)^2}\frac{1}{\mu^2-(p^{\prime}_3 - p_3)^2}\right].
\end{align}
In working out the variational derivative the following identity is used:
\begin{equation}
  \frac{\delta F_{S}(\p_{1..4})}{\delta F(\q_{1..4})} = \sum_{i_1, i_2, i_3, i_4}^{24} \de(\p_{i_1} - \q_1) \, \delta(\p_{i_2} - \q_2) \, \delta(\p_{i_3} - \q_3) \,  \delta(\p_{i_4} - \q_4)
  \label{EQ:VARIATIONAL2}
\end{equation}
where the summation is on the 24 permutation of the indices $i_1, i_2, i_3$ and $i_4$. 

The relativistic Yukawa and cubic interaction kernels for the the four-particle trial state (\ref{EQ:TRIALFOUR}) are
\begin{align}
  \cY_{4,4} & \, (\p^{\prime}_{1..4}, \p_{1..4}) = -\frac{g^2}{16(2\pi)^3} \nn \\
  & \H! \times \sum_{i_1, i_2, i_3, i_4}^{24} \frac{\de(\p^{\prime}_{1}+\p^{\prime}_{2}-\p_{i_1}-\p_{i_2})\, \de(\p^{\prime}_{3}-\p_{i_3}) \, \de(\p^{\prime}_{4}-\p_{i_4})}{\ds\sqrt{\omega_{\p^{\prime}_1} \omega_{\p^{\prime}_{2}} \omega_{\p_{i_1}}\omega_{\p_{i_2}}}} \left[\frac{1}{\mu^2-(p^{\prime}_{1}-p_{i_1})^2}\right], \\
  \cC_{4,4} & \, (\p^{\prime}_{1..4}, \p_{1..4}) = - \frac{g^3\eta}{8(2\pi)^6} \, \sum^{24}_{i_1, i_2, i_3,i_4} \frac{ \de(\p^{\prime}_1+\p^{\prime}_2+\p^{\prime}_3-\p_{i_1}-\p_{i_2}-\p_{i_3}) \,\de(\p^{\prime}_4-\p_{i_4})}{\ds\sqrt{\omega_{\p^{\prime}_1}\omega_{\p^{\prime}_2}\omega_{\p^{\prime}_3}\omega_{\p_{i_1}}\omega_{\p_{i_2}}\omega_{\p_{i_3}}}} \nn \\
  & \H! \times \, \left[\frac{1}{\mu^2-(p^{\prime}_1+p^{\prime}_2-p_{i_1}-p_{i_2})^2}\frac{1}{\mu^2-(p^{\prime}_1-p_{i_1})^2}\frac{1}{\mu^2-(p^{\prime}_2-p_{i_2})^2}\right].
\end{align}
The relativistic quartic kernel for the four-particle trial state is given in equation (\ref{EQ:Q44}).
In the non-relativistic limit the interaction kernels for the four-particle trial state reduce to
\begin{align}
  Y & (\p^{\prime}_{1..4}, \p_{1..4}) = \frac{g^2}{4(2\pi)^3m^2} \ds\left[\frac{\de(\p^{\prime}_1 + \p^{\prime}_2 - \p_1 - \p_2) \, \de(\p^{\prime}_3 - \p_3) \, \de(\p^{\prime}_4 - \p_4)}{\mu^2+(\p^{\prime}_1-\p_1)^2} \right. \nn \\
    & + \left. \frac{\de(\p^{\prime}_1 + \p^{\prime}_3 - \p_1 - \p_3) \, \de(\p^{\prime}_2 - \p_2) \, \de(\p^{\prime}_4 - \p_4)}{\mu^2+(\p^{\prime}_1-\p_1)^2} + \frac{\de(\p^{\prime}_1 + \p^{\prime}_4 - \p_1 -\p_4) \, \de(\p^{\prime}_2 - \p_2) \, \de(\p^{\prime}_3 - \p_3)}{\mu^2+(\p^{\prime}_1 - \p_1)^2} \right. \nn \\
    & + \left. \frac{\de(\p^{\prime}_2 + \p^{\prime}_3 - \p_2 - \p_3) \, \de(\p^{\prime}_1 - \p_1) \, \de(\p^{\prime}_4 - \p_4)}{\mu^2+(\p^{\prime}_2 - \p_2)^2} + \frac{\de(\p^{\prime}_2 + \p^{\prime}_4 - \p_2 - \p_4) \, \de(\p^{\prime}_1 - \p_1) \, \de(\p^{\prime}_3 - \p_3)}{\mu^2+(\p^{\prime}_2-\p_2)^2} \right. \nn \\
    & + \left. \frac{\de(\p^{\prime}_3 + \p^{\prime}_4 - \p_3 - \p_4) \, \de(\p^{\prime}_1 - \p_1) \, \de(\p^{\prime}_2 - \p_2)}{\mu^2 + (\p^{\prime}_3- \p_3)^2}\right], \\
  C & (\p^{\prime}_{1..4}, \p_{1..4}) = \frac{3 \, g^3 \, \eta}{4 (2\pi)^6 m^3} \ds \left[\frac{\delta(\p^{\prime}_1 + \p^{\prime}_2 + \p^{\prime}_3 - \p_1 - \p_2 - \p_3) \, \de(\p^{\prime}_4 - \p_4)}{(\mu^2+(\p^{\prime}_1-\p_1)^2)(\mu^2+(\p^{\prime}_2 - \p_2)^2)(\mu^2+(\p^{\prime}_3 - \p_3)^2)}\right. \nn \\
    & \F!\F!\F! + \frac{\de(\p^{\prime}_1 + \p^{\prime}_2 + \p^{\prime}_4 - \p_1 - \p_2 - \p_4) \, \de(\p^{\prime}_3 - \p_3)}{(\mu^2+(\p^{\prime}_1 - \p_1)^2)(\mu^2+(\p^{\prime}_2 - \p_2)^2)(\mu^2+(\p^{\prime}_4 - \p_4)^2)} \nn \\
    & \F!\F!\F! + \frac{\de(\p^{\prime}_1 + \p^{\prime}_3 + \p^{\prime}_4 - \p_1 - \p_3 - \p_4) \, \de(\p^{\prime}_2 - \p_2)}{(\mu^2+(\p^{\prime}_1 - \p_1)^2)(\mu^2+(\p^{\prime}_3 - \p_3)^2)(\mu^2 + (\p^{\prime}_4 - \p_4)^2)} \nn \\
    & \F!\F!\F! + \left.\frac{\de(\p^{\prime}_2 + \p^{\prime}_3 + \p^{\prime}_4 - \p_2 - \p_3 - \p_4) \, \de(\p^{\prime}_1 - \p_1)}{(\mu^2+(\p^{\prime}_2 - \p_2)^2)(\mu^2+(\p^{\prime}_3 -\p_3)^2)(\mu^2+(\p^{\prime}_4 - \p_4)^2)}\right], \\
  Q & (\p^{\prime}_{1..4}, \p_{1..4})  = - \frac{3 \, g^4 \, \sigma}{2(2\pi)^9m^4} \nn \\
  & \F! \times \left[\frac{\de(\p^{\prime}_1 + \p^{\prime}_2 + \p^{\prime}_3 + \p^{\prime}_4 - \p_1 - \p_2 - \p_3 - \p_4)}{(\mu^2+(\p^{\prime}_1-\p_1)^2)(\mu^2+(\p^{\prime}_2-\p_2)^2)(\mu^2+(\p^{\prime}_3-\p_3)^2)(\mu^2+(\p^{\prime}_4-\p_4)^2)}\right].
\end{align}
%
%
\section*{Gaussian parametrization}
An alternative method for evaluating the cubic and the quartic potential contributions $V_C$ and $V_Q$ for the four identical particle state is presented in this section. The invariance under rotations and translations of the coordinates becomes explicit in this method. 

The denominators of equations (\ref{EQ:V_C1234}) and (\ref{EQ:V_Q1234}) can be written using a technique which enables one to perform Gaussian integration over the momentum variables. The relevant identity and the Gaussian integration formula are given in equations (\ref{EQ:WILSON}) and (\ref{EQ:GAUSS}).

The four-particle trial state (\ref{EQ:TRIALFOUR}) yields the cubic interaction kernel (\ref{EQ:V_C1234}). The calculation of $V_C$ for the four-particle trial state is identical to that of the three-particle trial state. There are basically four copies of the three-particle results with different inter-particle distances involved. The matrix $K$ and its inverse $K^{-1}$ are identical for all four terms and the same as for the three-particle trial state equation (\ref{EQ:MK3}). The vectors $L_i$ pertaining to each term in equation (\ref{EQ:V_C1234}) are
\begin{align}
  L_1 = & \, i \, \left[
    \begin{array}{cc}
      \x_{21}, & \x_{31} \end{array}
    \right],\,
  L_2 = i \, \left[
    \begin{array}{cc}
      \x_{21}, & \x_{41} \end{array}
    \right],\nn \\
  L_3 = & \, i \, \left[
    \begin{array}{cc}
      \x_{31}, & \x_{41} \end{array}
    \right],\,
  L_4 = i \, \left[
    \begin{array}{cc}
      \x_{32}, & \x_{42} \end{array}
    \right]
\end{align}
where the subscript indicates the corresponding term in equation (\ref{EQ:V_C1234}). The constant $W$ is the same as in equation (\ref{EQ:VLM}). Application of the Gaussian integration formula (\ref{EQ:GAUSS}) leads to the result
\begin{multline}
  V_C(\x_{ij}\, \mu > 0) = -\alpha_{\eta} \, \pi^3 \, \int^{\infty}_0 d\b_{1,2,3} \, \frac{\e^{-\mu^2(\b_1+\b_2+\b_3)}}{(\b_1\b_2+\b_1\b_3+\b_2\b_3)^{3/2}} \\
  \times \, \Bigg\{\exp\left(-\frac{\b_1\x^2_{12}+\b_2\x^2_{13}+\b_3\x^2_{23}}{4(\b_1\b_2+\b_1\b_3+\b_2\b_3)}\right) + \exp\left(-\frac{\b_1\x^2_{12}+\b_2\x^2_{14}+\b_3\x^2_{24}}{4(\b_1\b_2+\b_1\b_3+\b_2\b_3)}\right) \\
  \F!\; \exp\left(-\frac{\b_1\x^2_{13}+\b_2\x^2_{14}+\b_3\x^2_{34}}{4(\b_1\b_2+\b_1\b_3+\b_2\b_3)}\right) + \exp\left(-\frac{\b_1\x^2_{23}+\b_2\x^2_{24}+\b_3\x^2_{34}}{4(\b_1\b_2+\b_1\b_3+\b_2\b_3)}\right)\Bigg\}.
  \label{EQ:V_C1234C}
\end{multline}
From equation (\ref{EQ:V_C1234C}) it is evident that $V_C$ depends on the inter-particle distances $x_{ij} = |\x_i - \x_j|$ only.

The four-particle trial state yields the quartic interaction kernel equation (\ref{EQ:V_Q1234}). The calculation follows the same steps as for the cubic interaction kernel. Applying the identity of equation (\ref{EQ:WILSON}) leads to a 9 dimensional Gaussian integral. The matrix $K$ and its inverse $K^{-1}$ in block diagonal form are
\begin{align}
  K = & 2 \,\left[
    \begin{array}{ccc}
      (\b_1+\b_4) & \b_4         & \b_4   \\
      \b_4        & (\b_2+\b_4)  & \b_4   \\ 
      \b_4        & \b_4         & (\b_3+\b_4) 
    \end{array}
    \right], \\
  K^{-1}= & \frac{1}{2\b_{1234}}\left[
    \begin{array}{ccc}
      \b_{234}     & -\b_{3}\b_{4} & -\b_{2}\b_{4}  \\
      -\b_3\b_4    & \b_{134}      & -\b_1\b_4      \\
      -\b_2\b_4    & -\b_1\b_4     & \b_{124}      
    \end{array}
    \right],
\end{align}
where $\b_{1234}=\b_1\b_2\b_3+\b_1\b_3\b_4+\b_1\b_2\b_4+\b_2\b_3\b_4$ and $\b_{ijk}=\b_{ij}+\b_{ik}+\b_{jk}$. The vector $L$ and the constant $W$ of the Gaussian integration are 
\begin{equation}
  L= i \, \left(\x_{41}, \, \x_{42}, \, \x_{43}\right), \F! W = \left(\b_1+\b_2+\b_3+\b_4\right)\mu^{2}.
\end{equation}
Applying the Gaussian integration formula (\ref{EQ:GAUSS}) and after some algebra one ends up with the expression
\begin{multline}
  V_Q(\x_{ij}, \mu > 0) = \alpha_{\sigma} \pi^{9/2} \, \int^{\infty}_{0} d\b_{1..4} \, \frac{\e^{-\mu^2(\b_1+\b_2+\b_3+\b_4)}}{(\b_{1234})^{3/2}} \\
  \times \, \exp\left(-\frac{\b_3\b_4\x^2_{12}+\b_2\b_4\x^2_{13}+\b_2\b_3\x^2_{14}+\b_1\b_4\x^2_{23}+\b_1\b_3\x^2_{24}+\b_1\b_2\x^2_{34}}{4\b_{1234}}\right).
  \label{EQ:V_Q1234C}
\end{multline}
This expression shows explicitly that the quartic potential term $V_Q$ depends only on the inter-particle distances. Numerical integrations over the parameters $\beta_i$ are required to complete the calculation of $V_C$ and $V_Q$.
%

\subsection*{Derivation of the cubic potential energy $V_Q$ for $\x_1 = \x_3$ and $\x_2 = \x_4$ with $\mu > 0$}
%
%
The quartic inter-particle potential term for the four-particle trial state (\ref{EQ:TRIALFOUR}) in the case when $\mu > 0$, $\x_1 = \x_3$ and $\x_2 = \x_4$ is determined from equation (\ref{EQ:V_Q1234B}), which upon imposing the restrictions is the following expression:
\begin{equation}
  V_Q(\x_{1,2,1,2}, \mu = 0) = V_Q(x_{21}, \mu = 0) = \alpha_{\sigma}\pi^4 \, \int d\v \; \frac{\e^{-2 \, \mu \, |\v|}}{|\v|^2}\frac{\e^{- 2 \, \mu \, |\v + \x_{21}|}}{|\v + \x_{21}|^2}.
\end{equation}
%
Following similar steps as those leading to equation (\ref{EQ:STEP2}), one obtains the result 
\begin{equation}
  V_Q(\x_{1,2,1,2}, \mu > 0) = V_Q(x_{21}, \mu > 0) = \frac{2 \, \pi^5 \, \sigma}{x_{21}} \int \frac{dv}{v} \, \e^{-2\mu v}\int^{\varrho_1}_{\varrho_2} \frac{d\varrho}{\varrho} \, \e^{- 2\mu \varrho},
\end{equation}
where $\varrho_1$, $\varrho_2$ and $x_{21}$ are as before. The result of the integration over the variable $\varrho$ can be expressed in terms of the exponential integral (\ref{EQ:EXPINT}) and is given in equation (\ref{EQ:V_Q1212MU}). The remaining integrals have to be integrated numerically. 
%

\section{Improved Particle-Antiparticle State}
\label{SEC:scalar_two_four}
Some intermediate steps of the derivations with the improved particle-antiparticle trial state (\ref{EQ:TRIALTWOFOUR}) are presented in this section. All matrix elements are calculated using a Maple worksheet.

\subsection*{Derivation of the kernels}
The matrix element in the Schr\"odinger picture for the improved particle-antiparticle trial state (\ref{EQ:TRIALTWOFOUR}) is comprised of 
\begin{multline}
  \bra \Psi_t | \, \hat{H} - E \, | \Psi_t \ket =  \; \bra \Psi_2 | \, \hH_{\phi} - E | \Psi_2 \ket + \bra \Psi_2 | \, \hH_{I_1} \, | \Psi_2 \ket \\
  + \bra \Psi_2 |, \hH_{I_1} \, | \Psi_4 \ket +  \bra \Psi_2 | \hH_{I_2} | \Psi_4 \ket 
  + \bra \Psi_4 | \, \hH_{I_1} \, |  \Psi_2 \ket +  \bra \Psi_4 | \, \hH_{I_2} \, |  \Psi_2 \ket \\ 
  + \bra \Psi_4 | \, \hH_{\phi} - E \, |  \Psi_4 \ket + \bra  \Psi_4 | \, \hH_{I_1} \, |  \Psi_4 \ket   + \bra \Psi_4 | \, \hH_{I_2} \, |  \Psi_4 \ket 
  + \bra \Psi_4 | \, \hH_{I_3} \, | \Psi_4 \ket,
\end{multline}
where the kinetic energy contributions are 
\begin{align}
  \bra\Psi_2| \, \hat{H}_{\phi} - E \, |\Psi_2\ket  = & \, |C_F|^2\int d\p_{1,2} \; F^{\ast}(\p_{1,2}) \, F(\p_{1,2})  \; \ds\left(\omega_{\p_1} + \omega_{\p_2}  - E\right), \\
  \bra\Psi_4| \, \hat{H}_{\phi} - E \, |\Psi_4\ket  =  & \, |C_G|^2 \int d\p_{1..4} \; G^{\ast}(\p_{1..4}) \, G(\p_{1..4}) \; \ds\left( \omega_{\p_1} + \omega_{\p_2} + \omega_{\p_3} + \omega_{\p_4} - E\right),
\end{align}
and, the relevant interaction contributions are
\begin{multline}
  \bra \Psi_2 | \, \hat{H}_{I_1} \, | \Psi_2 \ket = -\frac{|C_F|^2 \, g^{2}}{8\,(2\pi)^{3}}\int d\p^{\prime}_{1,2} \, d\p_{1,2} \, F^{\ast}(\p^{\prime}_{1,2}) \, F(\p_{1,2}) \, \frac{\delta(\p^{\prime}_1 + \p^{\prime}_2 - \p_1 - \p_2)}{\ds\sqrt{\omega_{\p^{\prime}_1}\ds\omega_{\p^{\prime}_2}\ds\omega_{\p_1}\omega_{\p_2}}} \\
   \times \, \left[\frac{1}{\mu^{2}-(p^{\prime}_1-p_1)^{2}}+\frac{1}{\mu^{2}-(p^{\prime}_2-p_2)^{2}}+\frac{1}{\mu^{2}-(p_1+p_2)^{2}}+\frac{1}{\mu^{2}-(p^{\prime}_1+p^{\prime}_2)^{2}}\right], 
\end{multline}
\begin{align}
  \bra \Psi_2 | \, \hH_{I_1} & \, | \Psi_4 \ket = - \frac{C_F^\ast C_G \, g^2}{4(2\pi)^3} \int d\p^{\prime}_{1,2} \, d\p_{1..4} \, F^{\ast}(\p^{\prime}_{1,2}) \, G(\p_{1..4}) \nn \\ 
  \times & \, \Bigg\{\frac{\de(\p^{\prime}_1-\p_1) \, \de(\p^{\prime}_2 -\p_2 - \p_3 -\p_4)}{\ds\sqrt{\omega_{\p^{\prime}_2}\omega_{\p_2}\omega_{\p_3}\omega_{\p_4}}} \, \left[\frac{1}{\mu^2-(p^{\prime}_2 - p_2)^2} + \frac{1}{\mu^2-(p_3 + p_4)^2} \right] \nn \\
  & \H! + \frac{\de(\p^{\prime}_2-\p_2) \, \de(\p^{\prime}_1 -\p_1 - \p_3 -\p_4)}{\ds\sqrt{\omega_{\p^{\prime}_1}\omega_{\p_1}\omega_{\p_3}\omega_{\p_4}}} \, \left[\frac{1}{\mu^2-(p^{\prime}_1 - p_1)^2} + \frac{1}{\mu^2-(p_3 + p_4)^2} \right] \Bigg\},
\end{align}
\begin{align}
  \bra \Psi_2 | \, \hH_{I_2} \, | \Psi_4 \ket = & -\frac{C_F^\ast C_G \, g^3\eta}{4(2\pi)^6} \int d\p^{\prime}_{1,2} d\p_{1..4} \; F^{\ast}(\p^{\prime}_{1,2}) \, G(\p_{1..4}) \; \frac{\de(\p^{\prime}_1 + \p^{\prime}_2 - \p_1 - \p_2 - \p_3 - \p_4)}{\ds\sqrt{\omega_{\p^{\prime}_1}\omega_{\p^{\prime}_2}\omega_{\p_1}\omega_{\p_2}\omega_{\p_3}\omega_{\p_4}}} \nn \\
  \H! & \times \Bigg\{\frac{1}{\mu^2-(p^{\prime}_1 + p^{\prime}_2 - p_1 - p_2)^2}\frac{1}{\mu^2-(p^{\prime}_1 + p^{\prime}_2)^2}\frac{1}{\mu^2-(p_1 + p_2)^2} \nn \\
  & \H! + \frac{1}{\mu^2-(p^{\prime}_1 + p^{\prime}_2 - p_1 - p_2)^2}\frac{1}{\mu^2-(p^{\prime}_1 - p_1)^2}\frac{1}{\mu^2-(p^{\prime}_2 - p_2)^2}\nn \\
  & \H! +  \frac{1}{\mu^2-(p^{\prime}_1 - p_1 - p_3 - p_4)^2}\frac{1}{\mu^2-(p^{\prime}_1 - p_1)^2}\frac{1}{\mu^2-(p_3 + p_4)^2}\nn \\
  & \H! +  \frac{1}{\mu^2-(p^{\prime}_2 - p_2 - p_3 - p_4)^2}\frac{1}{\mu^2-(p^{\prime}_2 - p_2)^2}\frac{1}{\mu^2-(p_3 + p_4)^2}\nn \\
  & \H! + \frac{1}{2} \left[\frac{1}{\mu^2-(p_1 + p_2 + p_3 + p_4)^2}\frac{1}{\mu^2-(p_1 + p_2)^2}\frac{1}{\mu^2-(p_3 + p_4)^2}\right]\Bigg\}.
\end{align}

In working out the variational derivative expressions similar to those in equations (\ref{EQ:VARIATIONAL1}) and (\ref{EQ:VARIATIONAL2}) are used, except that here there is no permutation index to keep track of. The relevant interaction kernels are given below:
\begin{multline}
  \cY_{2,2}(\p^\prime_{1,2}, \p_{1,2}) = -\frac{|C_F|^2 \, g^{2}}{8\,(2\pi)^{3}}\int d\p^{\prime}_{1,2} \,F(\p^\prime_{1,2}) \, \frac{\delta(\p^{\prime}_1 + \p^{\prime}_2 - \p_1 - \p_2)}{\ds\sqrt{\omega_{\p^{\prime}_1}\ds\omega_{\p^{\prime}_2}\ds\omega_{\p_1}\omega_{\p_2}}} \\
   \times \, \left[\frac{1}{\mu^{2}-(p^{\prime}_1-p_1)^{2}}+\frac{1}{\mu^{2}-(p^{\prime}_2-p_2)^{2}}+\frac{1}{\mu^{2}-(p_1+p_2)^{2}}+\frac{1}{\mu^{2}-(p^{\prime}_1+p^{\prime}_2)^{2}}\right], 
\end{multline}
\begin{multline}
  \cY_{2,4}(\p^\prime_{1..4}, \p_{1,2}) = - \frac{C_F^\ast C_G \, g^2}{4(2\pi)^3} \int d\p^\prime_{1..4} \, G(\p^\prime_{1..4})  \\ 
  \times \, \Bigg\{\frac{\de(\p^{\prime}_1-\p_1) \, \de(\p^\prime_2 + \p^\prime_3 + \p^\prime_4 - \p_2)}{\ds\sqrt{\omega_{\p^{\prime}_2}\omega_{\p^\prime_3}\omega_{\p^\prime_4}\omega_{\p_2}}} \, \left[\frac{1}{\mu^2-(p^{\prime}_2 - p_2)^2} + \frac{1}{\mu^2-(p^\prime_3 + p^\prime_4)^2} \right] \\
  \H! + \frac{\de(\p^{\prime}_2-\p_2) \, \de(\p^\prime_1 + \p^\prime_3 + \p^\prime_4 - \p_1)}{\ds\sqrt{\omega_{\p^{\prime}_1}\omega_{\p^\prime_3}\omega_{\p^\prime_4}\omega_{\p_1}}} \, \left[\frac{1}{\mu^2-(p^{\prime}_1 - p_1)^2} + \frac{1}{\mu^2-(p^\prime_3 + p^\prime_4)^2} \right] \Bigg\}, 
\end{multline}
\begin{align}
    \cC_{2,4}(\p^\prime_{1..4}, \p_{1,2}) = & -\frac{C_F^\ast C_G \, g^3 \, \eta}{4(2\pi)^6} \int d\p^\prime_{1..4} \; G(\p^\prime_{1..4}) \; \frac{\de(\p^{\prime}_1 + \p^{\prime}_2 + \p^\prime_3 + \p^\prime_4 - \p_1 - \p_2)}{\ds\sqrt{\omega_{\p^{\prime}_1}\omega_{\p^{\prime}_2}\omega_{\p^\prime_3}\omega_{\p^\prime_4}\omega_{\p_1}\omega_{\p_2}}} \nn \\
  \H! & \times \Bigg\{\frac{1}{\mu^2-(p^{\prime}_1 + p^{\prime}_2 - p_1 - p_2)^2}\frac{1}{\mu^2-(p^\prime_1 + p^\prime_2)^2}\frac{1}{\mu^2-(p_1 + p_2)^2} \nn \\
  & \H! + \frac{1}{\mu^2-(p^{\prime}_1 + p^{\prime}_2 - p_1 - p_2)^2}\frac{1}{\mu^2-(p^{\prime}_1 - p_1)^2}\frac{1}{\mu^2-(p^{\prime}_2 - p_2)^2}\nn \\
  & \H! +  \frac{1}{\mu^2-(p^{\prime}_1 + p^\prime_3 + p^\prime_4 - p_1)^2}\frac{1}{\mu^2-(p^{\prime}_1 - p_1)^2}\frac{1}{\mu^2-(p^\prime_3 + p^\prime_4)^2}\nn \\
  & \H! +  \frac{1}{\mu^2-(p^{\prime}_2 + p^\prime_3 + p^\prime_4 - p_2)^2}\frac{1}{\mu^2-(p^{\prime}_2 - p_2)^2}\frac{1}{\mu^2-(p^\prime_3 + p^\prime_4)^2}\nn \\
  & \H! + \frac{1}{2} \left[\frac{1}{\mu^2-(p^\prime_1 + p^\prime_2 + p^\prime_3 + p^\prime_4)^2}\frac{1}{\mu^2-(p^\prime_1 + p^\prime_2)^2}\frac{1}{\mu^2-(p^\prime_3 + p^\prime_4)^2}\right]\Bigg\}.
\end{align}
The remaining interaction kernels are non-vanishing; they are just not included here.

\subsection*{Derivation of the term $V_{2,4}^{Y}$}
The derivation of the $V_{2,4}^Y$ terms, as given by equation (\ref{EQ:V24Y}), begins with the substitution of equations (\ref{EQ:ANSATZ1}) and (\ref{EQ:ANSATZ2}) into the appropriate terms in equation (\ref{EQ:EQT_POS}). This substitution reduces to the following expression:
\begin{multline}
  V^{Y}_{2,4}(x_{12}) =  R \, \int d\x_{3,4} \; Y_{2,4}(\x_{1..4}) \frac{G(\x_{1..4})}{F(\x_{1,2})} \\
  = - \frac{R \, \alpha_g}{a^3} \int d\x_3 \, \left(\frac{\e^{-\mu \, |\x_{13}|}}{|\x_{13}|} + \frac{\e^{-\mu \, |\x_{23}|}}{|\x_{23}|} \right) \, \exp{\left(- \ds\frac{|\x_{13}| + |\x_{23}|}{a}\right)} .
\end{multline}

The dummy variable $\x_3$ can be integrated out by shifting the integration variable $\x_1 - \x_3 = \z$ and $\x_2 - \x_3 = \z $ in the first and second terms respectively:
\begin{equation}
  V^{Y}_{2,4}(x_{12}) = \, - \frac{2 \, R \, \alpha_g}{a^3} \int d\z \, \frac{\e^{-\mu |\z|}}{|\z|} \, \exp{\left(- \frac{|\z| + |\z + \x_{12}|}{a} \right)}.   
\end{equation}
It is already evident that $V^{Y}_{2,4}$ depends only on the inter-coordinate vectors. It is convenient to process in the spherical coordinates. The azimuthal integration is trivial and yields $2 \, \pi$. The polar integral can be calculated with the substitution: 
\begin{equation}
  z^2 + x^2 + 2 \, z \, x \, \alpha = \beta^2,
  \label{POLAR_SUBS}
\end{equation}
where $z = |\z|$ is the radial coordinate, $x = |\x_{12}|$ is the length of the separation vector and $\alpha$ is the cosine of the polar angle. Upon integration the new variable $\beta$, one obtains the following radial integral: 
\begin{multline}
  V^{Y}_{2,4}(x_{12}) = - \frac{4 \, \pi \, R \, \alpha_g}{a^3 \, x} \int dz \, \e^{-\mu \, z} \, \exp{\left(- \frac{z}{a}\right)} \\ 
  \times \, \Bigg\{- a \, (\beta_2 + a) \, \exp{\left(-\frac{\beta_2}{a}\right)} + a \, (\beta_1 + a) \, \exp{\left(-\frac{\beta_1}{a}\right)}\Bigg\}
\end{multline}
where $\beta_1 = |z - x|$ and $\beta_2 = z + x$, are the familiar quantities from the previous calculations. To perform the radial integral in the parameter $z$, one has to split the interval of integration into two intervals in order to account for the absolute value in $\beta_1$, i.e. $\ds\int^\infty_0 = \int^x_0 + \int^\infty_x$. The radial integration is straightforward, although tedious, and can be calculated with Maple. The result comes out to be 
\begin{equation}
  V^{Y}_{2,4}(x_{12}) = - \frac{8 \, \pi \, R \, \e^{- x_{12} \, ( \mu \, a + 1 ) / a} }{x_{12} \, a^2 \, \mu^2 \, (\mu \, a + 2)^2} \, 
  \, (a \, x_{12} \, \mu^2 \, \e^{x_{12} \, \mu} + 2 \, x_{12} \, \mu \, \e^{x_{12} \, \mu} - 2 \, \e^{x_{12} \, \mu} + 2),
\end{equation}
where, recall, the parameter $a$ and $R$ have to be determined through a variational calculation. 
%
\subsection*{Derivation of the term $V_{2,4}^C$}
The derivation of the $V_{2,4}^C$ terms, as given by equation (\ref{EQ:V24C}), begins with the substitution of equations (\ref{EQ:ANSATZ1}) and (\ref{EQ:ANSATZ2}) into the appropriate term in equation (\ref{EQ:EQT_POS}). This substitution reduces the term to the following expression:
\begin{multline}
  V^C_{2,4}(x_{12}) = R \, \int d\x_{3,4} \, C_{2,4}(\x_{1..4}) \, \frac{G(\x_{1..4})}{F(\x_{1,2})} \\
  =  \frac{R \, \alpha_\eta}{a^3} \int d\x_3 \, d\q_{1,2} \, \frac{\e^{-i \, \x_3 \cdot (\q_1 + \q_2)} \e^{ i \, \x_1 \cdot \q_1} \e^{i \, \x_2 \cdot \q_2}}{\left(\mu^2 + \left(\q_1 + \q_2\right)^2\right) \left(\mu^2 + \q_1^2\right) \left(\mu^2 + \q_2^2\right)} \, \exp{\left(- \frac{|\x_{13}| + |\x_{23}|}{a}\right)}.
  \label{EQ:POT_CUB1}
\end{multline}
The momentum integrals in the expression above can be solved in the Gaussian parametrization. Leaving out the constants in front and using equation (\ref{EQ:WILSON}) in equation (\ref{EQ:POT_CUB1}), one obtains
\begin{multline}
  V^C_{2,4} ~ \sim \int d\x_3 \, d\q_{1,2} \, d\beta_{1,2,3} \, \e^{-i \, \q_1 \cdot\x_{31} } \, \e^{-i \, \q_2 \cdot \x_{32}}  \\
  \times \e^{- (\mu^2 + (\q_1 + \q_2)^2) \, \beta_1} \, \e^{- (\mu^2 + \q_1^2) \, \beta_2} \,  \e^{- (\mu^2 + \q_2^2) \, \beta_3} \, \exp{\left(- \frac{|\x_{13}| + |\x_{23}|}{a}\right)},
\end{multline}
where $\x_{ij} = \x_i - \x_j$ are the inter-particle vectors. Upon expanding the squares in the exponential and defining $\q = \q_1 + \q_2$, a six dimensional Gaussian integral is identified with the following parameters in block diagonal notation:
\begin{gather}
  K = \left[
    \begin{array}{cc}
      2(\beta_{1}+\beta_{2}) & 2\beta_{1}  \\
      2\beta_{1} & 2(\beta_{1}+\beta_{3})  \\ \end{array}
    \right] , \;  
  K^{-1}=\frac{1}{4 \, \beta_{123}}\left[
    \begin{array}{cc}
      2(\beta_{1}+\beta_{3}) & -2\beta_{1}  \\
      -2\beta_{1} & 2(\beta_{1}+\beta_{2})  \\ \end{array}
    \right], 
  \label{EQ:1}  \\
  L=  \; i \, \left[
    \begin{array}{cc}
      \x_{31}, & \x_{21} \end{array}
    \right],\F!
  W = \left(\beta_{1}+\beta_{2}+\beta_{3}\right)\mu^{2},
  \label{EQ:2}
\end{gather}
where $\beta_{123} = \beta_{1}\beta_{2}+\beta_{1}\beta_{3}+\beta_{2}\beta_{3}$. Applying equation (\ref{EQ:GAUSS}) and making the change of variable $\x_3 = \z + \x_1$ leads to the following expression:
\begin{align}
  V^C_{2,4}(x_{12}) \sim  & \int  \frac{d\z \, d\beta_{1,2,3}}{\beta_{123}^{3/2}} \, \e^{-\mu^2 \, (\beta_1 + \beta_2 + \beta_3)} \nn \\
    & \times\, \exp{\left(-  \frac{\beta_1 \, \x_{12}^2 + \beta_2 \, \z^2 + \beta_3 \, (\z + \x_{12})^2}{4 \, \beta_{123}} -\frac{|\z| + |\z + \x_{12}|}{a}\right)}.
    \label{EQ:POT_C24}
\end{align}
Already in its present form, it is evident that $V^C_{2,4}$ depends only on the inter-coordinate vector $\x_{12}$ and no angles. 

As before, length is expressed in units of the Bohr radius $\ds\frac{1}{m \, \alpha_g}$, and, correspondingly, energy in units of $m \, \alpha_g^2$. Consequently, the contribution of $V_{2,4}^C$ can be written in terms of the dimensionless variables $\r = \x_{12}\, m\, \alpha_g$ and $\w = \z \, m \, \alpha_g$, and the dimensionless variational parameter $A = a \, m \, \alpha_g$. Reinserting the constants in front and trivially integrating out the azimuthal coordinate of $\w$, one obtains:
\begin{align}
 V^C_{2,4}(x_{12}) = & - \, \frac{2 \, \pi \, R \, \alpha_\eta}{A^3}  \int \frac{d\w \, d\beta_{1,2,3} }{\beta_{123}^{3/2}} \, \e^{-\mu^2 \, (\beta_1 + \beta_2 + \beta_3)} \nn \\
  & \times \, \exp{\left(- \frac{\beta_1 \, \r^2 + \beta_2 \, \w^2 + \beta_3 \, (\w + \r)^2}{4 \, \beta_{123}} -\frac{|\w| + |\w + \r|}{A}\right)}.
  \label{EQ:POT_C24_DIM}  
\end{align}
There remains a five dimensional integral to perform which has to be numerically.

\chapter{Appendix B: QED and QCD}
\section{Particle-Antiparticle State in QED}
\label{SEC:QED_two}
This section is reserved for the details involved in and related to calculations dealing with Dirac spinors. 

The following multiplication identities involving a single type of the Dirac spinor and the gamma matrices have been used:
\begin{align}
  \ub(\p, s) \, \gamma_0 \, u(\pp, \sigma) = \, & \ds\sqrt{\frac{(\omega_\p + m) \,  (\omega_{\pp} + m)}{4 \, m^2}} \, \chi^\dagger_{s} \left[\frac{\bs\sigma\cdot\pp}{(\omega_{\pp} + m)} + \frac{\bs\sigma\cdot\p}{(\omega_\p + m)} \right] \chi_{\sigma}, \\
  \ub(\p, s) \, \gamma_i \, u(\pp, \sigma) = \, & \ds\sqrt{\frac{(\omega_\p + m) \, (\omega_{\pp} + m)}{4 \, m^2}} \, \chi^\dagger_{s} \, \left[\frac{\sigma_i \, \bs\sigma\cdot\pp}{(\omega_{\pp} + m)} + \frac{\bs\sigma\cdot\p \, \sigma_i}{(\omega_\p + m)}\right] \chi_{\sigma}, \\
  \vb(\p^\prime, s) \, \gamma_0 \, v(\p, \sigma) = \, & \ds\sqrt{\frac{{(\omega_\p + m) \, (\omega_{\pp} + m)}}{4 \, m^2}} \; \eta^\dagger_{s} \left[\frac{\bs\sigma\cdot\pp}{ (\omega_{\pp} + m)} + \frac{\bs\sigma\cdot\p}{(\omega_\p + m)} \right] \eta_{\sigma}, \\
  \vb(\p^\prime, s) \, \gamma_i \,  v(\p, \sigma) =  \, &  \ds\sqrt{\frac{(\omega_{\pp} + m)(\omega_{\p} + m)}{4 \, m^2}} \; \eta^\dagger_{s} \left[\frac{ \bs\sigma\cdot\pp \, \sigma_i}{(\omega_{\pp} + m)} + \frac{\sigma_i, \bs\sigma\cdot\p}{(\omega_\p + m)}\right] \eta_{\sigma}.
\end{align}
The set of identities below involves both types of the Dirac spinors together with the gamma matrices:
\begin{align}
  \ub(\p, s) \, \gamma^0 \, v(\pp, \sigma) = & \, \sqrt{\frac{(\omega_\p + m)(\omega_{\pp} + m)}{4 \, m^2}} \, \chi_{s} \left[ \frac{\bs\sigma\cdot\pp}{(\omega_\pp + m)} + \frac{\bs\sigma\cdot\p}{(\omega_\p + m)}\right]\eta_{\sigma}, \\
  \ub(\p, s) \, \gamma^i \, v(\p, \sigma) = & \, \sqrt{\frac{(\omega_\p + m)(\omega_{\pp} + m)}{4 \, m^2}} \, \chi_{s} \left[ \sigma_i + \frac{\bs\sigma\cdot\pp \, \sigma_i \, \bs\sigma\cdot\p}{(\omega_\pp + m) \, (\omega_\p + m)} \right] \eta_{\sigma}.
\end{align}
The complex conjugates of these equations are straightforward to obtain. 

In the strict non-relativistic limit all of the above expressions reduce to
\begin{align}
  \ub(\p, s) \, \gamma^0 \, u(\p^\prime, \sigma) = & \, \de_{s \, \sigma}, \\
  \ub(\p, s) \, \gamma^i \, u(\p^\prime, \sigma) = & \, 0,  \\
  \vb(\pp, s) \, \gamma_0 \, v(\p, \sigma) = & \, \de_{s \, \sigma}, \\ 
  \vb(\pp, s) \, \gamma_i \, v(\p, \sigma) = & \, 0, \\
  \ub(\p, s) \, \gamma^0 \, v(\p, \sigma) = & \, 0, \\
  \ub(\p, s) \, \gamma^i \, v(\p, \sigma) = & \, \chi^\dagger_{s} \, \sigma_i \, \eta_{\sigma}.
\end{align}

The fundamental and defining relation of the Pauli matrices is
\begin{equation}
  \sigma_i \, \sigma_j = \, i \, \epsilon_{ijk} \, \sigma_k.
  \label{EQ:IDPAULI}
\end{equation}

The following, and frequently used, identity involves three Pauli matrices and can be obtained with a repeated use of equation (\ref{EQ:IDPAULI}):
\begin{equation}
  \sigma_j \, \sigma_j \, \sigma_k = \delta_{ij} \, \sigma_k - \delta_{ik} \, \sigma_j + \delta_{jk} \, \sigma_i + i \, \epsilon_{ijk}.
\end{equation}

\section{Three Quark Trial State}
\label{SEC:QCD_three}
The contents of this section pertain to the derivations related to the three quark trial state (\ref{EQ:TRIALQCDTHREE}). 

The matrix element in the Schr\"odinger picture of the trial state (\ref{EQ:TRIALQCDTHREE}) with the reformulated QCD Hamiltonian (\ref{EQ:QCDHAMREFORM}) is comprised of 
\begin{align}
  \bra \Psi_3 | \, H_R - E \, | \Psi_3 \ket =   \bra \Psi_3 | \, H_{\psi} - E \, | \Psi_3 \ket  + \bra \Psi_3 | \, H_{\psi A}^R \, | \Psi_3 \ket 
  + \bra \Psi_3 | \, H_{3A}^R  \, | \Psi_3 \ket, 
\end{align}
where, upon the integration over the momentum variables and summation over the spin and colour indices, the contributions are
\begin{align}
  \bra \Psi_3 |\, H_{\psi} - E \, | \Psi_3 \ket = & \, \int d\p_{1,2,3} \; F^{\ast}_{s,r,t}(\p_{1..3}) \, F_{s,r,t}(\p_{1..3}) \left(\omega^A_{\p_1} + \omega^B_{\p_2} + \omega^C_{\p_3} - E\right), \\
  \bra \Psi_3 | \, H_{\psi A}^R \, | \Psi_3 \ket = & \; \frac{g^2}{2 \, (2\pi)^3} \sum_\tx{colour} \sum_\tx{spin}\int d\p_{1...6} \; F^{\ast}_{s_1, r_1, t_1}(\p_{1..3}) \, F_{s_2, r_2, t_2}(\p_{4..6}) \; \nn \\ 
  & \times \,  \Omega_{i_1, j_1, k_1} \, \Omega_{i_2, j_2, k_2}  \, (L_1 + L_2 + L_3 + L_4 + L_5 + L_6), \\
  \bra \Psi_3 | \, H_{3A}^R \, | \Psi_3 \ket = & \; \frac{g^4}{(2\pi)^6 \, i} \, f^{abc} \, m_A \, m_B \, m_C \sum_\tx{spin} \sum_{\tx{colour}} \int \frac{d\p_{1..6}}{(\omega^A_{\p_1}\omega^A_{\p_4}\omega^B_{\p_2}\omega^B_{\p_5}\omega^C_{\p_3}\omega^C_{\p_6})^{1/2}} \nn \\
  & \H! \times F^{\ast}_{s_1,r_1,t_1}(\p_{1..3}) \, F_{s_2,r_2,t_2}(\p_{4..6}) \; \Omega_{i_1, j_1, k_1} \, \Omega_{i_2, j_2, k_2} \nn \\
  & \H! \times \de(\p_1 + \p_2 + \p_3 - \p_4 -\p_5 - \p_6) \, \left(T_1 + T_2 + T_3 + T_4 + T_5 + T_6\right),
\end{align}
where, the $L$-type terms are
\begin{align}
  L_1 = & \; \de(\p_1 + \p_2 - \p_4 - \p_5) \, \de(\p_3 - \p_6) \, \frac{m_A m_B}{(\omega^A_{\p_1}\omega^A_{\p_4}\omega^B_{\p_2}\omega^B_{\p_5})^{1/2}} \nn \\  
    & \H! \times \ub(\p_1, s_1) \gamma_{\mu} u(\p_4, s_2) \ub(\p_2, r_1) \gamma^{\mu} u(\p_5, r_2) \, T^a_{i_1 i_2} T^a_{j_1 j_2} \, \frac{\de_{k_1 k_2} \, \de_{t_1 t_2}}{(p_2^B - p_5^B)^2}, \\
  L_2 = & \; \de(\p_1 + \p_2 - \p_4 - \p_5) \, \de(\p_3 - \p_6) \, \frac{m_A m_B}{(\omega^A_{\p_1}\omega^A_{\p_4}\omega^B_{\p_2}\omega^B_{\p_5})^{1/2}} \nn \\ 
    & \H! \times \ub(\p_2, r_1) \gamma_{\mu} u(\p_5, r_2) \ub(\p_1, s_1) \gamma^{\mu} u(\p_4, s_2) \, T^a_{j_1 j_2} T^a_{i_1 i_2} \, \frac{\de_{k_1 k_2} \, \de_{t_1 t_2}}{(p_1^A - p_4^A)^2}, \\
  L_3 = & \; \de(\p_1 + \p_3 - \p_4 - \p_6) \, \de(\p_2 - \p_5) \, \frac{m_A m_C}{(\omega^A_{\p_1}\omega^A_{\p_4}\omega^C_{\p_3}\omega^C_{\p_6})^{1/2}} \nn \\ 
    & \H! \times\ub(\p_1, s_1) \gamma_{\mu} u(\p_4, s_2) \ub(\p_3, t_1) \gamma^{\mu} u(\p_6, t_2) \, T^a_{i_1 i_2} T^a_{k_1 k_2} \, \frac{\de_{j_1 j_2} \, \de_{r_1 r_2}}{(p_3^C - p_6^C)^2}, \\
  L_4 = & \; \de(\p_1 + \p_3 - \p_4 - \p_6) \, \de(\p_2 - \p_5) \, \frac{m_A m_C}{(\omega^A_{\p_1}\omega^A_{\p_4}\omega^C_{\p_3}\omega^C_{\p_6})^{1/2}} \nn \\ 
    & \H! \times\ub(\p_3, t_1) \gamma_{\mu} u(\p_6, t_2) \ub(\p_1, s_1) \gamma^{\mu} u(\p_4, s_2) \, T^a_{k_1 k_2} T^a_{i_1 i_2} \, \frac{\de_{j_1 j_2} \, \de_{r_1 r_2}}{(p_1^A - p_4^A)^2}, \\
  L_5 = & \;  \de(\p_2 + \p_3 - \p_5 - \p_6) \, \de(\p_1 - \p_4) \, \frac{m_B m_C}{(\omega^B_{\p_2}\omega^B_{\p_5}\omega^C_{\p_3}\omega^C_{\p_6})^{1/2}} \nn \\ 
    & \H! \times\ub(\p_2, r_1) \gamma_{\mu} u(\p_5, r_2) \ub(\p_3, t_1) \gamma^{\mu} u(\p_6, t_2) \, T^a_{j_1 j_2} T^a_{k_1 k_2} \, \frac{\de_{i_1 i_2} \, \de_{s_1 s_2}}{(p_3^C - p_6^C)^2}, 
\end{align}
\begin{align}
  L_6 = & \; \de(\p_2 + \p_3 - \p_5 - \p_6) \, \de(\p_1 - \p_4) \, \frac{m_B m_C}{(\omega^B_{\p_2}\omega^B_{\p_5}\omega^C_{\p_3}\omega^C_{\p_6})^{1/2}} \nn \\ 
    & \H! \times\ub(\p_3, t_1) \gamma_{\mu} u(\p_6, t_2) \ub(\p_2, r_1) \gamma^{\mu} u(\p_5, r_2) \, T^a_{k_1 k_2} T^a_{j_1 j_2} \, \frac{\de_{i_1 i_2} \, \de_{s_1 s_2}}{(p_2^B - p_5^B)^2}. 
\end{align}
%
The colour factors are calculated, for instance, as follows:
\begin{align}
  \Omega^{\ast}_{i_1, j_1, k_1} \, \Omega_{i_2, j_2, k_2} \,  T^a_{i_1 i_2} T^a_{j_1 j_2} \, \de_{k_1 k_2} = & \, \frac{1}{6} \, \epsilon_{i_1, j_1 k_1} \, \epsilon_{k_1 j_2 i_2} \; T^a_{i_1 i_2} T^a_{j_1 j_2} \nn\\
  = & \, \frac{1}{6} \left( \de_{j_1 j_2} \, \de_{i_1 i_2} - \de_{j_1 i_2} \, \de_{i_1 j_2}\right) \, T^a_{i_1 i_2} T^a_{j_1 j_2} \nn \\
  = & \, \frac{1}{6} \left( T^a_{i_2 i_2} T^a_{j_2 j_2} - T^a_{j_2 i_2} T^a_{i_2 j_2} \right) = - \frac{1}{6} \, \tx{tr} \, \left[ T^a \, T^a \right] = - \frac{4}{3},
\end{align}
where the fact that the generators are traceless and the following identity have been used:
\begin{align}
  \epsilon_{ijk} \, \epsilon_{imn} = \de_{jm} \, \de_{kn} - \de_{jn} \, \de_{km}.
\end{align}
%
The $T$-type terms are
\begin{align}
  T_1 = & \, \ub(\p_1, s_1) \gamma_\nu u(\p_4, s_2) \ub(\p_2, r_1) \gamma^\mu u(\p_5, r_2) \ub(\p_3, t_1) \gamma^\nu u(\p_6, t_2) \nn \\
  & \times \! \frac{(p_1^A- p_4^A)_\mu}{(p_1^A- p_4^A)^2 (p_2^B - p_5^B)^2 (p_3^C - p_6^C)^2} \; T^a_{i_1 i_2} \, T^b_{j_1 j_2} \,  T^c_{k_1 k_2},  \\
  T_2 = & \, \ub(\p_1, s_1) \gamma_\nu u(\p_4, s_2) \ub(\p_3, t_1) \gamma^\mu u(\p_6, t_2) \ub(\p_2, r_1) \gamma^\nu u(\p_5, r_2) \nn \\
  & \times \! \frac{(p_1^A- p_4^A)_\mu}{(p_1^A- p_4^A)^2 (p_3^C - p_6^C)^2 (p_2^B - p_5^B)^2} \; T^a_{i_1 i_2} \, T^b_{k_1 k_2} \,  T^c_{j_1 j_2}, \\ 
  T_3 = & \, \ub(\p_2, r_1) \gamma_\nu u(\p_5, r_2) \ub(\p_1, s_1) \gamma^\mu u(\p_4, s_2) \ub(\p_3, t_1) \gamma^\nu u(\p_6, t_2) \nn \\
  & \times \! \frac{(p_2^B - p_5^B)_\mu}{(p_2^B - p_5^B)^2 (p_1^A- p_4^A)^2 (p_3^C - p_6^C)^2} \; T^a_{j_1 j_2} \, T^b_{i_1 i_2} \,  T^c_{k_1 k_2}, \\
  T_4 = & \, \ub(\p_2, r_1) \gamma_\nu u(\p_5, r_2) \ub(\p_3, t_1) \gamma^\mu u(\p_6, t_2) \ub(\p_1, s_1) \gamma^\nu u(\p_4, s_2) \nn \\
  & \times \! \frac{(p_2^B - p_5^B)_\mu}{(p_2^B - p_5^B)^2 (p_1^A- p_4^A)^2 (p_3^C - p_6^C)^2} \; T^a_{j_1 j_2} \, T^b_{k_1 k_2} \, T^c_{i_1 i_2}, \\
  T_5 = & \, \ub(\p_3, t_1) \gamma_\nu u(\p_6, t_2) \ub(\p_1, s_1) \gamma^\mu u(\p_4, s_2) \ub(\p_2, r_1) \gamma^\nu u(\p_5, r_2) \nn \\
  & \times \! \frac{(p_3^C - p_6^C)_\mu}{(p_3^C - p_6^C)^2 (p_1^A- p_4^A)^2 (p_2^B - p_5^B)^2} \; T^a_{k_1 k_2} \, T^b_{i_1 i_2} \,  T^c_{j_1 j_2}, 
\end{align}
\begin{align}
  T_6 = & \, \ub(\p_3, t_1) \gamma_\nu u(\p_6, t_2) \ub(\p_2, r_1) \gamma^\mu u(\p_5, r_2) \ub(\p_1, s_1) \gamma^\nu u(\p_4, s_2) \nn \\
  & \times \! \frac{(p_3^C - p_6^C)_\mu}{(p_3^C - p_6^C)^2 (p_2^B - p_5^B)^2 (p_1^A- p_4^A)^2} \; T^a_{k_1 k_2} \, T^b_{j_1 j_2} \,  T^c_{i_1 i_2}. 
\end{align}

\section{Products of $SU(2)$ and $SU(3)$ Representations}
\label{SEC:YOUNG}
The Young tableaux method provides a way to find the dimensions of products of group representations. The general rules of the Young tableaux method for particle physicists are described in the book by Georgi~\cite{Georgi1999}. In this section, the Young tableaux for the product of $SU(2)$ and $SU(3)$ representations along with the corresponding dimensions of the resulting representations are shown. The products of the $SU(2)$ representations refer to the addition of spin while those of $SU(3)$ refer to colour.

The Young tableaux and the dimensions of the representations corresponding to the product of the fundamental and conjugate $SU(2)$ representations are
\begin{align}
 \Yvcentermath1 \; \yng(1) \; \otimes \; \yng(1) \; = & \; \Yvcentermath1 \; \yng(2) \; \oplus \; \yng(1,1) \nn \\
 2 \; \otimes \; \bar{2} \; = \; & \; 3 \; \oplus \; 1. 
\end{align}
Adding a fundamental $SU(2)$ representation to the above yields:
\begin{align}
  \Yvcentermath1\; \left( \; \yng(2) \; \oplus \; \yng(1,1) \;\, \right) \; \otimes \; \yng(1) \; = & \; \Yvcentermath1 \; \yng(2,1) \; \oplus \; \yng(2,1) \; \oplus \; \yng(3) \nn \\
  2 \; \otimes \; \bar{2} \; \otimes 2 \; = \; & \; 2 \; \oplus \; 2 \; \oplus \; 4.
\end{align}
Lastly, adding a conjugate representation to the the above yields:
\begin{multline}
  \Yvcentermath1 \; \left( \;\, \yng(2,1) \; \oplus \; \yng(2,1) \; \oplus \; \yng(3) \;\, \right) \; \otimes \; \yng(1) \; \\ =  \;  \Yvcentermath1 \; \yng(2,2) \; \oplus \; \yng(2,2) \; \oplus \yng(3,1) \; \oplus \; \yng(3,1) \; \oplus \; \yng(3,1) \; \oplus \; \yng(4) \; \nn \\
\end{multline}
\begin{align}
  2 \; \otimes \; \bar{2} \; \otimes 2 \; \otimes \; \bar{2} \; = \;  & \; 1 \; \oplus \; 1 \; \oplus \; 3 \; \oplus \; 3 \; \oplus \; 3 \; \oplus 5. 
\end{align}

The Young tableaux and the dimensions of the res-presentations corresponding to the product of the fundamental and conjugate $SU(3)$ representations are
\begin{align}
 \Yvcentermath1 \; \yng(1) \; \otimes \; \young(a,b) \; = & \; \Yvcentermath1 \; \young(\hfil a,b) \; \oplus \; \young(\hfil ,a,b) \nn \\
 3 \; \oplus \; \bar{3} \; = \; & \; 8 \; \oplus \; 1. 
\end{align}
Adding a fundamental $SU(3)$ representation to the above yields:
\begin{align}
  \Yvcentermath1\; \left( \; \Yvcentermath1 \; \yng(2,1) \; \oplus \; \yng(1,1,1) \;\; \right) \; \otimes \; \yng(1) \; = & \; \Yvcentermath1 \; \yng(2,1,1) \; \oplus \; \yng(2,1,1) \; \oplus \; \yng(2,2) \; \oplus \; \yng(3,1) \nn \\
  3 \; \otimes \; \bar{3} \; \otimes \; 3 \;  = \; & \; 3 \; \oplus \; 3 \; \oplus \; 6 \; \oplus \; 15.
\end{align}
Lastly, adding a conjugate $SU(3)$ representation to the above yields:
\begin{align}
  & \Yvcentermath1 \; \left( \; \yng(2,1,1) \; \oplus \; \yng(2,1,1) \; \oplus \; \yng(2,2) \; \oplus \; \yng(3,1) \;\; \right) \; \otimes \; \young(a,b) \; \nn \\
  & \F! = \Yvcentermath1 \; 2 \;\; \young(\hfil\hfil,\hfil a,\hfil b) \; \oplus \; 4 \;\; \young(\hfil\hfil\hfil,\hfil a,b) \; \oplus \; \young(\hfil\hfil a,\hfil\hfil b) \; \oplus \; \young(\hfil\hfil\hfil a,\hfil,b) \; \oplus \; \young(\hfil\hfil\hfil a,\hfil b) \nn \\
  3 \; \otimes \; \bar{3} \; \otimes & \; 3 \; \otimes \bar{3} \; = \; 1 \; \oplus \; 1 \; \oplus \; 8 \; \oplus \; 8 \; \oplus \; 8 \; \oplus \; 8 \; \oplus \; 10 \; \oplus \; 10 \; \oplus \; 27.
\end{align}
The labels $a$ and $b$ are included because the representations associated with multiple boxes, in this case $\bar{3}$, must be incorporated with other representations with the assistance of these labels. This is described in detail in the book by Georgi~\cite{Georgi1999}.

The Young tableaux and the dimensions of the representations corresponding to the product of two fundamental $SU(3)$ representations are
\begin{align}
 \Yvcentermath1 \; \yng(1) \; \otimes \; \yng(1) \; = & \; \Yvcentermath1 \; \yng(2) \; \oplus \; \yng(1,1) \nn \\
 3 \; \otimes \; 3 \; = \; & \; 3 \; \oplus \; 6. 
\end{align}
Adding another fundamental $SU(3)$ representation to the above yields:
\begin{align}
  \; \Yvcentermath1 \; \left( \; \yng(2) \; \oplus \; \yng(1,1) \;\; \right) \; \otimes \; \yng(1) \; = \; & \Yvcentermath1 \; \yng(3) \; \oplus \; \yng(2,1) \; \oplus \; \yng(2,1) \; \oplus \; \yng(1,1,1) \; \nn \\
  3 \; \otimes \; 3 \; \otimes \; 3 \; = \; & \; 1 \; \oplus \; 8 \; \oplus \; 8 \; \oplus \; 10. 
\end{align}

\section{Multi-Component Quark-Antiquark Trial State}
\label{SEC:QQ}
The matrix element in the Schr\"odinger picture of the trial state (\ref{EQ:QCDTRIAL24}) with the reformulated Hamiltonian operator (\ref{EQ:QCDHAMREFORM}) is comprised of the following contributions:
\begin{align}
  & \bra \Psi_t | \, H_R - E \, | \Psi_t \ket = \, \bra \Psi_t | \, H_{\psi} - E \, | \Psi_t \ket  + \bra \Psi_2 | \, H_{\psi A}^R \, | \Psi_2 \ket + \bra \Psi_2 | \, H_{\psi A}^R \, | \Psi_4 \ket  + \bra \Psi_2 | \, H_{3A}^R  \, | \Psi_4 \ket \nn \\
  & \; + \bra \Psi_4 | \, H_{\psi A}^R \, | \Psi_4 \ket + \bra \Psi_4 | \, H_{\psi A}^R \, | \Psi_2 \ket  + \bra \Psi_4 | \, H_{3A}^R  \, | \Psi_2 \ket + \bra \Psi_4 | \, H_{3A}^R  \, | \Psi_4 \ket + \bra \Psi_4 | \, H_{4A}^R  \, | \Psi_4 \ket.
\end{align}
The free and the linear contributions to the matrix element are
\begin{align}
  \bra \Psi_{2} | & \, H_\psi - E \, | \Psi_{2} \ket = \, |C_F|^2 \, \sum_{\kappa_1 \lambda_1}\int d\p_{1,2} \, F^{\ast}_{\kappa_1, \lambda_1}(\p_{1,2}) \, F_{\kappa_1, \lambda_1}(\p_{1,2}) \left(\omega^A_{\p_1} + \omega^A_{\p_2} - E \right), \nn \\
  \bra \Psi_{2} | & \, H_\psi \,| \Psi_{2} \ket = \, - \frac{|C_F|^2 \,  g^2 \,  m_A^2}{2(2\pi)^3} \nn\\
  & \times \sum_{\kappa_{1,2} \; \lambda_{1,2}} \, \int \frac{d\p_{1..4} \,  \de(\p_1 + \p_2 - \p_3 - \p_4)}{\left(\omega^A_{\p_1} \omega^A_{\p_2} \omega^A_{\p_3} \omega^A_{\p_4} \right)^{1/2}} \, F^\ast_{\kappa_1, \lambda_1}(\p_{1,2}) F_{\kappa_2, \lambda_2}(\p_{3,4}) \nn \\
  & \times \left(\ub(\p_1, \kappa_1) \gamma_\mu u(\p_3, \kappa_2)\, \vb(\p_4, \lambda_2) \gamma^\mu v(\p_2, \lambda_1) \frac{\tx{Tr}(T^a T^a)}{(p_4 - p_2)^2} \right. \nn \\
  & \H! -\left. \ub(\p_1, \kappa_1) \gamma_\mu v(\p_2, \lambda_1)\, \vb(\p_4, \lambda_2) \gamma^\mu u(\p_3, \kappa_2) \frac{\tx{Tr}(T^a) \tx{Tr}(T^a)}{(p_1 + p_2)^2}\right. \nn \\
  & \H! -\left. \ub(\p_1, \kappa_1) \gamma_\mu v(\p_2, \lambda_1)\, \vb(\p_4, \lambda_2) \gamma^\mu u(\p_3, \kappa_2) \frac{\tx{Tr}(T^a) \tx{Tr}(T^a)}{(p_3 + p_4)^2}\right. \nn \\
  & \H! +\left. \vb(\p_4, \lambda_2) \gamma_\mu v(\p_2, \lambda_1)\, \ub(\p_1, \kappa_1) \gamma^\mu u(\p_3, \kappa_2) \frac{\tx{Tr}(T^a T^a)}{(p_3 - p_1)^2}\right),
\end{align}
where the two virtual annihilation terms are set to zero by the colour factors (i.e. the factor coming from the summation of the colour indices) since the $T$ matrices are traceless. The gluon exchange terms have the colour factor of $\tx{Tr}(T^a T^a) = \ds\frac{4}{3}$.

The contribution to the matrix element from the linear cross term is
\begin{align}
  \bra \Psi_2 | & \, H_\psi \, | \Psi_4 \ket = \,- \frac{C_F^\ast \, C_G \, g^2 \, m_A \, m_B}{2(2\pi)^3} \; \tx{Tr}\left(T^a T^a\right) \nn \\
  & \F! \times \sum_{\kappa_{1,2} \; \lambda_{1,2} \; \mu_2, \nu_2} \, \int d\p_{1..6} \, F^\ast_{\kappa_1, \lambda_1}(\p_{1,2}) G_{\kappa_2, \lambda_2, \mu_2, \nu_2}(\p_{3..6}) \nn \\
  & \times  \Bigg\{\frac{\de(\p_1 - \p_3 - \p_5 - \p_6) \, \de(\p_2 - \p_4) \, \de_{\lambda_1 \lambda_2}}{\left(\omega^A_{\p_1} \omega^A_{\p_3} \omega^B_{\p_5} \omega^B_{\p_6} \right)^{1/2}} \,  \nn \\
  & \F!\F! \times \ub(\p_1, \kappa_1) \gamma_\nu u(\p_3, \kappa_2) \, \vb(\p_6, \nu_2) \gamma^\nu u(\p_5, \mu_2) \, \left(\frac{1}{(p_1 - p_3)^2} - \frac{1}{(p_5 + p_6)^2}\right) \nn \\
& \H! + \frac{\de(\p_2 - \p_4 - \p_5 - \p_6) \, \de(\p_1 - \p_3) \, \de_{\kappa_1 \kappa_2}}{\left(\omega^A_{\p_2} \omega^A_{\p_4} \omega^B_{\p_5} \omega^B_{\p_6} \right)^{1/2}} \nn \\
  & \F!\F! \times \vb(\p_4, \lambda_2) \gamma_\nu v(\p_2, \lambda_1) \, \vb(\p_6, \nu_2) \gamma^\nu u(\p_5, \mu_2) \, \left(\frac{1}{(p_2 - p_4)^2} - \frac{1}{(p_5 + p_6)^2}\right) \Bigg\},
  \label{EQ:QCDY24}
\end{align}
where the colour factor is calculated as before.

The contribution to the matrix element from the cubic cross term is 
\begin{align}
  \bra & \Psi_2 | \, H_{3A} \, | \Psi_4 \ket  = \, \frac{i \, C_F^\ast \, C_G \, g^4 \, m_A^2 \, m_B}{(2\pi)^6} \nn \\
  & \, \times \sum_{\substack{ \kappa_{1,2} \; \lambda_{1,2} \\ \mu_2, \nu_2}} \, \int \frac{d\p_{1..6} \; \de(\p_1 + \p_2 - \p_3 - \p_4 - \p_5 - \p_6)}{\left(\omega_{\p_1}^A \omega_{\p_2}^A \omega_{\p_3}^A \omega_{\p_4}^A \omega_{\p_5}^B \omega_{\p_6}^B\right)^{1/2}} \; F^\ast_{\kappa_1, \lambda_1}(\p_{1,2}) \, G_{\kappa_2, \lambda_2, \mu_2, \nu_2}(\p_{3..6}) \nn \\
  & \times \Bigg\{(p_3 - p_1)_\mu \, \ub(\p_1, \kappa_1) \gamma_\nu u(\p_3, \kappa_2) \, \vb(\p_4, \lambda_2) \gamma^\mu v(\p_2, \lambda_1) \, \vb(\p_6, \nu_2) \gamma^\nu u(\p_5, \mu_2) \; f^{abc} \, \tx{Tr}(T^a T^c T^b) \nn \\
  & \H! + \, (p_3 - p_1)_\mu \, \ub(\p_1, \kappa_1) \gamma_\nu u(\p_3, \kappa_2) \, \vb(\p_6, \nu_2) \gamma^\mu u(\p_5, \mu_2) \, \vb(\p_4, \lambda_2) \gamma^\nu v(\p_2, \lambda_1) \; f^{abc} \, \tx{Tr}(T^a T^b T^c) \nn \\
  & \H! + \,  (p_4 - p_2)_\mu \, \vb(\p_4, \lambda_2) \gamma_\nu v(\p_2, \lambda_1) \, \vb(\p_6, \nu_2) \gamma^\mu u(\p_5, \mu_2) \, \ub(\p_1, \kappa_1) \gamma^\nu u(\p_3, \kappa_2) \; f^{abc} \, \tx{Tr}(T^a T^c T^b) \nn \\
  & \H! + \, (p_4 - p_2)_\mu \, \vb(\p_4, \lambda_2) \gamma_\nu v(\p_2, \lambda_1) \, \ub(\p_1, \kappa_1) \gamma^\mu u(\p_3, \kappa_2) \, \vb(\p_6, \nu_2) \gamma^\nu u(\p_5, \mu_2) \; f^{abc} \, \tx{Tr}(T^a T^b T^c) \nn \\
  & \H! + \, (p_5 + p_6)_\mu \, \vb(\p_6, \nu_2) \gamma_\nu u(\p_5, \mu_2) \, \ub(\p_1, \kappa_1) \gamma^\mu u(\p_3, \kappa_2) \, \vb(\p_4, \lambda_2) \gamma^\nu v(\p_2, \lambda_2) \; f^{abc} \, \tx{Tr}(T^a T^c T^b) \nn \\
  & \H! + \, (p_5 + p_6)_\mu \, \vb(\p_6, \nu_2) \gamma_\nu u(\p_5, \mu_2) \, \vb(\p_4, \lambda_2) \gamma^\mu v(\p_2, \lambda_1) \, \ub(\p_1, \kappa_1) \gamma^\nu u(\p_3, \kappa_2) \; f^{abc} \, \tx{Tr}(T^a T^b T^c) \Bigg\}\nn \\
  & \F!\F!\F!\F!\F!\F!\F!\F!\F!\F! \times \frac{1}{(p_1 - p_3)^2} \frac{1}{(p_2 - p_4)^2} \frac{1}{(p_5 + p_6)^2} 
  \label{EQ:QCDC24}
\end{align}

The colour factor, for instance, can be evaluated as follows:
\begin{align}
  N_C \equiv & \,  f^{abc} \, \tx{Tr}(T^a T^b T^c) \nn \\
  = & \, \frac{1}{2} \, f^{abc} \, \tx{Tr}(T^aT^bT^c + T^aT^bT^c) = \frac{1}{2} \, f^{abc} \, \tx{Tr}(T^aT^bT^c - T^aT^cT^b) \nn \\ 
  = & \, \frac{1}{2} \, f^{abc} \, \tx{Tr}(T^a\left[T^b, T^c\right]) = \frac{i}{2} \, f^{abc} \, \tx{Tr}(T^a \, f^{bcd} T^{d}) \nn \\
  = & \, \frac{i}{4} \, f^{abc} \, f^{bcd} \, \de^{ab} = \frac{i}{4} \, f^{abc} \, f^{abc}.
\end{align}
where the anti-symmetric property of $f^{abc}$ and the commutator definition of the generating matrices (\ref{EQ:SUNGENERATORS}) has been used.

Similarly, one can obtain the other index combination
\begin{align}
  f^{abc} \, \tx{Tr}(T^a T^c T^b) = - \, \frac{i}{4} \, f^{abc} f^{abc}.
\end{align}
Notice that a factor of $i$ in the matrix element (\ref{EQ:QCDC24}) and the $i$ in $N_C$ multiply out to give a real result as is expected in an energy calculation. The product $f^{abc} f^{abc}$ is a positive definite quantity whose value can be absorbed into the variational parameter $R = \ds\frac{C_F}{C_G}$. Therefore, it is actually not required to calculate $R$. 

The contributions $\bra \Psi_4 | \, H_{\psi A}^R \, | \Psi_2 \ket$ and $\bra \Psi_4 | \, H_{3A}^R \, | \Psi_2 \ket$ to the matrix element are just the complex conjugates of equations (\ref{EQ:QCDY24}) and (\ref{EQ:QCDC24}) respectively. Without actually making any effort one can see immediately that $\bra \Psi_4 | \, H_{3A}^R \, | \Psi_4 \ket$ vanishes because of the colour factor as in equations (\ref{EQ:QCDTHREEC33}). An extra delta function arising from the ``spectator'' quark would be of no consequence for the colour factor calculation in $\bra \Psi_4 | \, H_{3A}^R \, | \Psi_4 \ket$. The remaining contributions $\bra \Psi_4 | \, H_{\psi A}^R \, | \Psi_4 \ket$ and $\bra \Psi_4 | \, H_{4A}^R \, | \Psi_4 \ket$ are irrelevant and therefore are not included here. 

The calculation leading up to equation (\ref{EQ:EQFORSMALLF}) involves the non-relativistic versions of kernels (\ref{EQ:RELY22}) and (\ref{EQ:RELC24}) and the spin index function $\Theta$ of equation (\ref{EQ:SPINTHETA}). In particular, there are extensive summations over the spin and spinor indices. In order to perform them it is prudent to use Maple rather to do these summations by hand. The results are obtained in Maple and presented below where the $\sim$ symbol refers to the fact that only factor carrying the spin and spinor indices have been included.

The calculation involving the linear kernel $\cY_{2,2}$ is
\begin{align}
  \Theta_{\kappa_1, \lambda_1} & \, \left(\cY_{2,2}\right)^{\kappa_2, \lambda_2}_{\kappa_1, \lambda_1}(\p_{1..4}) \, \Theta_{\kappa_2, \lambda_2} \nn \\
  & \ds\sim \, \Theta_{\kappa_1, \lambda_1} \, \bigg\{ \ub\left(\p_1, \kappa_1\right) \, \gamma_\mu \, u\left(\p_3, \kappa_2\right) \; \vb\left(\p_4, \lambda_2\right) \, \gamma^\mu \, v\left(\p_2, \lambda_1\right) \bigg\} \, \Theta_{\kappa_2, \lambda_2}\nn \\
  & = \, \Theta_{\kappa_1, \lambda_1}  \, \bigg\{ u^{\dagger}(\p_1, \kappa_1) \, u(\p_3, \kappa_2) \, v^{\dagger}(\p_4, \lambda_2) \, v(\p_2, \lambda_1) \nn \\
   & \F!\F!\F!\F! - u^{\dagger}(\p_1, \kappa_1) \, \gamma^0 \, \gamma^i \, u(\p_3, \kappa_2) \; v^{\dagger}(\p_4, \lambda_2) \, \gamma^0 \, \gamma^i \, v(\p_2, \lambda_1) \bigg\} \, \Theta_{\kappa_2, \lambda_2}\nn \\
  & \approx \, \Theta_{\kappa_1, \lambda_1} \, \bigg\{ \delta_{\kappa_1 \, \kappa_2} \, \de_{\lambda_1, \lambda_2} \nn \\
  & \F!\F! - \frac{1}{4 \, m_A^2} \; \chi^{\dagger}_{\kappa_1} \left(\sigma_i \, \bs\sigma\cdot\p_3 + \bs\sigma\cdot\p_1 \, \sigma_i\right) \, \chi_{\kappa_2} \, \eta^\dagger_{\lambda_2} \, \left(\bs\sigma\cdot\p_4 \, \sigma_i + \sigma_i \, \bs\sigma\cdot\p_2\right) \, \eta_{\lambda_1} \bigg\} \; \Theta_{\kappa_2, \lambda_2} \nn \\
  & = \, 1 - \frac{\epsilon_{\kappa_1 \, \lambda_1} \, \epsilon_{\kappa_2 \, \lambda_2}}{8 \, m^2_A} \; \chi^\dagger_{\kappa_1} \left(\sigma_i \, \bs\sigma\cdot\p_3 + \bs\sigma\cdot\p_1 \, \sigma_i\right) \, \chi_{\kappa_2} \, \eta^\dagger_{\lambda_2} \, \left(\bs\sigma\cdot\p_4 \, \sigma_i + \sigma_i \, \bs\sigma\cdot\p_2\right) \, \eta_{\lambda_1} \nn \\
  & = \, 1 - \frac{1}{4 \, m^2_A} \Big( 3 \, \left(\p_1\cdot\p_2 + \p_3\cdot\p_4\right) - \left(\p_1\cdot\p_4 + \p_2\cdot\p_3 \right)\Big).
  \label{EQ:Y22_SPN}
\end{align}

The contribution involving summations in the kernel $\cY_{2,4}$ has two terms each distinguished by a Roman numeral subscript:
\begin{align}
  \ds\Bigl( \Theta_{\kappa_1, \lambda_1} & \left(\cY_{2,4}\right)^{\kappa_2, \lambda_2, \mu_2, \nu_2}_{\kappa_1, \lambda_1}(\p_{1..6}) \; \Xi_{\kappa_2, \lambda_2, \mu_2, \nu_2} \Bigr)_\tx{I}\nn \\
  & \ds \sim \, \Theta_{\kappa_1, \lambda_1} \; \bigg\{ \ub(\p_1, \kappa_1) \, \gamma_\nu \, u(\p_3, \kappa_2) \, \vb(\p_6, \nu_2) \, \gamma^\nu \, u(\p_5, \mu_2) \, \de_{\lambda_1, \lambda_2} \bigg\} \,  \Xi_{\kappa_2, \lambda_2, \mu_2, \nu_2} \nn \\
  & \ds = \, \Theta_{\kappa_1, \lambda_1} \bigg\{ u^{\dagger}(\p_1, \kappa_1) \,  u(\p_3, \kappa_2) \, v^{\dagger}(\p_6, \nu_2) \, u(\p_5, \mu_2) \nn \\
  & \F!\F! - u^{\dagger}(\p_1, \kappa_1) \, \gamma^0 \, \gamma^i \, u(\p_3, \kappa_2) \, v^{\dagger}(\p_6, \nu_2) \, \gamma^0 \, \gamma^i \, u(\p_5, \mu_2) \bigg\} \,  \, \de_{\lambda_1 \, \lambda_2} \,  \Xi_{\kappa_2 \, \lambda_2, \mu_2, \nu_2} \nn \\
  & = \, \Theta_{\kappa_1, \lambda_1} \; \Bigg\{ \frac{\de_{\lambda_1 \, \lambda_2} \, \de_{\kappa_1 \, \kappa_2}}{2 \, m_B} \; \eta^{\dagger}_{\nu_2} \left( \bs\sigma\cdot\p_6 + \bs\sigma\cdot\p_5\right) \, \chi_{\mu_2} \nn \\
  & \F!\F! - \frac{\de_{\lambda_1 \, \lambda_2} }{2 \, m_A} \, \chi^{\dagger}_{\kappa_1} \; \left(\sigma_i \, \bs\sigma\cdot\p_3 + \bs\sigma\cdot\p_1 \, \sigma_i\right) \; \chi_{\kappa_2} \nn\\
  & \F!\F!\F!\F! \times \eta^{\dagger}_{\nu_2} \, \left( \sigma_i + \frac{\bs\sigma\cdot\p_6 \, \sigma_i \, \bs\sigma\cdot\p_5}{4 \, m^2_B}\right) \, \chi_{\mu_2} \Bigg\} \; \Xi_{\kappa_2 \, \lambda_2, \mu_2, \nu_2} \nn \\
  & \ds = \; - \, \Theta_{\kappa_1, \lambda_1} \; \frac{\de_{\lambda_1 \, \lambda_2} }{8 \, m^2_B \, m_A} \, \chi^{\dagger}_{\kappa_1} \, \left(\sigma_i \, \bs\sigma\cdot\p_3 + \bs\sigma\cdot\p_1 \, \sigma_i\right) \, \chi_{\kappa_2} \nn \\
  & \F!\F!\F!\F! \times \eta^{\dagger}_{\nu_2} \,  \left( \bs\sigma\cdot\p_6 \, \sigma_i \, \bs\sigma\cdot\p_5\right) \,  \chi_{\mu_2} \; \Xi_{\kappa_2, \lambda_2, \mu_2, \nu_2} \nn \\
  & = - \frac{i}{96 \, m^2_B \, m_A} \Big(\p_3 \cdot (\p_6 \times \p_5) + \p_1 \cdot (\p_6 \times \p_5)\Big) + {\cal O}(m^{-4}),
  \label{EQ:ABCD1}
\end{align}
\begin{align}
  \ds\Bigl( \Theta_{\kappa_1, \lambda_1} & \left(\cY_{2,4}\right)^{\kappa_2, \lambda_2, \mu_2, \nu_2}_{\kappa_1, \lambda_1}(\p_{1..6}) \Xi_{\kappa_2, \lambda_2, \mu_2, \nu_2} \Bigr)_\tx{II} \nn \\
  & \sim \, \Theta_{\kappa_1, \lambda_1} \; \vb(\p_4, \lambda_2) \, \gamma_\nu \, v(\p_2, \lambda_1) \, \vb(\p_6, \nu_2) \, \gamma^\nu \, u(\p_5, \mu_2) \de_{\kappa_1, \kappa_2} \; \Xi_{\kappa_2, \lambda_2, \mu_2, \nu_2} \nn \\
  & \approx \,  - \, \Theta_{\kappa_1, \lambda_1} \, \frac{\de_{\lambda_1 \, \lambda_2} }{8 \, m^2_B \, m_A} \; \eta^{\dagger}_{\lambda_2} \, \left(\bs\sigma\cdot\p_4 \, \sigma_i + \sigma_i \, \bs\sigma\cdot\p_2 \right) \, \eta_{\lambda_1} \, \eta^{\dagger}_{\nu_2} \, \left( \bs\sigma\cdot\p_6 \, \sigma_i \, \bs\sigma\cdot\p_5\right) \, \chi_{\mu_2} \; \Xi_{\kappa_2, \lambda_2, \mu_2, \nu_2} \nn \\
  & = - \frac{i}{96 \, m^2_B \, m_A} \, \Big(\p_4 \cdot (\p_6 \times \p_5) + \p_2 \cdot (\p_6 \times \p_5)\Big) + {\cal O}(m^{-4}).
  \label{EQ:ABCD2}
\end{align}
where similar steps have been used to obtained equation (\ref{EQ:ABCD2}) as for (\ref{EQ:ABCD1}).

The contribution involving summations in the kernel $\cC_{2,4}$ has six terms each distinguished by a Roman numeral index:
\begin{align}
  \Bigl( \Theta_{\kappa_1, \lambda_1} &  \left(\cC_{2,4}\right)^{\kappa_2, \lambda_2, \mu_2, \nu_2}_{\kappa_1, \lambda_1}(\p_{1..6}) \; \Xi_{\kappa_2, \lambda_2, \mu_2, \nu_2} \Bigr)_\tx{I} \nn \\
  & \ds \sim \, \Theta_{\kappa_1, \lambda_1} \; (p_3 - p_1)_\mu \, \ub(\p_1, \kappa_1) \, \gamma_\nu \, u(\p_3, \kappa_2) \nn \\ 
  & \F!\F!\F! \times \vb(\p_4, \lambda_2) \, \gamma^\mu \, v(\p_2, \lambda_1) \; \vb(\p_6, \nu_2) \, \gamma^\nu \, u(\p_5, \mu_2) \; \Xi_{\kappa_2, \lambda_2, \mu_2, \nu_2} \nn \\
  & \approx \Theta_{\kappa_1, \lambda_1} \; \Biggl( (p_3 - p_1)_0 \, \delta_{\lambda_1, \lambda_2} - \frac{1}{2 \, m_A}(p_3 - p_1)_i \, \eta^{\dagger}_{\lambda_2} \left(\bs\sigma\cdot\p_4 \, \sigma_i + \sigma_i \, \bs\sigma\cdot\p_2 \right) \eta_{\lambda_1}  \Biggr) \nn \\
  & \F! \times \Biggl(\frac{\de_{\kappa_1 \, \kappa_2}}{2 \, m_B} \; \eta^{\dagger}_{\nu_2} \left( \bs\sigma\cdot\p_6 + \bs\sigma\cdot\p_5\right) \, \chi_{\mu_2} \nn \\
  & \F!\F! - \frac{1}{2 \, m_A} \; \chi^{\dagger}_{\kappa_1} \, \left(\sigma_i \, \bs\sigma\cdot\p_3 + \bs\sigma\cdot\p_1 \, \sigma_i\right) \, \chi_{\kappa_2} \nn \\
  & \F!\F!\F!\F!\F!\F! \times \eta^{\dagger}_{\nu_2} \, \left( \sigma_i + \frac{\bs\sigma\cdot\p_6 \, \sigma_i \, \bs\sigma\cdot\p_5}{4 \, m^2_B}\right) \, \chi_{\mu_2}\Biggr) \; \Xi_{\kappa_2, \lambda_2, \mu_2, \nu_2} \nn \\
  & = \frac{i}{2 \, m_A \, m_B} \, (\p_3 - \p_1) \cdot\Bigl((\p_5 + \p_6) \times (\p_4 -\p_2)\Bigr) \nn \\
  & \F! + \, \frac{i}{2 \, m^2_A} (\p_3 - \p_1) \cdot \Bigl((\p_3 - \p_1) \times (\p_4 - \p_2) - 2 \, \p_1 \times (\p_4 - \p_2)\Bigr) + {\cal O}(m^{-3}).
\end{align}
The remaining terms can be determined using analogous steps:
\begin{align}
  \Bigl(\Theta_{\kappa_1, \lambda_1} & \left(\cC_{2,4}\right)^{\kappa_2, \lambda_2, \mu_2, \nu_2}_{\kappa_1, \lambda_1}(\p_{1..6}) \; \Xi_{\kappa_2, \lambda_2, \mu_2, \nu_2} \Bigr)_{\tx{II}} \nn \\
  & = \frac{3 \, i}{2 \, m_B^2} \; (\p_3 - \p_1) \cdot (\p_5 \times \p_6) + \frac{i}{ 2 \, m_A^2} \; (\p_3 - \p_1) \cdot \Bigl( 3 \, (\p_3 \times \p_4) + (\p_2 \times \p_3) \nn \\ 
  & \F!\F!\F!\F!\F!\F!\F! - (\p_1 \times \p_2) - (\p_1 \times \p_4) \Bigr) + {\cal O}(m^{-3}),
\end{align}
\begin{align}
  \Bigl( \Theta_{\kappa_1, \lambda_1} & \left(\cC_{2,4}\right)^{\kappa_2, \lambda_2, \mu_2, \nu_2}_{\kappa_1, \lambda_1}(\p_{1..6}) \; \Xi_{\kappa_2, \lambda_2, \mu_2, \nu_2} \Bigr)_{\tx{III}} \nn \\
  & = \frac{3 \, i}{ 2 \, m_B^2} \; (\p_4 - \p_2) \cdot (\p_5 \times \p_6)  + \frac{i}{ 2 \, m_A^2} \; (\p_4 - \p_2) \cdot \Bigl( 3 \, (\p_3 \times \p_4) + (\p_2 \times \p_3) \nn \\
  & \F!\F!\F!\F!\F!\F!\F! - (\p_1 \times \p_2) - (\p_1 \times \p_4) \Bigr) + {\cal O}(m^{-3}),
\end{align}
\begin{align}
  \Bigl( \Theta_{\kappa_1, \lambda_1} & \left(\cC_{2,4}\right)^{\kappa_2, \lambda_2, \mu_2, \nu_2}_{\kappa_1, \lambda_1}(\p_{1..6}) \; \Xi_{\kappa_2, \lambda_2, \mu_2, \nu_2} \Bigr)_{\tx{IV}} \nn \\
  & = \frac{i}{2 \, m_A \, m_B} \, (\p_4 - \p_2) \cdot \Bigr( (\p_6 + \p_5) \times (\p_1 - \p_3) \Bigl) \nn \\
  & \F! + \frac{i}{2 \, m^2_A} \, (\p_4 - \p_2) \cdot \Bigl( - \, (\p_4 - \p_2) \times (\p_3 - \p_1) + 2 \, \p_2 \times (\p_3 - \p_1) \Bigr) + {\cal O}(m^{-3}),
\end{align}
\begin{align}
  \Bigl( \Theta_{\kappa_1, \lambda_1} & \left(\cC_{2,4}\right)^{\kappa_2, \lambda_2, \mu_2, \nu_2}_{\kappa_1, \lambda_1}(\p_{1..6}) \; \Xi_{\kappa_2, \lambda_2, \mu_2, \nu_2} \Bigr)_\tx{V} \nn \\
  & = \frac{3 \, i}{2 \, m_A \, m_B} \; (\p_2 + \p_4) \cdot (\p_5 \times \p_6) \nn \\
  & \F! + \frac{i}{2 \, m_A^2} \,(\p_1 - \p_3) \cdot \Bigl(2 \, \p_2 \times (\p_5 + \p_6) + (\p_2 - \p_4) \times (\p_5 + \p_6) \Bigr) + {\cal O}(m^{-3}),
\end{align}
\begin{align}
  \Bigl(  \Theta_{\kappa_1, \lambda_1} & \left(\cC_{2,4}\right)^{\kappa_2, \lambda_2, \mu_2, \nu_2}_{\kappa_1, \lambda_1}(\p_{1..6}) \; \Xi_{\kappa_2, \lambda_2, \mu_2, \nu_2} \Bigr)_\tx{VI} \nn \\
  & = \frac{3 \, i}{2 \, m_A \, m_B} \; (\p_1 + \p_3) \cdot (\p_5 \times \p_6) \nn \\
  & \F! + \frac{i}{2 \, m_A^2} \; (\p_2 - \p_4) \cdot \Bigl(- \, 2 \, \p_1 \times (\p_5 + \p_6) - (\p_1 - \p_3) \times (\p_5 + \p_6)\Bigr) + {\cal O}(m^{-3}).
\end{align}

\section{The Determination of the Optimal Variational Parameters}
\label{SEC:PARAMETER}
This section shows the details of the calculation in equation (\ref{EQ:ENERGYEXPRESSION}). Recall that the function $f$ as given by ansatz (\ref{EQ:HYDROGEN}), which is a properly normalized function. The calculation on the right hand side of (\ref{EQ:ENERGYEXPRESSION}) is straightforward:
\begin{align}
   \int d\p \, {\cal E}\,  f^\ast(\p) \, f(\p) \; = {\cal E},
\end{align}
and, the kinetic energy contribution, calculated in spherical coordinates, is
\begin{align}
  \int d\p \, f^\ast(\p) \, f(\p) \; \frac{\p^2}{m_A} = & \frac{32 \, a^3}{m_A \, \pi}   \int^\infty_0 dp \, p^2 \, \frac{p^2}{(p^2 \, a^2 + 1)^4} \nn \\  
  = & \, \frac{32}{m_A \, \pi \, a^2} \int^\infty_0 dp \, \frac{p^4}{(p^2 + 1)^4} = \frac{32}{m_A \, \pi \, a^2} \ds\left(\frac{\pi}{32}\right) = \frac{m_A \, \alpha_s^2}{A^2}, 
\end{align}
where the substitution for a dimensionless variable of integration has been made and the consequent radial integral can be calculated using software like Maple or Mathematica, and the constants are $A = a \, m_A \, \alpha_s$ and $\alpha_s = \ds\frac{g^2}{4 \, \pi}$.

The contribution to the energy from the linear interaction term $H^R_{\psi \, A}$ is divided into three terms ${\cal T}_1$, ${\cal T}_2$ and ${\cal T}_3$. The calculation of the first term is as follows:
\begin{align}
  {\cal T}_1 = & \, \int d\p_{1,3} \, f^\ast(\p_1) \, \Big\{\cY_{2,2}(\p_{1,3}) \Big\}_\tx{I} \, f(\p_3) \nn \\
  & = \, \frac{32 \, a^3 \, g^2}{3 \, \pi^2 \, (2 \, \pi)^3} \int d\p_{1,3} \frac{1}{(p^2_1 \, a^2 + 1)^2} \, \frac{1}{(p^2_3 \, a^2 + 1)^2} \, \frac{1}{(\p_3 - \p_1)^2} \nn \\
  & = \frac{256 \, a^3 \, \alpha_s}{3 \, (2 \, \pi)^3 } \, \int d\p_1 \, dp_3 \, p^2_3 \, \frac{1}{(p^2_1 \, a^2 + 1)^2} \, \frac{1}{(p^2_3 \, a^2 + 1)^2} \int^{+1}_{-1} d\mu \, \frac{1}{(p_3^2 + p_1^2 - 2 \, p_1 \, p_2 \, \mu)}.
\end{align}
The evaluation of the $\mu$ integral can be done with the substitution $v = p^2_3 + p^2_1 - 2 p_1 p_3 \mu$, where the limits of integration become: $v_1 = (p_3 + p_1)^2$ corresponds to $\mu = -1$ and $v_2 = (p_3 - p_1)^2$ corresponds to $\mu = +1$. The expression, upon the integration over the variable $\mu$ and the angular coordinates of the vector $\p_1$, becomes:
\begin{align}
  {\cal T}_1 = & \, \frac{1024 \, a^3 \, \alpha_s}{3 \, (2 \, \pi)^2 } \int dp_{1,3} \; p_1 \, p_3 \; \frac{1}{(p_1^2 \, a^2 + 1)^2} \, \frac{1}{(p_3^2 \, a^2 + 1)^2} \, \ds\ln\left|\frac{p_3 + p_1}{p_3 - p_1}\right| \nn \\
  = & \; \frac{1024 \, \alpha_s}{3 \, a \, (2 \, \pi)^2 } \int dp_1 \, dp_3 \; \frac{p_1 \, p_3}{(p^2_1 + 1)^2 (p^2_3 + 1)^2} \, \ds\ln\left|\frac{p_3 + p_1}{p_3 - p_1}\right| \nn \\
  = & \; \frac{128 \, \alpha_s}{3 \, a \, \pi}\int dp_1 \; \frac{p^2_1}{3 \, p^2_1 + 3 \, p_1^4 + p_6 + 1} = \frac{4 \, \alpha_s}{3 \, a} = \frac{4}{3}\, \frac{m_A \, \alpha_s^2}{A},
  \label{EQ:QCDT1}
\end{align}
where the substitution for dimensionless variables of integration have been made and the radial integrals over the variables $p_1$ and $p_3$ have been found in Maple. 

The second terms ${\cal T}_2$ is calculated as follows:
\begin{align}
  {\cal T}_2 = & \, \int d\p_{1,3} \, f^\ast(\p_1) \, \Big\{\cY_{2,2}(\p_{1,3}) \Big\}_\tx{II} \, f(\p_3) \nn \\
  = & \, \frac{4 \, a^3 \, \alpha_s}{3 \, m^2_A \, \pi^4} \int d\p_1 \; \frac{p^2_1}{(p_1^2 \, a^2 + 1)^2} \int d\p_3 \; \frac{p^2_3}{(p_3^2 \, a^2 + 1)^2} \nn \\
  = & \, \frac{64 \, \alpha_s}{3 \, m^2_A \, a^3 \, \pi^2} \int dp_1 \; \frac{p_1^2}{(p_1^2 + 1)^2} \int dp_3 \;\frac{p_3^2}{(p_3^2 + 1)^2} = \frac{4}{3} \, \frac{\alpha_s}{m_A^2 \, a^3} = \frac{4}{3}\frac{m_A \,\alpha_s^4}{A^3},
\end{align}
where, once again, one can easily evaluate the radial integrals over the variables $p_1$ and $p_3$ in Maple. 

Finally, the calculation of the third term ${\cal T}_3$ is as follows:
\begin{align}
  {\cal T}_3 = & \, \int d\p_{1,3} \, f^\ast(\p_1) \, \Big\{\cY_{2,2}(\p_{1,3}) \Big\}_\tx{III} \, f(\p_3) \nn \\
  = & \; \frac{8 \, a^3 \, \alpha_s}{3 \, m^2_A \, \pi^4} \int d\p_{1,3} \; \frac{p^2_1}{(p_1^2 \, a^2 + 1)^2} \, \frac{p^2_3}{(p_3^2 \, a^2 + 1)^2} \, \frac{(p_1^2 + p_3^2)}{(\p_3 - \p_1)^2} \nn \\
  = & \; \frac{64 \, a^3 \, \alpha_s}{3 \, m^2_A \, \pi^2} \int dp_{1,3} \; p_1 \, p_3 \; \frac{(p_1^2 + p_3^2)}{(p^2_1 + 1)^2 (p^2_3 + 1)^2} \, \ds\ln\left|\frac{p_3 + p_1}{p_3 - p_1}\right| \nn \\
  = & \; \frac{4}{3} \, \frac{16 \, C_1}{m_A^2 \, a^3 \, \pi^2} = \frac{4}{3} \, m_A \, \alpha_s^4 \, \frac{16 \, C_1}{\pi^2},
  \label{EQ:QCDT3}
\end{align}
where similar steps as those leading up to equation (\ref{EQ:QCDT1}) has been taken to obtain this result, and the evaluation of the double-quadrature in Maple yields the constant
\begin{align}
  C_1 = \int dp\, dq \; \frac{p\, q \, (p^2 + q^2)}{(p^2 + 1)^2 \, (q^2 + 1)^2} \; \ds\ln\bigg|\frac{p+q}{p-q}\bigg| \approx 1.85055.
\end{align}
%

The contribution to the energy from the cubic interaction term $H^R_{3 \, \psi}$ is discussed in the course of the dissertation.
%


\spacing{1}
\bibliographystyle{ieeetr} 
\bibliography{references} 

\end{document}